\documentclass{emulateapj}

\usepackage{amsmath}

\def\H2{{{\rm H}_2}}

\def\CO{{\rm CO}}
\def\UMW{U_{\rm MW}}
\def\MW{{\rm MW}}

\def\xm#1{ x_{#1} }
\def\Lsob{L_\mathrm{sob}}

\shorttitle{The X-factor dependence on environment and scale}
\shortauthors{Feldmann, Gnedin \& Kravtsov}

\begin{document}

%=================================================
\title{The X-Factor in galaxies: I. Dependence on environment and scale}
%=================================================

\author{Robert Feldmann\altaffilmark{1,2},
Nickolay Y. Gnedin\altaffilmark{1,2,3} and Andrey V. Kravtsov\altaffilmark{2,3,4}}  
\altaffiltext{1}{Particle Astrophysics Center, 
Fermi National Accelerator Laboratory, Batavia, IL 60510, USA; gnedin@fnal.gov}
\altaffiltext{2}{Kavli Institute for Cosmological Physics and Enrico
  Fermi Institute, The University of Chicago, Chicago, IL 60637 USA;
  andrey@oddjob.uchicago.edu} 
\altaffiltext{3}{Department of Astronomy \& Astrophysics, The
  University of Chicago, Chicago, IL 60637 USA} 
\altaffiltext{4}{Enrico Fermi Institute, The University of Chicago,
Chicago, IL 60637}

\begin{abstract}
Characterizing the conversion factor between $\CO$ emission and column density of molecular hydrogen, $X_\CO$, is crucial in studying the gaseous content of galaxies, its evolution, and relation to star formation.  In most cases the conversion factor is assumed to be close to that of giant molecular clouds (GMCs) in the Milky Way, except possibly for mergers and star-bursting galaxies. However, there are physical grounds to expect that it should also depend on the gas metallicity, surface density, and strength of the interstellar radiation field. The $X_\CO$ factor may also depend on the scale on which $\CO$ emission is averaged due to effects of limited resolution. 
We study the dependence of $X_\CO$ on gas properties and averaging scale using a model that is based on a combination of results of sub-pc scale magneto-hydrodynamic simulations and on the gas distribution from self-consistent cosmological simulations of galaxy formation. Our model predicts $X_\CO\approx 2-4\times 10^{20}\rm\ K^{-1}\,cm^{-2}\,km^{-1}\,s$, consistent with the Galactic value, for interstellar medium conditions typical for the Milky Way. For such conditions the predicted $X_\CO$ varies by only a factor of two for gas surfaced densities in the range $\Sigma_{\rm H_2}\sim 50-500\ \rm M_{\odot}\,pc^{-2}$. However, the model also predicts that more generally on the scale of GMCs, $X_\CO$ is a strong function of metallicity, and depends on the column density and the interstellar UV flux.  We show explicitly that neglecting these dependencies in observational estimates can strongly bias the inferred distribution of $\H2$ column densities of molecular clouds to have a narrower and offset range compared to the true distribution. We find that when averaged on $\sim$ kpc scales the X-factor depends only weakly on radiation field and column density, but is still a strong function of metallicity. The predicted metallicity dependence can be approximated as $X_\CO\propto Z^{-\gamma}$ with $\gamma\approx 0.5-0.8$. 
\end{abstract}


%----------------------
\section{Introduction}
\label{sect:intro}
%----------------------
Molecular hydrogen ($\H2$), the major constituent of cold clouds in the interstellar medium (ISM), is playing a major role in shaping the visible Universe around us. Cooling by $\H2$ aided the formation of the first stars (e.g., \citealt{2002Sci...295...93A,2002ApJ...564...23B}) and  star formation in nearby galaxies takes place in molecular and giant molecular clouds\footnote{The distinction between a molecular cloud and a giant molecular cloud is somewhat blurry and several non-equivalent definitions are in use in the literature. We will use the term ``molecular cloud'' to denote a molecular region of $\gtrsim{}1$ pc diameter within galaxies that is embedded in a non-molecular component of the ISM. A ``giant molecular cloud'' is then simply a large molecular cloud with a diameter of $\gtrsim{}50$ pc.} (GMCs). Furthermore, the properties of molecular gas in galaxies, e.g., its distribution, formation and destruction are intricately linked to many other phenomena occurring in galaxies, such as the formation of stars, various stellar feedback processes, and the physics of dust grains.

Unfortunately, detecting $\H2$ in emission is difficult because the lowest excited levels in the rotational ladder are hardly populated at the low temperatures ($\sim{}10$ K) of molecular clouds. Moreover, the $\H2$ molecule, lacking a permanent electric dipole moment, radiates via the much slower quadrupole transition (e.g., \citealt{1982ARA&A..20..163S,2005fost.book.....S}). Therefore, a practical solution has been to measure the emission from some other molecule, that is assumed to trace $\H2$, and convert its line intensity into an $\H2$ column density. The conversion factor is called the X-factor for the particular emission line. 

Cooling lines from the carbon monoxide ($\CO$) isotope $^{12}\CO$ are often used as $\H2$ tracers, because $^{12}\CO$ is the second most abundant molecule in molecular clouds and can be observed even in extragalactic objects at high redshifts \citep{1991AJ....102.1956B, 2005ARA&A..43..677S, 2006ApJ...640..228T, 2008ApJ...680..246T, 2010ApJ...713..686D, 2010ApJ...714L.118D, 2010Natur.463..781T, 2010MNRAS.407.2091G, 2011ApJ...734L..25E, 2011arXiv1106.2098G}. In this study we will focus on the $^{12}\CO$ $J=1\rightarrow{}0$ transition (and on the corresponding X-factor $X_\CO$), i.e., the transition from the first excited rotational level to the ground state, that is widely used in the literature, particularly in studies of nearby galaxies or molecular clouds in the Milky Way (\citealt{1970ApJ...161L..43W,1987ASSL..134...21S,1987ApJ...319..730S, 1991ARA&A..29..581Y, 1995ApJS...98..219Y, 2001ApJ...561..218R, 2003ApJS..145..259H, 2007PASJ...59..117K, 2007prpl.conf...81B}). 

$X_\CO$ has been measured for molecular clouds in the Milky Way and in a few other nearby galaxies with a variety of techniques. These methods compare the directly measured $\CO$ luminosity (or intensity) with an independently determined $\H2$ mass (or column density). The most common approaches: (i) estimates of virial masses from line-width and cloud size (e.g., \citealt{1987ApJ...319..730S}),  (ii) measurement of $\H2$ masses using a different tracer molecule (e.g., $^{13}\CO$), or a higher order transition, with its own X-factor (e.g., \citealt{1975ApJ...202...50D, 1978ApJS...37..407D, 2009ApJ...699.1092H}), (iii) measurement of proton number density from gamma rays produced in cosmic ray - proton collision (e.g., \citealt{1986A&A...154...25B,1996A&A...308L..21S, 1997ApJ...481..205H, 2010ApJ...710..133A}), (iv) measurement of total hydrogen column density using infrared emission (e.g., \citealt{2001ApJ...547..792D, 2007ApJ...663..866D}), (v) estimates of visual extinction, $A_V$, and thus total gas density from IR star counts \citep{1923AN....219..109W} or the near-infrared color excess \citep{1994ApJ...429..694L}. The latter three approaches have to be supplemented with maps of atomic hydrogen unless most of the hydrogen is molecular. 

The consensus is that for GMCs in the Milky Way all these different methods give similar values of $X_\CO\sim{}1.5-4\times10^{20}$ cm$^{-2}$ K$^{-1}$ km$^{-1}$ s. The question why the X-factor is remarkably constant in molecular clouds in the Milky Way is not yet entirely settled, but it has been suggested to be a consequence of the narrow range of $\H2$ column densities and temperatures of Milky Way clouds \citep{2011MNRAS.412.1686S}.

On the other hand, there is increasing evidence that the X-factor may be different in other galaxies (e.g., \citealt{2008ApJ...686..948B, 2011arXiv1102.4618L}). Specifically, there has been a long debate as to whether the X-factor is \emph{larger} in low metallicity galaxies (e.g., \citealt{1995ApJ...448L..97W, 1996PASJ...48..275A, 2002A&A...384...33B, 2005A&A...438..855I}, but cf. \citealt{2007prpl.conf...81B}), and possibly \emph{smaller} in the central regions of disk galaxies (e.g., \citealt{1988ApJ...325..389M, 1998ApJ...493..730O}), in local starbursts and (ultra-)luminous infrared galaxies (e.g., \citealt{1992A&A...265..447W, 1994ApJ...433L...9S, 1996A&A...305..421M, 1997ApJ...478..144S, 1998ApJ...507..615D, 1999AJ....117.2632B, 2010AJ....140.1294M}), or in submillimeter galaxies at higher redshifts (e.g., \citealt{2008ApJ...680..246T}). In a simple toy model in which the $\CO$ emission stems from an ensemble of optically thick, virialized clumps, the X-factor should scale as $n_{H}^{1/2}/T$ \citep{1986ApJ...309..326D}. However, one arrives at a more complex scaling of the X-factor if the complex geometry of the supersonically turbulent molecular gas and effects of radiative transfer are taken into account (e.g., \citealt{2011MNRAS.tmp..943S}).

A further important aspect is that many extragalactic surveys do not resolve individual GMCs, but measure $\CO$ emission on $\sim{}$ kpc or even larger scale. A study of the properties of the X-factor on such large scales, however, requires galaxy models with reasonably realistic density distributions. The large dynamical scale necessary to resolve individual $\CO$ emitting gas clumps ($\sim{}0.1-1$ pc) in a self-consistent, preferentially cosmological, simulation poses a serious challenge.

In this paper, we study the behavior of the X-factor as a function of gas metallicity, UV radiation field, and as a function of scale. We analyze the impact of the X-factor on star formation laws in a follow-up paper (Feldmann et al, in prep). In section \ref{sect:methods} we present our novel numerical approach that aims at circumventing the problem of the large dynamical scale. The basic idea is that we combine cosmological simulations, in post-processing, with a model that is calibrated on results of small-scale magneto-hydrodynamic (MHD) simulations of the ISM. This approach allows us to study $\CO$ emission at GMC scales and above with an effective resolution of $~\sim{}0.1$ pc in a cosmological volume. In section \ref{sect:XfacScaleGMCs} we discuss the scaling of the X-factor with $\H2$ column density on GMC scales. Next, in section \ref{sect:envDepXfac}, we study the dependence of $X_\CO$ on the local environment (metallicity and UV radiation field). Subsequently, in section \ref{sect:SurfDensGMC}, we discuss the observational evidence of a constant surface density of molecular clouds. Finally, in section \ref{sect:spatav}, we address the effect of spatial averaging of the X-factor. We then summarize our results and conclude.

%---------------------
\section{Methods}
\label{sect:methods}

The approach we adopt in this study is to complement large scale simulations, in which the global structure of galaxies and their ISM is modeled self-consistently in a cosmological context, with an adequate subgrid model for CO emission. 

One possibility is to compute the X-factor based on PDR models (e.g., \citealt{1985ApJ...291..722T, 1988ApJ...334..771V, 1995ApJS...99..565S, 2006ApJ...644..283K, 2010ApJ...716.1191W, 2011ApJ...731...25K}). Although the PDR approach is useful in constructing simple models to make predictions about the atomic-to-molecular transition or the X-factor, one may wonder how sensitive such predictions are to the specific assumptions of this method. It is thus useful to explore alternative models based on a different set of assumptions. To this end, we use a model based on tailored small-scale simulations that are able to resolve the turbulent structure of the ISM on $\sim{}$pc scales \citep{2011MNRAS.412..337G}.

Doing so has several key advantages. First, these small scale simulations rely on well-defined physically understood processes and
their incorporation into cosmological simulations fits well within an `ab initio' approach of understanding the evolution of galaxies and the ISM. Second, the small scale simulations follow the chemical evolution and the dynamics of the ISM in a presumably more realistic way than ad hoc sub-grid models. Finally, large scale simulations can provide the appropriate boundary conditions for the small scale simulations and thus, as long as one stays within the regime explored by the small scale simulations, this approach is potentially not plagued by free parameters that have to be tuned. Thus, if the model turns out to disagree with observations, we cannot simply adjust the parameters of the model. Instead, this would signal that important physics is missing which needs to be included in the small scale simulations. 

However, there are a few technical challenges to this optimistic perspective. First, since the ISM simulations of \cite{2011MNRAS.412..337G} assume a Milky-Way like UV radiation field \citep{1978ApJS...36..595D}, we have to model the UV dependence of the $\CO$ abundance. We do this using a simple but approximate assumption that $\CO$ is in photodissociation equilibrium. Second, we do not perform line radiative transfer (RT). Instead, we use the photon escape probability formalism that has been shown to provide a good fit to observations \citep{2008ApJ...679..481P} and compares reasonably well with line RT calculations, but may differ in some detail \citep{2011MNRAS.412.1686S}. Third, we can neither model the $\CO$ line-width nor the excitation temperature self-consistently at our numerical resolution. We address the first point by considering two extreme cases of the scaling of the line width with column density finding some quantitative, but not qualitative, changes in our results. We assume that the excitation temperature is constant, which is roughly correct for many GMCs in the MW and moderately star forming galaxies (e.g., \citealt{2009AJ....137.4670L, 2009ApJ...699.1092H}), but this assumption may break in galaxies undergoing massive starbursts or in high redshift mergers (see, e.g., \citealt{2011arXiv1104.4118N} and references therein). Hence, while our approach allows a precise modeling of $\CO$ and $\H2$ abundances, it does rely on assumptions regarding the gas temperature and the scaling of the velocity dispersion to infer $\CO$ intensities. This makes it an orthogonal ansatz, with its own strengths and weaknesses, compared with other approaches (e.g., \citealt{2011MNRAS.412.1686S, 2011MNRAS.tmp..943S, 2011arXiv1104.4118N}). Many of these limitations may be circumvented in the future when a larger set of small-scale ISM simulations is available.

A main limitation of our approach is that it relies on the assumption that the small scale simulations are a reasonable approximation to the real ISM, even though several physical mechanisms, such as star formation and stellar feedback, have not been included. Furthermore, possible feedback from small to large scales is not accounted for in our approach. Modeling the ISM below the scales of GMCs is a very challenging problem due to the variety of physical processes and feedback mechanisms that are operating such as stellar winds, HII gas pressure, protostellar jets or radiation pressure (e.g., \citealt{2002ApJ...566..302M, 2004RvMP...76..125M, 2006ApJ...640L.187L, 2006ApJ...653..361K, 2007ApJ...662..395N, 2007ApJ...668.1028B, 2009A&A...504..883B, 2009ApJ...703.1352K, 2010ApJ...709...27W, 2010ApJ...709..191M, 2011piim.book.....D, 2011ApJ...731...91L, 2011ApJ...735...66M, 2011arXiv1105.6097G}). On super GMC ($\gtrsim{}100$ pc) scales feedback from supernovae and (potentially) active galactic nuclei are likely important feedback mechanisms (e.g., \citealt{1998A&A...331L...1S,2004RvMP...76..125M,2010A&A...518L.155F, 2011ApJ...735...66M}).

A reasonable scale for separating sub- and super-grid physics lies therefore in the range $\sim{}10-100$ pc and the resolution of our simulations (formal grid resolution is $\sim{}60$ pc) is chosen to fall in this window. The resolution roughly matches the box-sizes (20 pc) of the ISM simulation presented in \cite{2011MNRAS.412..337G} that we employ to extend the resolution of our simulations to smaller scales.

Note that  we use $Z_\odot\equiv{}0.02$ ($12+\log_{10}(O/H)=8.92$) throughout the paper, the metallicity of the solar neighborhood, which is somewhat larger than the metallicity of the Sun according to recent estimates \citep{2001ApJ...556L..63A, 2004A&A...417..751A, 2009ARA&A..47..481A}.

We present our X-factor model in \S\ref{sect:Xmodel}. We give a short description of the suite of simulations that we use in this paper in \S\ref{sect:sims}. In \S\ref{sect:XfactorSims} we discuss how we add the X-factor model as a post-processing step to our simulations.

\subsection{Modeling of the X-factor}
\label{sect:Xmodel}

In this section we will discuss a simple model for the X-factor  that is calibrated using small scale ($\sim{}$few tens of parsecs) MHD simulations. The aim is to predict $X_\CO$ on these scales as a function of metallicity $Z$, mean density $\bar{n}_H$, and the UV radiation field $\UMW$. The latter is given in units of the local interstellar radiation field \citep{1978ApJS...36..595D, 1983A&A...128..212M}, i.e. 
\[
\UMW=J_{1000 \textrm{\AA}}/J_{\rm MW},
\]
where $J_{1000 \textrm{\AA}}$ is the mean intensity at 1000 \AA{} and $J_{\rm MW}=10^6$ photons cm$^{-2}$ s$^{-1}$ sr$^{-1}$ eV$^{-1}$. 

The X-factor is defined as the ratio between the molecular hydrogen column density $N_\H2$ and the velocity integrated intensity of the $J=1\rightarrow0$ transition of carbon monoxide along a line of sight:
\begin{equation}
\label{eq:X}
X_\CO = \frac{N_\H2}{W_\CO}.
\end{equation}
A canonical value of $X_\CO$ of molecular clouds in the Milky Way is $X_{\CO,{\rm MW}}=2\times{}10^{20}$ cm$^{-2}$ K$^{-1}$ km$^{-1}$ s, which we will refer to as the galactic X-factor.

Computing $W_\CO$ in a simulation is non-trivial unless the gas is optically thin to its own line radiation. We use an approximate treatment based on the escape probability formalism to account, in a crude fashion\footnote{The approach assumes strictly local photon absorption
and plan-parallel geometry.}, for line radiative transfer effects \citep{2011MNRAS.412..337G}
\begin{equation}
\label{eq:W}
  W_\CO = T_B \Delta{}v \int_0^{\tau_{10}} 2\beta(\tau)d\tau.
\end{equation}
Here, $\Delta{}v$ is the width of the $\CO$ line, $\tau_{10}$ is the optical depth of the $^{12}\CO$ $J=1\rightarrow0$ transition,
$\beta(\tau)$ is the photon escape probability \citep{2005pcim.book.....T}
\begin{equation}
\label{eq:beta}
\beta(\tau) = 
\left\{
	\begin{array}{ll}
		[1-e^{-2.34\tau}]/(4.68\tau)  & \mbox{if } \tau\leq{}7 \\
		1/(4\tau[\ln(\tau/\sqrt\pi)]^{1/2}) & \mbox{if } \tau > 7,
	\end{array}
\right.
\end{equation}
and $T_B$ is the brightness temperature of the line, which we compute based on the gas temperature $T$ and the CMB temperature 
$T_{\rm CMB} = 2.725(1+z)$ as
\begin{equation}
\label{eq:TB}
T_B = 5.5 K  \left[ \frac{1}{e^{5.5/T}-1} - \frac{1}{e^{5.5/T_{\rm CMB}}-1}\right].
\end{equation}
Despite the simplicity of the method, its predictions compare reasonably well with more elaborate line radiative transfer calculations \citep{2011MNRAS.412.1686S} and it provides a good fit to observations \citep{2008ApJ...679..481P}.

In local thermal equilibrium (LTE) the optical depth of the  $J=1\rightarrow0$ line is given by \citep{2005pcim.book.....T, 2005fost.book.....S}
\begin{equation}
\label{eq:tau10}
\tau_{10} \simeq 1.4\times{}10^{-16}\left( 1 - e^{-5.5/T} \right)
\left[\frac{10 \textrm{ K}}{T}\right]\left[\frac{3 \textrm{ km s$^{-1}$}}{\Delta{}v}\right] \left[ \frac{N_\CO}{\textrm{cm$^{-2}$}} \right].
\end{equation}
Given estimates of the $\CO$ and $\H2$ column densities, and with some assumptions about the temperature and line width (see \S\ref{sect:XfactorSims}), we can compute the X-factor using (\ref{eq:X}-\ref{eq:tau10}).

Table 2 of \cite{2011MNRAS.412..337G} contains the mass-weighted mean abundance
of $\CO$ and $\H2$ at the end of each of their simulations. The mass-weighted mean abundance $\xm{s}$ of species $s$ (either $\CO$ or $\H2$) in a box with total mass $M=\sum{\rho{}\Delta{}V}$ is defined as
\[
\xm{s}  = \frac{\alpha_s}{M}\sum_i{\frac{n_s(i)}{n_{\rm H}(i)}\rho(i)\Delta{}V(i)} = \alpha_s\frac{\bar{n}_s}{\bar{n}_{\rm H}}.
\]
Here, the index $i$ runs over all grid cells in the box, while $n_{\rm H}$ and $n_{\rm s}$ denote the local number densities of hydrogen nuclei and molecules of species $s$, respectively. The convenient choice $\alpha_\H2=2$ (while $\alpha_\CO=1$) implies that $\xm{\H2}$ spans the range $0-1$.
The equivalent formulation on the right hand side gives the mass-weighted mean abundance in terms of the mean number density of molecules of species $s$, $\bar{n}_s$, and the mean number density of hydrogen nuclei $\bar{n}_H$.

We compute the average surface density $N_s$ of species $s$ along a line-of-sight of length $L$ as
\begin{equation}
\label{eq:Ns}
N_s = \frac{\xm{s}}{\alpha_s} \bar{n}_H L.
\end{equation}

The visual extinction along a line-of -sight is proportional to the column density of dust grains. Since the gas-to-dust ratio scales with the metallicity of the gas at least down to metallicities of $\sim{}1/10$-th solar (e.g., \citealt{2011arXiv1102.4618L}), we compute the mean (averaged over lines-of-sight) visual extinction as
\begin{equation}
\label{eq:AV}
  A_\mathrm{V} = \gamma_Z\,\frac{Z}{Z_\odot}\,\bar{n}_H\,L,
\end{equation}
with  $ \gamma_Z=5.348\times{}10^{-22}$ cm$^{2}$ $Z_\odot^{-1}$. 

Most of the simulations of \cite{2011MNRAS.412..337G} assume $\UMW=1$. However, the $\CO$ fraction in molecular clouds should depend to some degree on the strength of the incident UV radiation (e.g., \citealt{1988ApJ...334..771V}, \citealt{2010ApJ...716.1191W}). In highly star forming environments, such as gas-rich galaxy mergers or young galaxies at high redshifts (e.g., \citealt{1991ApJ...373..423S, 2010A&A...518L..35I}), or in low dust galaxies such as low metallicity local dwarfs (e.g, \citealt{2009AJ....138..130S}), the radiation field is expected to be much larger than that incident onto a typical GMC in the Milky Way.  In order to proceed we will make a number of simplifying assumptions based on the simulation results of \cite{2011MNRAS.412..337G} (see their Fig. 1 \& 2).
First, we assume that $\xm{\H2}$ depends on $\bar{n}_H$, $L$ and $Z$ only through the combination
$\bar{n}_HZ$, i.e. the number density of dust grains available to form $\H2$. Specifically, we will assume
\begin{equation}
\label{eq:xH2}
\xm{\H2}(\bar{n}_H,Z,L,\UMW) = \xm{\H2}(Z\bar{n}_H,\UMW).
\end{equation}
Several prescriptions exist in the literature to compute the mean molecular hydrogen abundance \citep{1988ApJ...332..400S,2008ApJ...689..865K, 2011ApJ...728...88G}. An alternative is to use the $\H2$ abundance that is estimated in the simulations of \cite{2011MNRAS.412..337G} and ignore the explicit radiation field dependence which is expected to be only logarithmic  \citep{2008ApJ...689..865K}. This latter approach is not as crude as it sounds, because the gas will be predominantly molecular whenever there is significant $\CO$ emission and hence self-shielded from UV radiation.

Second, we assume that $\xm{\CO}$ depends mainly on the mean extinction and the radiation field:
\begin{equation}
\label{eq:xCO}
\xm{\CO}(Z,\bar{n}_H,L,\UMW)= \xm{\CO}(A_V(Z\bar{n}_HL),\UMW,Z),
\end{equation}
yet we have also included an explicit dependence of the $\CO$ fraction on metallicity, which is required because the $\CO$ abundance can never be greater than the abundance of $C$ nuclei $x_C$ in the gas. We model this by enforcing an upper limit $x_C = 1.41\times{}10^{-4}Z/Z_\odot$ for $\xm{\CO}(A_V,\UMW,Z)$. The crucial point is that the simulations presented in \citet{2011MNRAS.412..337G} provide $\xm{\CO}(A_V,\UMW=1,Z)$.

In order to model the dependence of $\xm{\CO}(A_V,\UMW,Z)$ on the radiation field we assume
equilibrium between the rates of photodissociation of \CO{} and its creation by chemical reactions, i.e.
\begin{equation}
\label{eq:photoDissA}
x_\CO \, \UMW \, S_\mathrm{dust}(A_V)\, S_\H2(N_\H2)\, S_\CO(N_\CO) = \bar{n}_H\sum_{i,j}\mathcal{R}_{ij} x_i x_j,
\end{equation}
where $S_\mathrm{dust}$, $S_\H2$, $S_\CO$ are the shielding factors\footnote{Each shielding factor varies between 0 and 1, where $S=0$ corresponds to perfect shielding.} due to dust, $\H2$ and $\CO$, respectively, and the r.h.s. is a sum over all relevant chemical reactions that produce \CO. 

We determine $\xm{\CO}(A_V,\UMW\neq{}1,Z)$ in the following way. Consider a box of size $L$, mean density $\bar{n}_H$, metallicity $Z$, and visual extinction $A_V\propto{}Z\bar{n}_HL$ exposed to a radiation field $\UMW\neq{}1$. Consider now a second box of the same size $L'=L$, same metallicity $Z'=Z$, but with possibly different mean density $\bar{n}'_H$ and thus visual extinction $A'_V$ that is exposed to a MW-like radiation field $\UMW=1$. Note that $\bar{n}_H/\bar{n}'_H = A_V/A'_V$. We assume that photo-dissociation equilibrium holds for this second box, too, i.e.
\begin{equation}
\label{eq:photoDissB}
x'_\CO \, \UMW' \, S_\mathrm{dust}(A'_V)\, S_\H2(N'_\H2)\,S_\CO(N'_\CO) = \bar{n}'_H\sum_{i,j}\mathcal{R}_{ij} x'_i x'_j.
\end{equation}
The question is now for which value of the extinction $A'_V$ (or equivalently for which mean density $\bar{n}'_H$) do both boxes have the same CO abundance $x'_\CO=x_\CO$? Once we know $A'_V$ such that $x'_\CO=x_\CO$ (without necessarily knowing $x_\CO$ or $x'_\CO$) we can then use (\ref{eq:xCO}) to obtain $x_\CO$ and, hence, $x'_\CO$. 
Since $N'_\CO = x'_\CO \bar{n}'_H L' \propto{} A'_V x'_\CO/Z'$ and $N_\CO \propto{} A_V x_\CO/Z$, we get $N_\CO=(A_V/A'_V)N'_\CO$. The $\H2$ column densities $N_\H2$ and $N'_\H2$ are determined by the chosen $\H2$ mean abundance model (\ref{eq:xH2}) and the values  of $A_V$ and $A'_V$, respectively.

In order to determine $A'_V$ s.t. $x'_\CO=x_\CO$ we demand that the solution reduces to $A'_V=A_V$  for $\UMW=\UMW'$ and assume that sum involving the rate coefficients on the r.h.s. of  (\ref{eq:photoDissA}) and (\ref{eq:photoDissB}) is not strongly affected by a change in the UV field. The latter assumption is based on a plausibility argument, namely that since $xÔ_\CO = x_\CO$ for $A'_V$, also the abundances of other species $i$ likely satisfy $x'_i \sim{} x_i$.  

Dividing (\ref{eq:photoDissB}) by (\ref{eq:photoDissA}) we obtain
\begin{equation}
\label{eq:AVUMW2}
\frac{1}{\UMW}=\frac{A'_V}{A_V}\frac{S_{\rm dust}(A_V)}{S_{\rm dust}(A'_V)}\frac{S_\H2(N_\H2)}{S_\H2(N'_\H2)}\frac{S_\CO(N'_\CO{}A_V/A'_V)}{S_\CO(N'_\CO{})},
\end{equation}

This equation has to be solved iteratively (see below) to obtain $A'_V$ and $N_\CO=(A_V/A'_V)N'_\CO$ for a given $A_V$ and $\UMW$.
The shielding factors are tabulated in \cite{1996A&A...311..690L}, see also appendix \ref{sect:ShieldingFunctions}.

$\CO$ self-shielding and $\H2$ cross-shielding are due to a large number of individual absorption lines in the near UV. Hence, we multiply the column densities that enter the shielding functions by $L_{\rm c}/L$ to take into account that the bulk velocity of the gas changes of the order of the velocity width of the shielding line over a finite coherence length $L_{\rm c}$. We use $L_{\rm c}=1$ pc, similar to the coherence length for $\H2$ self-shielding \citep{2011ApJ...728...88G}, to convert mean abundances into shielding column densities. We show in appendix \ref{sect:ShieldingFunctions} that shielding of $\CO$ by dust grains is the dominant shielding mechanisms if $L_{\rm c}\lesssim{}1$ pc. Hence, none of our results depend on the precise value of  $L_{\rm c}$ provided it is sufficiently small.

If we ignore self-shielding and shielding by $\H2$, and approximate the dust shielding factor as a plain exponential
$S_\mathrm{dust} = e^{-\gamma_0A_V}$,
we arrive at a simpler version of equation (\ref{eq:AVUMW2}) that contains only the unknown $A'_V$:
\begin{equation}
\label{eq:AVsimple}
\frac{1}{\UMW}\approx\frac{A'_V}{A_V} e^{\gamma_0(A_V'-A_V)}.
\end{equation}

We first solve (\ref{eq:AVsimple}) with the exponent\footnote{This value provides a reasonable approximation, for our purposes, of the dust shielding function over the range $A_V=0-9$. We found that using a more sophisticated approximation in (\ref{eq:AVsimple}), or using the tabulated shielding function directly, does not affect the final solution of (\ref{eq:AVUMW2}) that we are trying to solve for. The solution of (\ref{eq:AVsimple}) approximates the final solution reasonably well, except at low metallicities $Z\lesssim{}0.1$, since dust shielding is the dominant shielding factor for $\CO$ (see appendix \ref{sect:ShieldingFunctions}).} $\gamma_0=3.4$, and then use the obtained approximate solution as a trial value to solve (\ref{eq:AVUMW2}) for $A'_V$. 

\begin{figure}
\begin{tabular}{c}
\includegraphics[width=80mm]{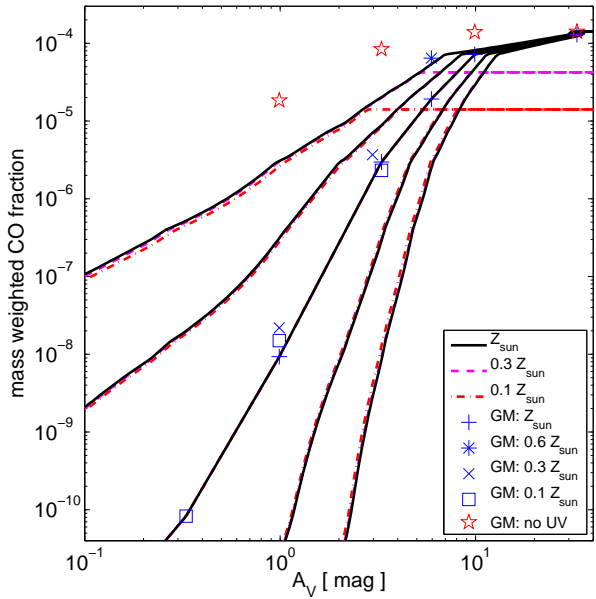} % figures/AV_vs_xCO_L20_2.eps
\end{tabular}
\caption{Mass-weighted mean $\CO$ abundance $\xm{\CO}$ as function of mean extinction $A_V$. Symbols show the results of driven turbulence, magneto-hydrodynamics simulations presented in \cite{2011MNRAS.412..337G}. The star symbols correspond to a simulation without UV radiation, the other symbols are for $\UMW=1$. Lines show the predictions of our model (see text) that is calibrated using the simulation results of \cite{2011MNRAS.412..337G}. The lines correspond to simulations with different metallicities: $Z=Z_\odot$ (solid black lines), 0.3 $Z_\odot$ (dashed magenta lines), and 0.1 $Z_\odot$ (dot-dashed red lines), and different UV radiation fields: $\UMW=10^{-4}$, $10^{-2}$, 1, $10^2$, $10^4$ (set of lines from top to bottom). Our model predicts that at fixed $A_V$ the $\CO$ abundance varies strongly with the interstellar radiation field, but not with metallicity -- except at high visual extinction, where the number density of carbon atoms in the gas, which scales with $Z$, imposes an upper limit on the CO abundance.
\label{fig:xmCO}}
\end{figure}

In Fig.~\ref{fig:xmCO} we show the model predictions for the mean $\CO$ abundance for $\UMW=10^{-6}-10^6$. For a Milky Way like radiation field ($\UMW=1$) the $\CO$ abundance changes strongly with $A_V$, but not with metallicity. The latter is by construction because the simulations of \cite{2011MNRAS.412..337G}  do not show a significant metallicity dependence. A stronger (weaker) UV field lowers (raises) the abundance of $\CO$ at a given mean extinction as expected. For $\UMW\neq{}1$ the $\CO$ abundance depends (even at fixed $A_V$) on metallicity. This behavior is a consequence of $\H2$ cross-shielding and can be understood from  (\ref{eq:AVUMW2}) as follows.

At a given $A_V$ a larger UV flux lowers the $\CO$ abundance, hence $A'_V<A_V$. The corresponding shielding columns $N_\H2$ and $N'_\H2$ are larger at lower metallicity (since $A_V$ is kept fixed) and, due to the shape of the shielding function $S_\H2$, the ratio $S_\H2(N'_\H2)/S_\H2(N_\H2)>1$ is larger at lower metallicity. From (\ref{eq:AVUMW2}) it can be seen that a larger ratio $S_\H2(N'_\H2)/S_\H2(N_\H2)$ can be compensated by increasing $\UMW$, while $A_V$ and $A'_V$ remain fixed. Hence, at a given $A_V$, lowering of the $\CO$ abundance requires a larger UV field in the low metallicity case than at high metallicity. Thus, raising the UV field ($\UMW>1$) leads to a larger $\CO$ abundance when the metallicity is lower. Analogous argument shows that lowering the UV field ($\UMW<1$) leads to a smaller $\CO$ abundance when metallicity is lower. If dust were the only shielding component, the $\CO$ abundance would only depend  on $A_V$ and not explicitly on metallicity.

Combining (\ref{eq:tau10},\ref{eq:Ns},\ref{eq:AV}) one gets (with $T=10$ K):
\begin{equation}
\label{eq:t10AV}
\begin{split}
\tau_{10}&\simeq{}5.9\times{}10^{-17}\xm{\CO}\left[ \frac{\bar{n}_HL}{\textrm{cm$^{-2}$}}\right] \left[\frac{3 \textrm{ km$^{-1}$ s}}{\Delta{}v}\right] \\
                &\simeq{}1.1\times{}10^5 \left[\frac{3 \textrm{ km s$^{-1}$}}{\Delta{}v}\right]  \left[\frac{Z_\odot}{Z}\right] \xm{\CO}A_V. 
\end{split}
\end{equation}
The $\CO$ optical depth $\tau_{10}$ increases rapidly with $A_V$ due to the strong dependence of $\xm{\CO}$ on $A_V$.
Specifically, for $\Delta{}v\sim{}3$K km s$^{-1}$ and $\UMW\sim{}1$, $\tau_{10}$ varies between $\sim{}10^{-3}Z_\odot/Z$, $\sim{}Z_\odot/Z$, and $\sim{}10^2Z_\odot/Z$ for $A_V=1$, $3$, and $10$, respectively. We see that in this case the gas becomes optically thick to the $\CO$ line emission for $A_V\sim{}3$, almost independent of metallicity. This is a significantly larger visual extinction compared with the results from the PDR calculation by \cite{2010ApJ...716.1191W} who predict that $\tau_{10}=1$ should occur around $A_V\sim{}1$.

\begin{figure*}
\begin{tabular}{cc}
\includegraphics[width=80mm]{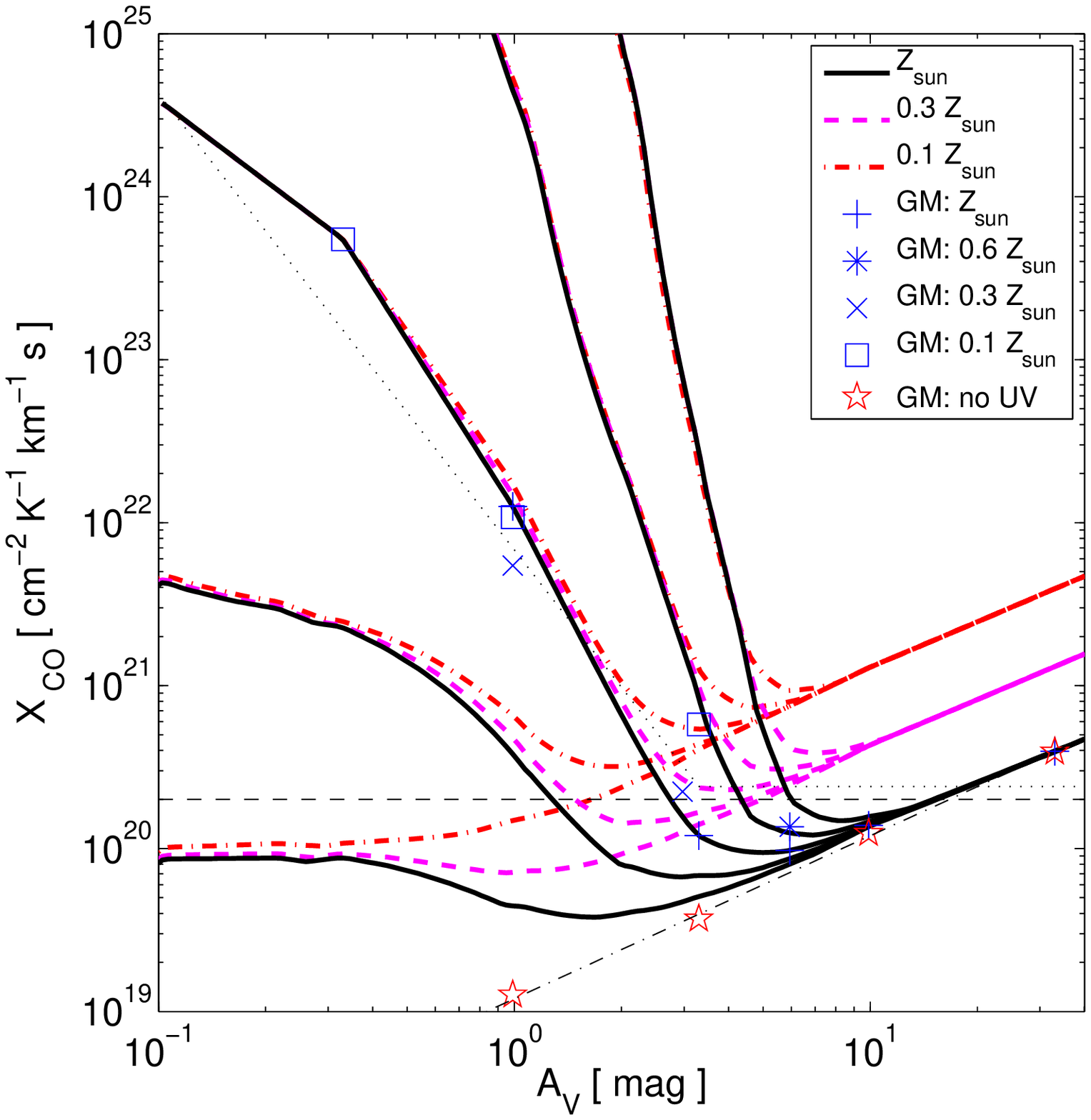} &
\includegraphics[width=80mm]{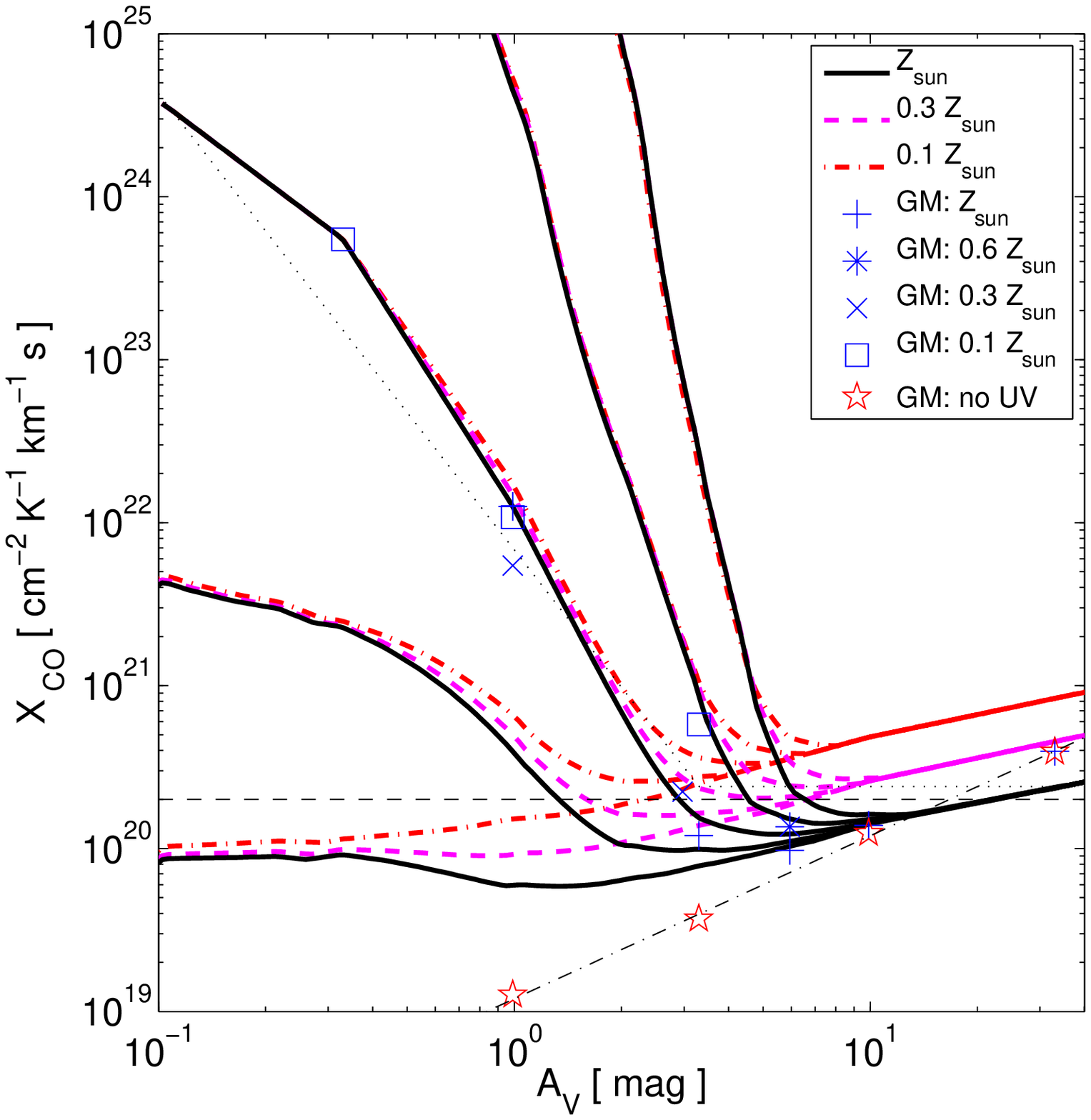} \\
\end{tabular}
\caption{X-factor of the $J=1\rightarrow{}0$ $^{12}\CO$ transition as function of mean extinction as predicted by our model (see text). (Left) Model predictions for a constant $\CO$ line width $\Delta{}v=3$ km s$^{-1}$, (Right) for a virial scaling of the line width (see text). Symbols and lines are as in Fig.~\ref{fig:xmCO}. Solid black lines show the predictions for different radiation fields ($\UMW=10^{-4}$, $10^{-2}$, 1, $10^2$, $10^4$) in \emph {increasing} order from bottom to top. The horizontal, dashed line indicates the canonical galactic X-factor of $X_{\CO,{\rm MW}}=2\times{}10^{20}$ cm$^{-2}$ K$^{-1}$ km$^{-1}$ s. The dotted line indicates the trend of X-factor with $A_V$ found in \cite{2011MNRAS.412..337G} for $\UMW=1$. The thin dot-dashed line in the bottom right shows the expected scaling $X_\CO\propto{}A_V$ when the 1-0 line is optically thick and the line width fixed.
\label{fig:FigXfactorModel}}
\end{figure*}

With $x_\CO$ and $x_\H2$ at hand, the column densities $N_\CO$ and $N_\H2$, and the optical depth of the $\CO$ emission line $\tau_{10}$ can be derived (\ref{eq:tau10}).
The $\CO$ velocity integrated intensity $W_\CO$ can be calculated via (\ref{eq:W}), if the brightness temperature and the $\CO$ line width is given. We discuss the brightness temperature in section \ref{sect:XfactorSims}. Our simulations do not resolve the velocity field within molecular clouds, hence we have to make a sensible choice for line width $\Delta{}v$. We consider two possibilities.

\begin{itemize}
\item The first case (\emph{constant line width}) assumes that the velocity width is constant and has a value typical of molecular clouds in the Milky Way $\Delta{}v=3$ km s$^{-1}$. This ansatz is motivated by the fact that our X-factor model is based on ISM simulations of a fixed box-size ($20$ pc) and 3 km s$^{-1}$ is consistent with the observed size-linewidth relation of molecular clouds of such size \citep{1981MNRAS.194..809L, 1987ApJ...319..730S, 2009ApJ...699.1092H}.

\item The second case (\emph{virial line width}) assumes:
\[
\Delta{}v = \left[f_{\rm grav}G\Sigma_{\rm bar}\frac{L}{2}\right]^{1/2}, 
\]
which is based on assumption that massive molecular clouds are bound and close to energy equipartition or even virialisation \citep{1987ApJ...319..730S, 2001ApJ...551..852H, 2006MNRAS.372..443B}. Here, $f_{\rm grav}$ is a constant of order unity that depends on the density structure and geometry of the cloud, its degree of virialization, and line-of-sight projections. We assume $f_{\rm grav}=1$.  
\end{itemize}

Equations (\ref{eq:X}), (\ref{eq:W}) imply that $X_\CO\propto{} \Sigma_{\H2}$ (for a constant line width) and $X_\CO\propto{} \Sigma_{\H2}^{1/2}$ (for a virial line width) in the optically thick regime. These scalings bracket those computed from detailed radiation transfer calculations of MHD GMC simulations \citep{2011MNRAS.tmp..943S}.

In Fig.~\ref{fig:FigXfactorModel} we show the X-factor $N_\H2/W_\CO$ as function of mean extinction, for different metallicities and radiation fields, and for the case of a constant and virial line width scaling, respectively. The main predictions of the model are:

\begin{itemize}

\item The X-factor depends on mean extinction $A_V$. It reaches a minimum at $A_V=2-10$ (depending on metallicity and UV radiation field), roughly where the $\CO$ line becomes optically thick ($\tau_{10}\approx{}1$). It increases at higher and lower values of  $A_V$. The increase at low $A_V$ is a consequence of the much faster decline of the abundance of $\CO$ (compared to $\H2$) with decreasing dust column. At large $A_V$, both hydrogen and carbon are fully molecular, the $\CO$ emission is saturated, and the X-factor increases simply due to the increase of the $\H2$ column density with $A_V$. In fact, the X-factor increases linearly with $A_V/Z$, if we assume that the $\CO$ line width and the excitation temperature remain fixed, and with the square-root of $A_V/Z$ if we assume a virial scaling of the line width.

\item The X-factor at fixed $A_V$ increases with increasing UV field in the optically thin and moderately optically thick regime of the $\CO$ line, e.g., for $A_V\lesssim{}4$ at $Z=0.1 Z_\odot$ and for $A_V\lesssim{}8$ at $Z=1 Z_\odot$, but it is insensitive to the UV field in the highly optically thick regime.

\item For $T=10$ K, $\Delta{}v=3$ km s$^{-1}$ and a UV field in the range $\UMW=0.01-100$ the minimum of the X-factor is at $0.7-1.1\times{}10^{20}$ cm$^{-2}$ K$^{-1}$ km$^{-1}$ (for $Z=Z_\odot$), $1.7-2.8\times{}10^{20}$ cm$^{-2}$ K$^{-1}$ km$^{-1}$ (for $Z=0.3 Z_\odot$), and $4.5-6.3\times{}10^{20}$ cm$^{-2}$ K$^{-1}$ km$^{-1}$ (for $Z=0.1 Z_\odot$). The minimum of the X-factor can be lowered further by increasing the metallicity to super-solar.

\item The X-factor at fixed $A_V$ does not strongly depend on metallicity at sufficiently low visual extinction. On the other hand, $X_\CO$ depends on metallicity at large $A_V$ in our model because $W_\CO$ is approximately constant in the optically thick regime and a change in metallicity at fixed $A_V$ corresponds to a change in $N_\H2$ (the gas is predominantly molecular), and thus in $X_\CO$.

\end{itemize}

\subsection{Simulations}
\label{sect:sims}

A detailed description of the simulations can be found in \cite{2011ApJ...728...88G}. Here we repeat the main points for the convenience of the reader, following closely \cite{2011ApJ...732..115F}.

All simulations are run with the Eulerian hydrodynamics + N-body code ART \citep{1997ApJS..111...73K, 2002ApJ...571..563K}, that uses an adaptive mesh refinement (AMR) technique to achieve high spatial resolution in the regions of interested (here: regions of high baryonic density). First, we ran an initial cosmological, hydrodynamical simulation down to $z=4$. This simulation follows a Lagrangian region that encloses five virial radii of a typical $L_*$ galaxy (halo mass $\sim{}10^{12}$ $M_\odot$ at $z=0$) within a box of 6 comoving Mpc/$h$. The mass of dark matter particles in the high resolution Lagrangian patch is $1.3\times{}10^6 M_\odot$. We adopt the following cosmological parameters $\Omega_{\rm matter}=0.3$, $\Omega_\Lambda=0.7$, $h=0.7$, $\Omega_{\rm baryon}=0.043$, $\sigma_8=0.9$. 

This initial, self-consistent simulation is consequently continued for additional $\sim{}600$ Myr before it is analyzed, but now with metallicities and UV fields  fixed to a specific, spatially uniform value. At the end of each simulation the high resolution Lagrangian region
contains one massive disk galaxy (halo mass $\sim{}4.2\times{}10^{11}$ $M_\odot$) and several smaller galaxies. The spatial resolution is $\sim{}60$ pc in physical coordinates.

We have run a grid of simulations with six different metallicities $Z/Z_\odot=$0.01, 0.03, 0.1, 0.3, 1.0, 3.0 and three different values of the interstellar radiation field $\UMW=0.1,1,100$. We continued one of our simulations ($Z/Z_\odot=1$, $\UMW=1$) for additional 400 Myr and found no significant changes in the ISM properties. This indicates that the predictions of our simulations should also hold for redshifts $z\lesssim{}3$, at least unless/until ISM properties change radically. In fact, in the computation of the brightness temperature of the $\CO$ line we use the $z=0$ CMB temperature in order to compare the model predictions with observations in the local universe. We have also run additional simulations at solar metallicity and $\UMW=1$, but with 2 times better and 2 times worse peak spatial resolution confirming that none of our results suffer strongly from numerical resolution effects.

The simulations include a sub-grid model for star formation, metal enrichment, supernova feedback (type Ia and type II) and cooling by metal lines and $\H2$. The molecular hydrogen fraction $f_\H2{}$ is computed self-consistently, including a chemical network comprised of 6 species and radiative transfer of the UV continuum and the Lyman-Werner bands \citep{2011ApJ...728...88G}.

\subsection{Modeling the X-factor on subgrid scales}
\label{sect:XfactorSims}

We include the X-factor model in our numerical simulations as follows.

First, we compute the mean visual extinction for a cell given its metallicity, density and an estimate of its line-of-sight depth, see (\ref{eq:AV}). We then predict the mass weighted $\CO$ abundance $x_\CO$ based on $A_V$, metallicity and UV radiation field as described in \S\ref{sect:Xmodel}. The $\H2$ mass fraction can either be taken directly from the cell (computed self-consistently with the non-equilibrium chemical network and radiative transfer within ART) or it can be computed from the table of \cite{2011MNRAS.412..337G} (ignoring its dependence on UV field). While the first approach is our default method, we arrive at similar predictions for the X-factor in either case, see \S~\ref{sect:XfacScaleGMCs}.
Next, we convert the mean abundances  $\xm{\CO}$ and $\xm{\H2}$ into column densities (see,  eq.~\ref{eq:Ns}) by multiplying the abundances by the Sobolev-like length $\Lsob=\rho/\vert{}2\nabla{}\rho\vert$. This procedure leads to more reliable column density estimates compared with the use of the cell size as the line-of-sight depth and also avoids  a few spurious artifacts related to the numerical resolution, see \cite{2011ApJ...728...88G}.
Finally, we compute the $\CO$ intensity and the X-factor of the cell via eqs.~(\ref{eq:X}-\ref{eq:tau10}).

The gas temperature $T$ and the velocity width $\Delta{}v$, which both may vary in space and time, enter the prediction of the X-factor through equations (\ref{eq:W}), (\ref{eq:TB}), (\ref{eq:tau10}). The temperature that is measured in individual grid cells is an average temperature over the unresolved thermal structure of the gas and thus not necessarily a good estimator of the interior temperature of a molecular cloud, for which observations  indicate $T\sim{}10$ K (e.g., \citealt{2009ApJ...699.1092H}). We therefore fix the gas temperature to $T=10$ K. 
Temperatures may increase in galaxies with strong star bursts and with densities high enough to allow for heat transfer between dust and gas via collisions (e.g. \citealt{1991ApJ...373..423S,2011arXiv1104.4118N}), but exploring this possibility is beyond the scope of this paper. For $\Delta{}v$ we consider the cases of a \emph{constant line width} and that of a \emph{virial line width scaling}, see \S\ref{sect:Xmodel}.

\section{The X-factor on the scales of GMCs}

\subsection{Scaling with $\H2$ column density}
\label{sect:XfacScaleGMCs}

\begin{figure*}
\begin{tabular}{cc}
\includegraphics[width=80mm]{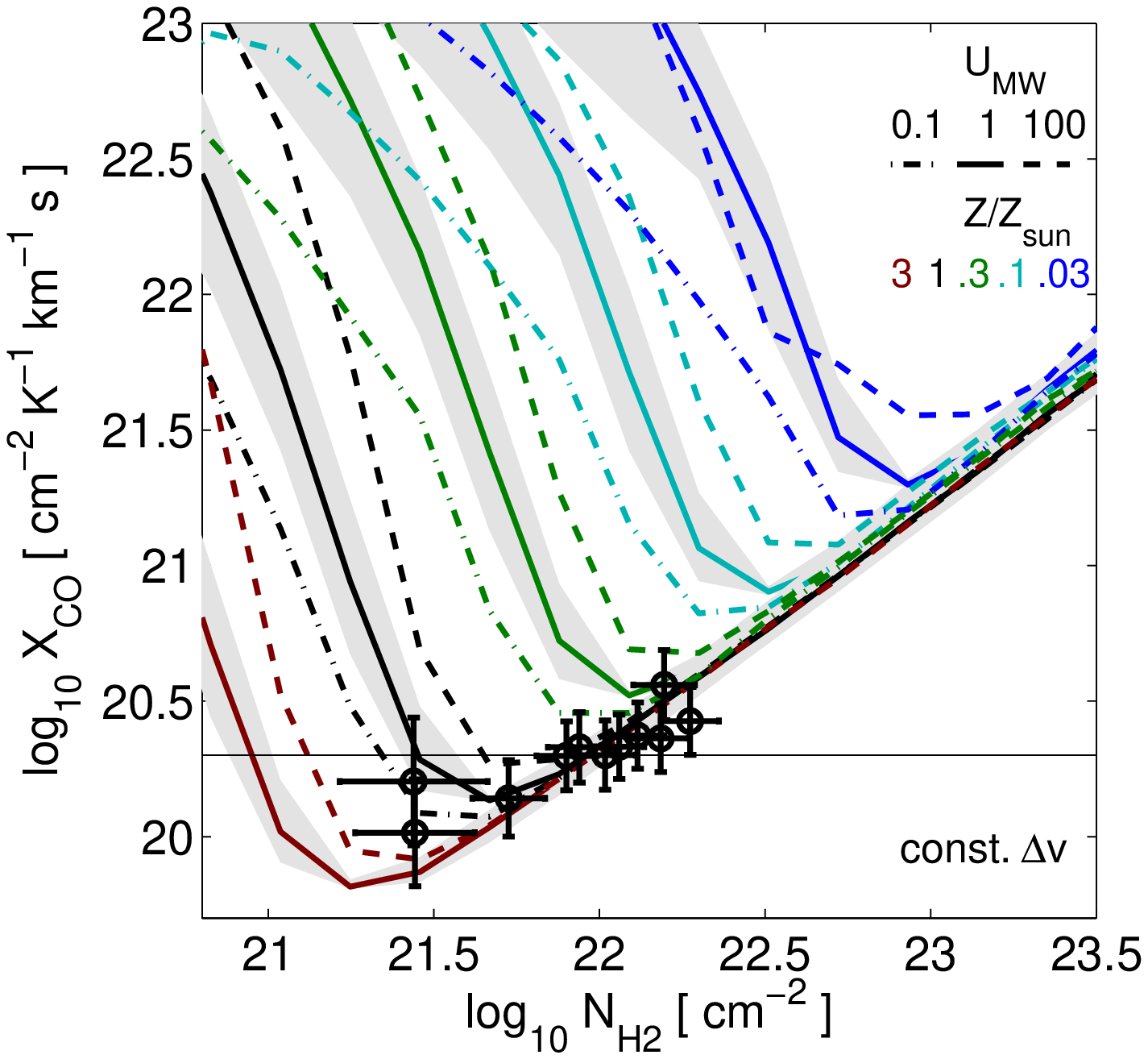} & % analysis/X_vs_NH2_m0_l9_CO_new3_LFIX20_CLUMP_Lsob_CO1_lr8
\includegraphics[width=80mm]{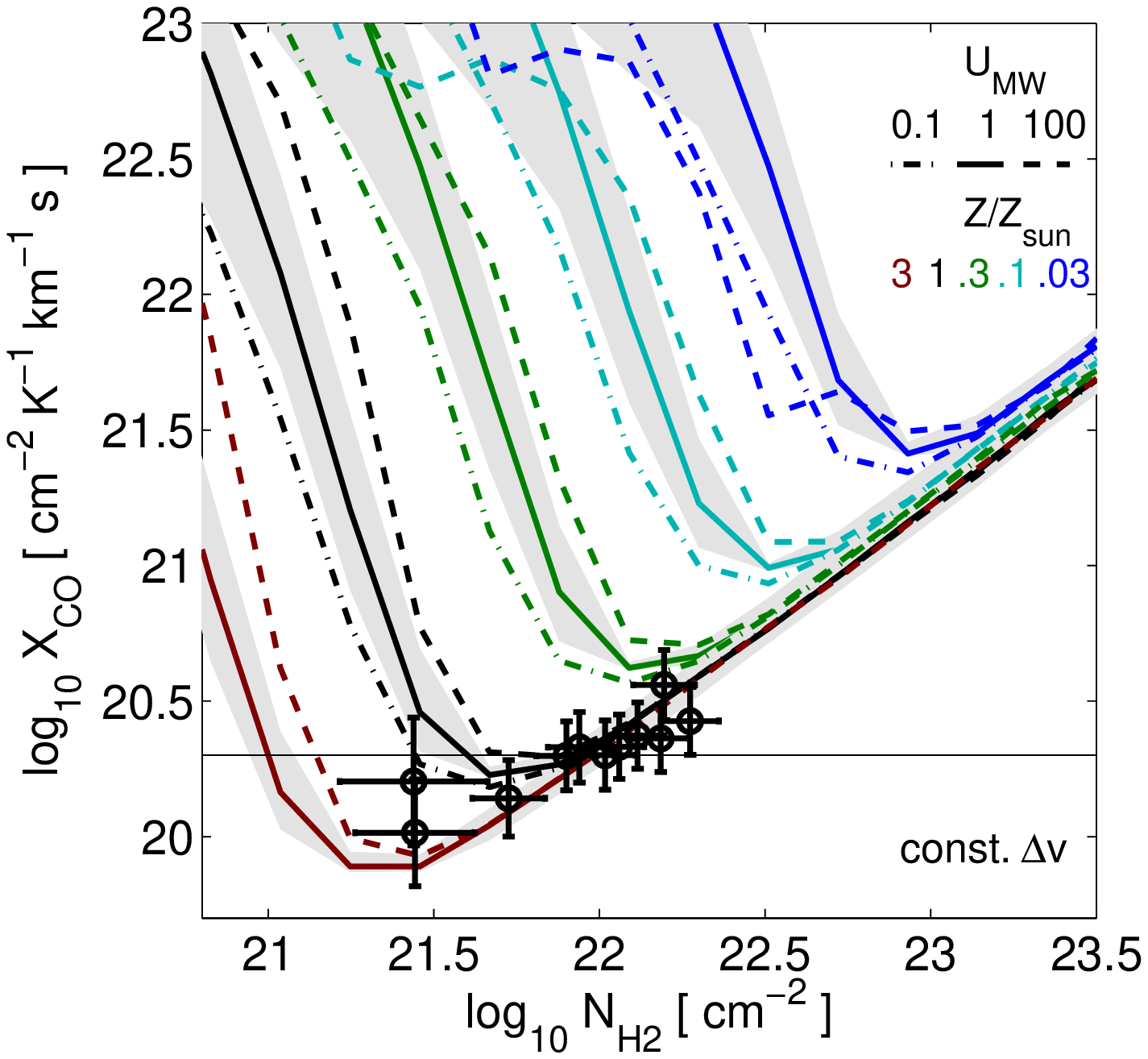} \\  % analysis/X_vs_NH2_m0_l9_CO_new3_H2ART_LFIX20_CLUMP_Lsob_CO1_lr8
\includegraphics[width=80mm]{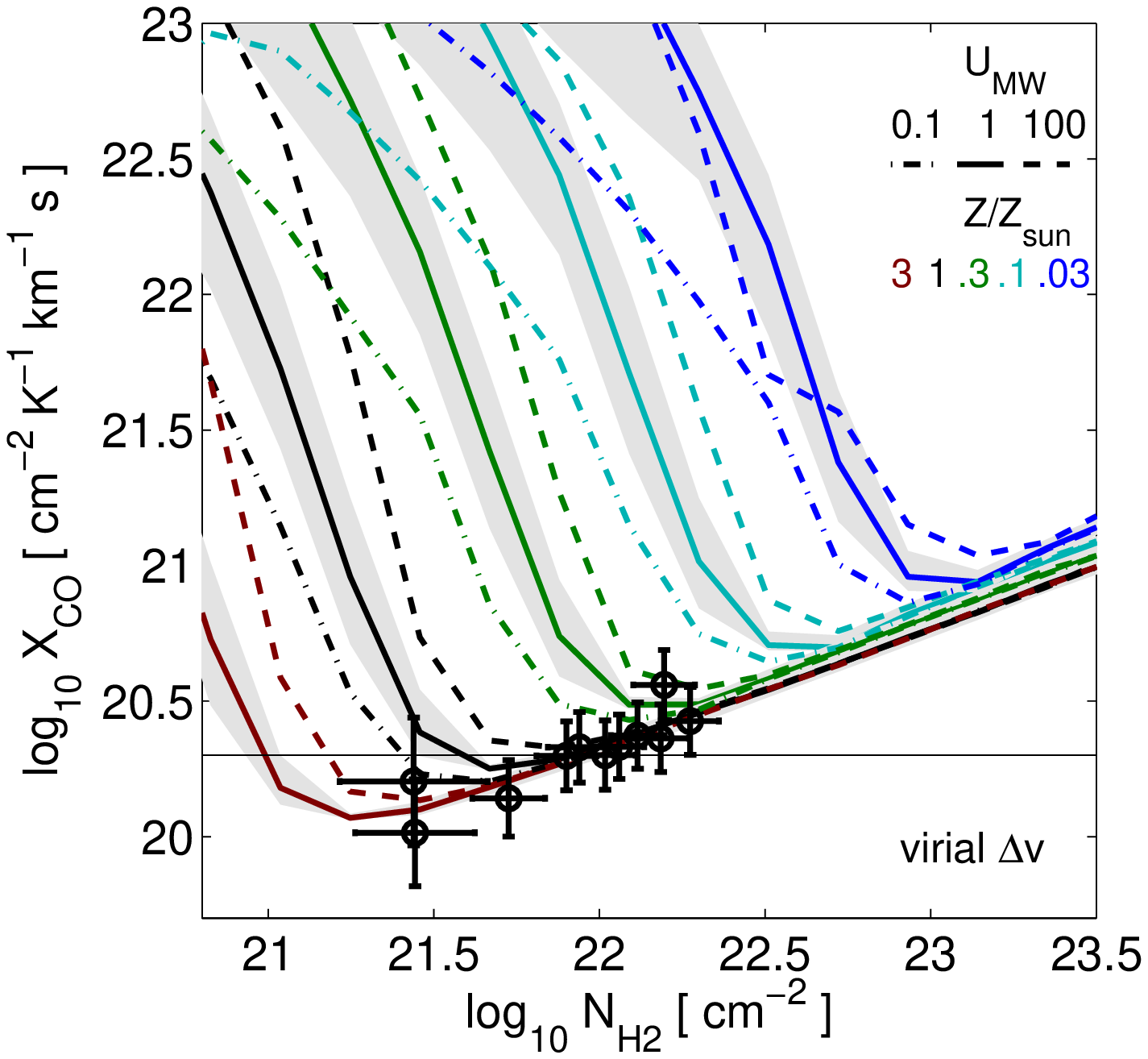} &  % analysis/X_vs_NH2_m2_l9_CO_new3_LFIX20_CLUMP_Lsob_CO1_lr8
\includegraphics[width=80mm]{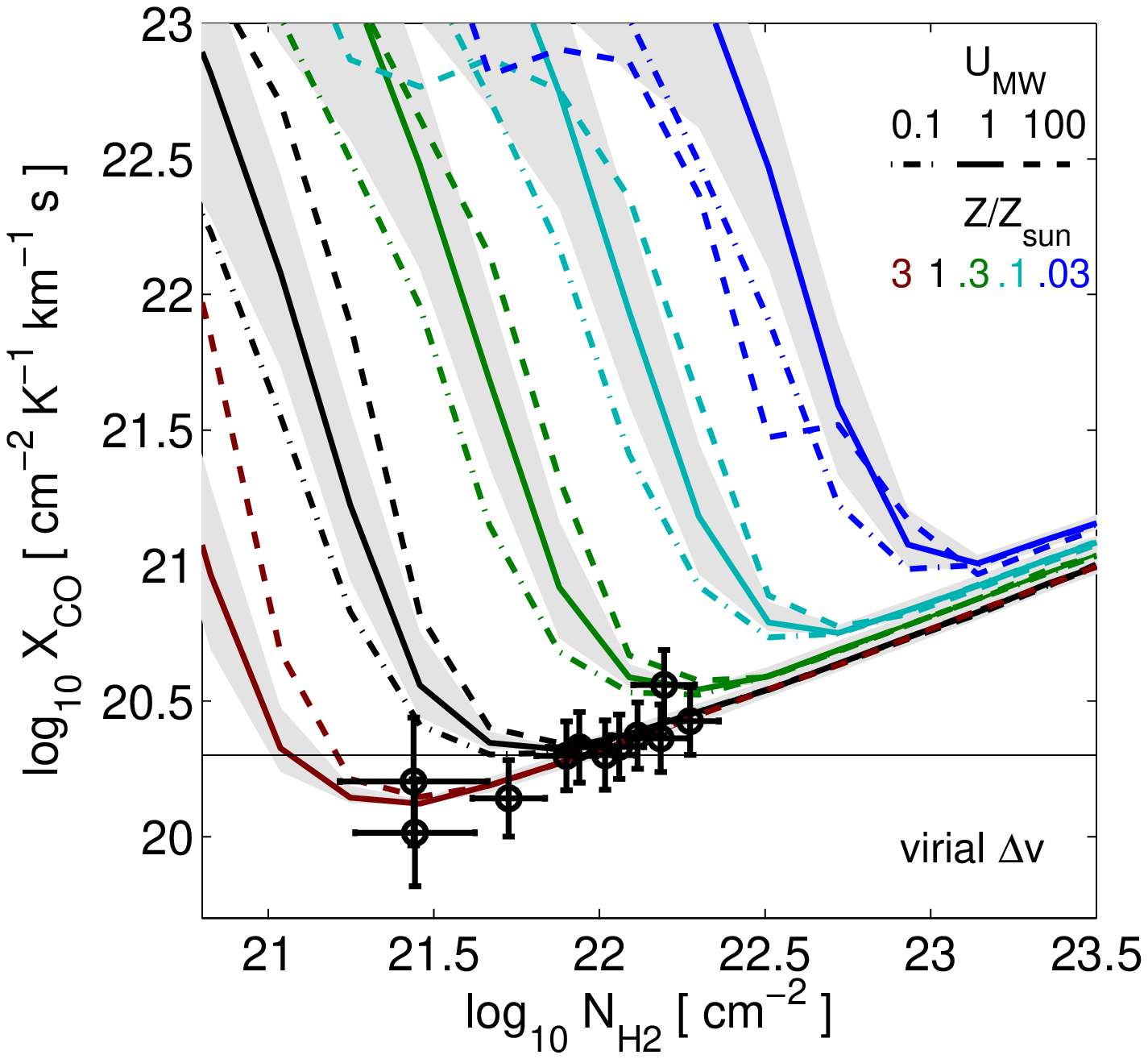}     % analysis/X_vs_NH2_m2_l9_CO_new3_H2ART_LFIX20_CLUMP_Lsob_CO1_lr8
\end{tabular}
\caption{X-factor vs $\H2$ column density on $\sim{}$ 60 pc scales as a function of metallicity and UV flux. (Left column) model predictions using the tabulated $\H2$ fractions of \cite{2011MNRAS.412..337G}. (Right column) using $\H2$ fractions computed via the photo-chemical network within the simulation. (Top row) assuming a constant line width ($\Delta{}v=3$ km s$^{-1}$), (bottom row) analogous predictions assuming a virial line width scaling ($\Delta{}v\propto{}\Sigma^{1/2}$), see \S\ref{sect:XfactorSims}. 
The curves correspond to 12 cosmological simulations with constrained ISM properties. The UV radiation field varies between $\UMW=0.1$ (dot-dashed lines), $\UMW=1$ (solid lines), and $\UMW=100$ (dashed lines). The metallicity varies (from bottom to top; see legend) between $Z=3Z_\odot$, $Z=Z_\odot$, $Z=0.3Z_\odot$, $Z=0.1Z_\odot$, and $Z=0.03Z_\odot$.
Each curve connects the median $X_\CO$ for a given $\H2$ column density. The 16$^{\rm th}$ and 84${\rm th}$ percentiles of the $X_\CO$ distribution for $\UMW=1$ is shown by the gray shaded areas. The galactic X-factor $X_{\CO,\MW}=2\times{}10^{20}$ K$^{-1}$ cm$^{-2}$ km$^{-1}$ is indicated by a solid horizontal line.
Circles with error bars show results for the Perseus and Ophiuchus clouds in the Milky Way and should be compared with the $Z\sim{}Z_\odot$ lines (\citealt{2010ApJ...723.1019H}; their Table 5).}
\label{fig:XfacGMC}
\end{figure*}

Figure~\ref{fig:XfacGMC} shows results of applying the X-factor model described in the previous section to 12 numerical simulations with constrained ISM properties ranging from $Z=0.03Z_\odot$ to $3 Z_\odot$ and $\UMW=0.1$ to 100.  All runs follow almost the same $X_\CO-N_\H2$ relation at \emph{high} column densities, but break off from this asymptotic relation at a specific $\H2$ column density. The scaling of this asymptotic relation is $X_\CO\propto{} \Sigma_{\H2}$ (constant line width) and $X_\CO\propto{} \Sigma_{\H2}^{1/2}$ (virial scaling) as expected. Note that the dependence of $A_V$ on metallicity via equation (\ref{eq:AV}) is the reason why this figure looks very different from Fig.~\ref{fig:FigXfactorModel}.

Figure~\ref{fig:XfacGMC} also shows that the UV radiation field has an significant impact on the X-factor. However, it should be pointed out that at fixed $\H2$ column density the effect of a change of the UV field by a factor 1000 can be often be compensated by a less than factor 2 change in metallicity. 

%At low metallicities ($Z\leq{}0.1Z_\odot$) the X-factor \emph{decreases} with increasing UV field at fixed $\H2$ column density. We can understand this behavior as follows. A larger UV field requires a larger total gas column density to shield a fixed $\H2$ column density from the intense UV radiation. How big the required increase in the total gas column density is determines whether the X-factor at fixed $\H2$ column density increases or decreases with UV field. If the increase in total gas column density is relatively small $X_\CO$ increases with UV field. If the increase is sufficiently large the X-factor decreases. This can also be see in Figure~\ref{fig:FigXfactorModel}: Take a point on the optically thin branch of the $\UMW=1$ line and connect it with another point of larger extinction on the optically thin branch of the $\UMW=100$ line. Depending on the extinction value that you pick $X_\CO$ increases or decreases. Consequently, at sufficiently high metallicities the result is an increase in $X_\CO$, because shielding of $\H2$ is easy and hence the $\H2$ and the total gas column densities are similar even at large UV fields. At low metallicities $X_\CO$ decreases with UV field, because shielding of the fixed $\H2$ column density is difficult and requires a large total gas column density.

Both the $\H2$ column density at which the minimum of the X-factor occurs and the corresponding minimal $X_\CO$ value decrease with increasing metallicity. More specifically, for a simulation with $Z_\odot$ and a UV radiation field in the range $\UMW=0.1-100$ $X_\CO$ stays within a factor of 2 of the galactic value $X_{\CO,\MW}=2\times{}10^{20}$ K$^{-1}$ cm$^{-2}$ km$^{-1}$ s over a wide range of $\H2$ surface densities:  $2.5\times{}10^{21}$ cm$^{-2}\lesssim\, N_\H2 \lesssim\,2\times{}10^{22}$ cm$^{-2}$ for a constant $\CO$ line width and $2.5\times{}10^{21}$ cm$^{-2}\lesssim\, N_\H2 \lesssim\,5\times{}10^{22}$ cm$^{-2}$ for a virial scaling, respectively. Assuming the gas is fully molecular, the latter range in $\H2$ column density corresponds to a range in total gas mass surface density (incl. He) $\sim{}50-1000$ $M_\odot$ pc$^{-2}$. This covers the range of surface densities of GMCs found in a variety of studies, including those that determine gas surface densities without the use of $^{12}\CO$ emission \citep{1981MNRAS.194..809L, 2009ApJ...699.1092H, 2010ApJ...723..492R, 2010ApJ...723.1019H} and which are thus not biased due to variations in $X_\CO$. 

Hence, we conclude that the near constancy of the X-factor among the GMC population in our Galaxy is caused by the coincidence that for near solar metallicity $X_\CO$ changes only weakly within the typical range of GMC surface densities. However, the figure clearly shows that $X_\CO$ changes strongly with $\H2$ surface density if it is outside the range of $\sim{}50-1000$ $M_\odot$ pc$^{-2}$, and for higher or lower metallicities even within that range. Generally, at any metallicity, the X-factor is expected to vary with $\H2$ column density unless either the $\H2$ surface density distribution in the ISM is very narrow (so that the dependence cannot be studied) or limited to the nearly flat part in the $X_\CO-N_\H2$ curve (the Milky Way coincidence; also cf. \citealt{2011MNRAS.tmp..943S}).

While the dependence of $X_\CO$ on metallicity is observationally confirmed, see section \ref{sect:envDepXfac}, the increase of the X-factor with $N_\H2$ at large column densities is not yet clearly established. Column density observations that are based on dust extinction maps should provide a good way to test these predictions. For instance, table 5 of \cite{2010ApJ...723.1019H} provides $\H2$ column densities estimates both from $^{12}\CO$ and dust extinction maps. If we assume that the latter is a reliable measure of the true gas column density then the ratio of the two column densities provides us with a measure of the X-factor in units of $X_{\CO,{\rm MW}}$. We include the X-factor derived in this way in Figure~\ref{fig:XfacGMC}. A weighted linear regression of such $X_{\CO}-N_\H2$ correlation\footnote{We only include data with extinction based $\H2$ column densities in excess of $10^{21.5}$ cm$^{-3}$ in order to avoid including the ``upturn'' of the $X_\CO-N_\H2$ relation at low column densities in the fit, see Fig.~\ref{fig:XfacGMC}.} results in an exponent of $0.53\pm{}0.14$. Observations thus indicate an increase of $X_\CO$ with $N_\H2$, consistent with a $\CO$ line width that scales as $\Delta{}v\propto{}\Sigma^{1/2}$, although the data does not exclude a somewhat steeper scaling.

Figure~\ref{fig:XfacGMC} also shows the predicted scatter of the X-factor at fixed $\H2$ column density. This scatter
arises due to the degeneracy between gas density and $\H2$ fraction that result in the same $\H2$ column density, but which lead to different predictions for $\xm{\CO}$, see (\ref{eq:xCO}). This scatter should be treated as a lower limit on the actual scatter on such scales. The scatter increases towards lower $\H2$ column densities and varies, for Milky-Way like ISM conditions, between 0.3 dex at $N_\H2\sim{}2\times{}10^{21}$ cm$^{-2}$, 0.2 dex at  $N_\H2\sim{}3\times{}10^{21}$ cm$^{-2}$, and $<0.1$ dex at $N_\H2\gtrsim{}5\times{}10^{21}$ cm$^{-2}$.
This relatively small lower limit ($\lesssim{}0.3$ dex) on the scatter of the X-factor is also consistent with the notion of a nearly constant X-factor among galactic GMCs.

Finally, comparing the left with the right column of Fig.~\ref{fig:XfacGMC} we can check how the X-factor predictions depend on the chosen $\H2$ modeling.
The left column uses the tabulated $\H2$ fractions of \cite{2011MNRAS.412..337G} and ignores any changes with UV radiation field, while in the right column the $\H2$ fractions and column densities are computed self-consistently with the photo-chemical network in ART. 

Differences in the $\H2$ modeling lead to shifts along the 45 degree line since both $X_\CO$ and $N_\H2$ depend on the $\H2$ fraction. Specifically, the use of ART-based $\H2$ fractions vs using the tabulated $\H2$ fractions of \cite{2011MNRAS.412..337G} results in shifts of $\sim{}0.2$ dex. However, these differences are relatively minor and in any case do not lead to qualitative changes of the results in the paper.
 
\subsection{A parametrization of the X-factor}
\label{sect:ParamModel}

\begin{figure*}
\begin{tabular}{cc}
\includegraphics[width=80mm]{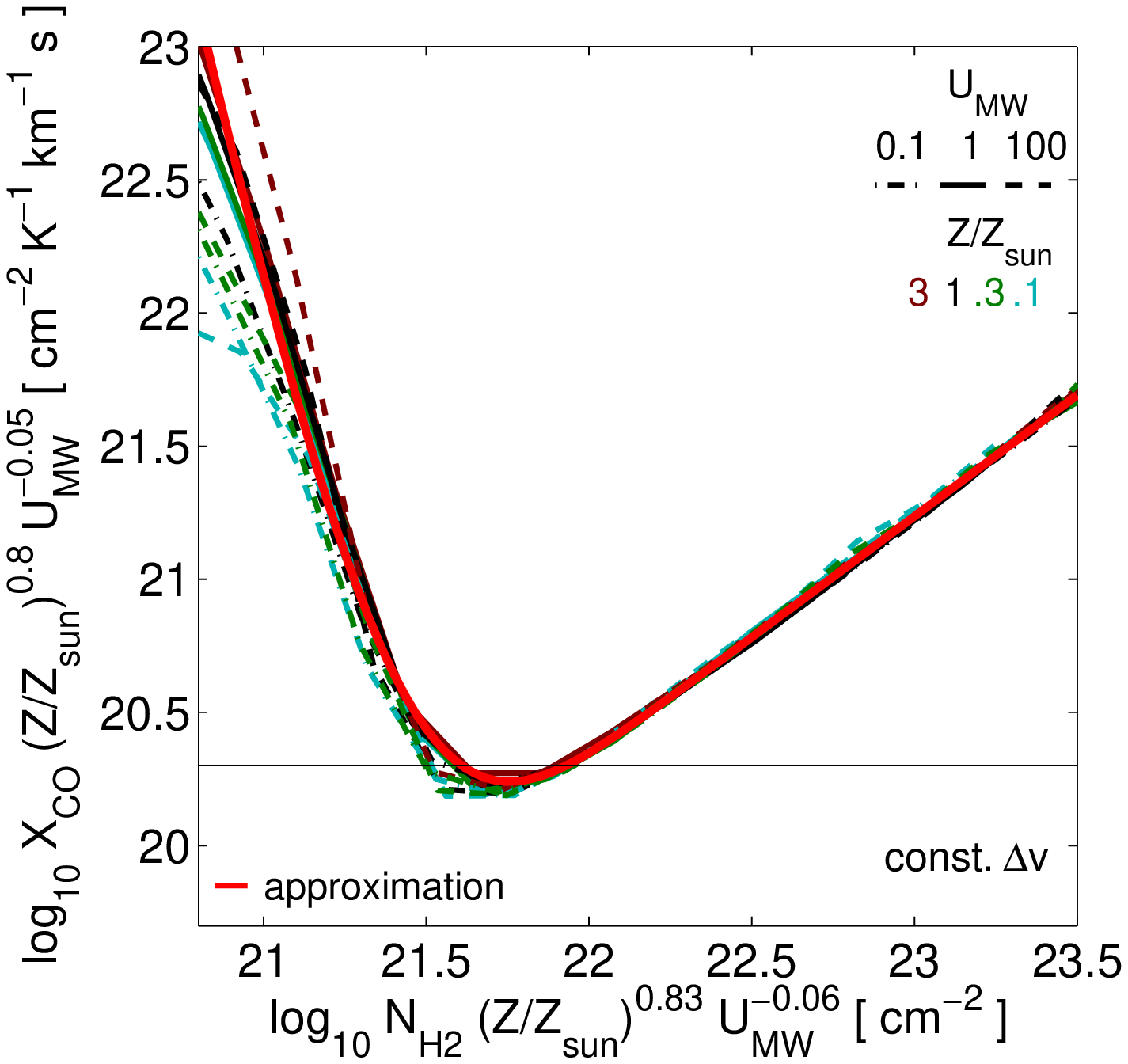} & % analysis/X_vs_AV_rescale_l9_m0.eps
\includegraphics[width=80mm]{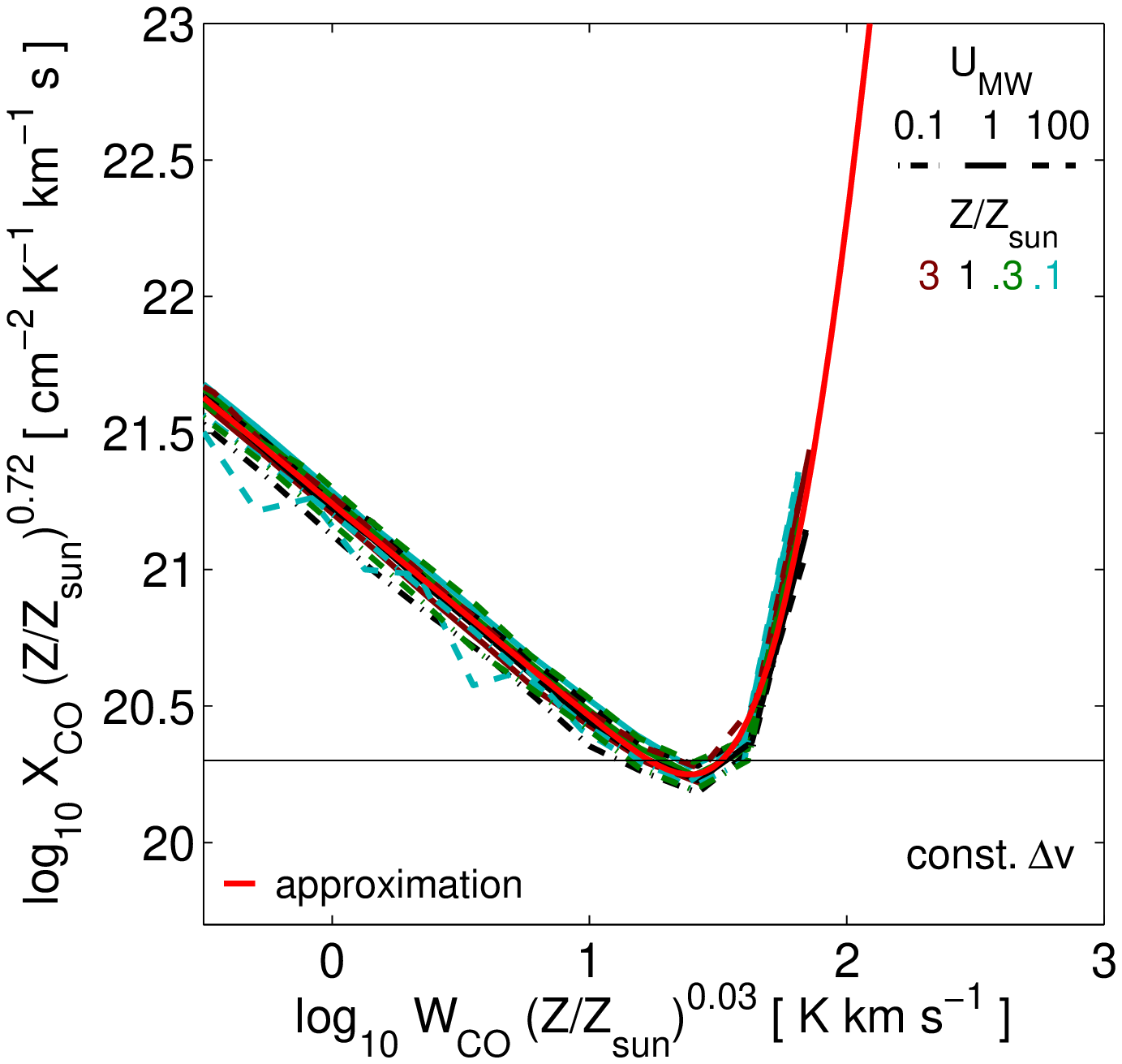} \\  % analysis/X_vs_WCO_2_rescale_l9_m0.eps
\includegraphics[width=80mm]{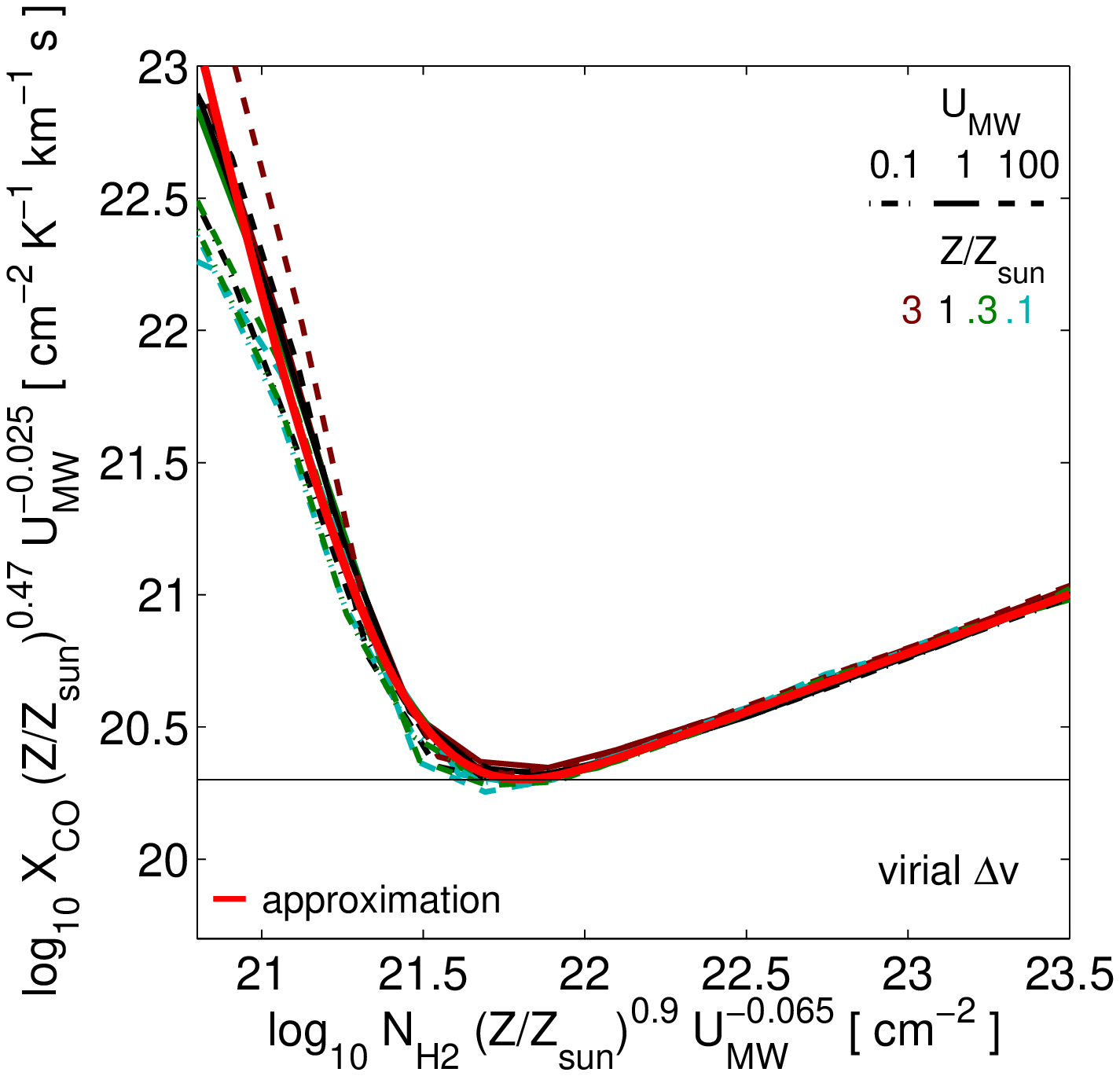} &  % analysis/X_vs_AV_rescale_l9_m2.eps
\includegraphics[width=80mm]{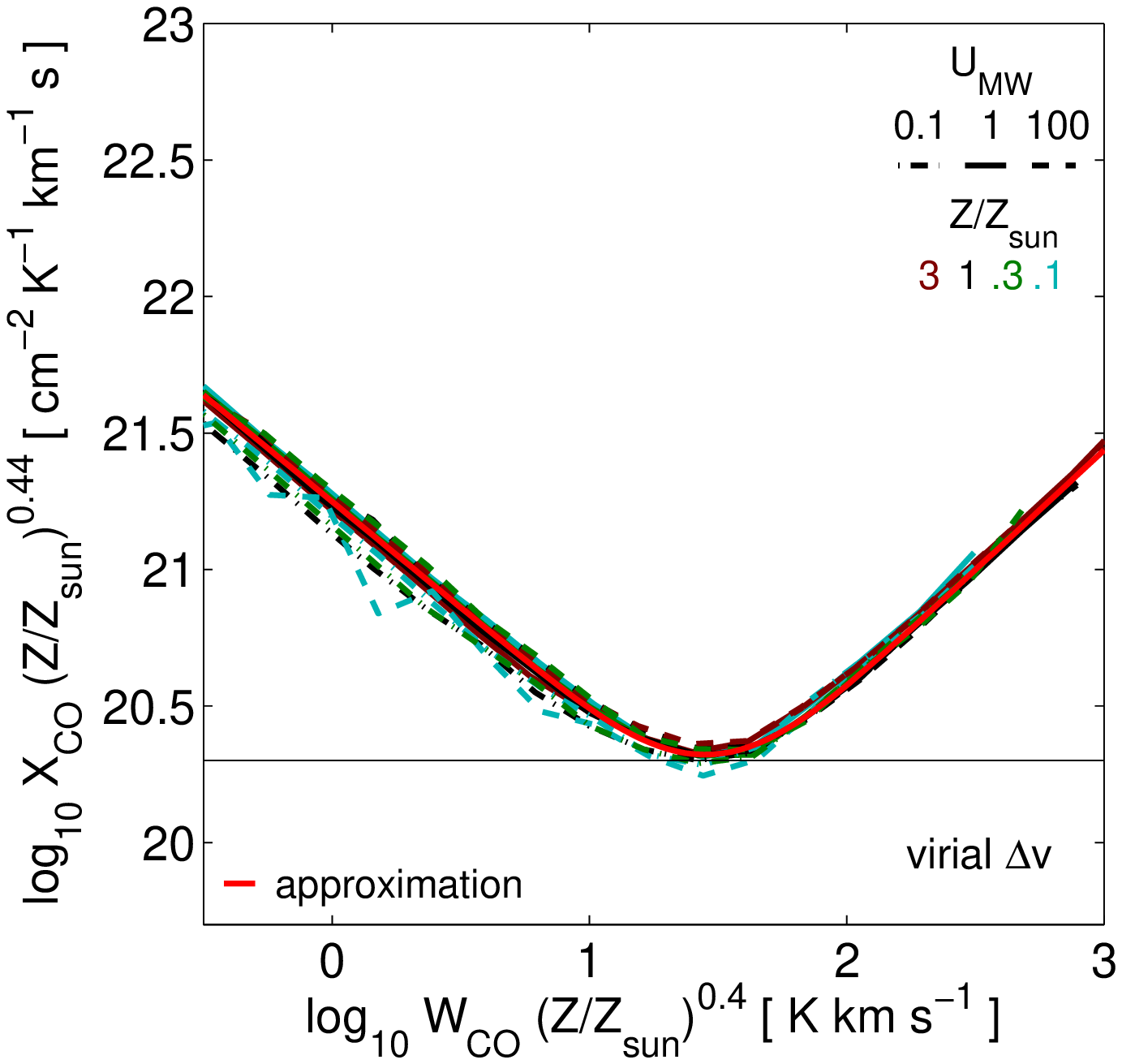}     % analysis/X_vs_WCO_2_rescale_l9_m2.eps
\end{tabular}
\caption{Rescaled X-factor vs rescaled $\H2$ column density (left column) and rescaled $\CO$ intensity (right column) on $\sim{}$ 60 pc scales as a function of metallicity and UV flux. (Top row) model predictions of the X-factor assuming a constant line width ($\Delta{}v=3$ km s$^{-1}$), (bottom row) analogous predictions assuming a virial line width scaling ($\Delta{}v\propto{}\Sigma^{1/2}$). The solid red line shows an empirical approximation of the relation between the rescaled quantities. It approaches power law behavior at low and high $\H2$ column densities and $\CO$ intensities, respectively, see text. The rescaling of the X-factor, $\H2$ column density and $\CO$ intensity with a metallicity dependent factor removes most of the explicit metallicity dependence. }
\label{fig:XfacGMCRescale}
\end{figure*}

The dependence of $X_\CO$ on $\H2$ column density as well as on $W_\CO$ can be captured reasonably accurately (to typically within $<30\%$, except at low metallicities and/or $\H2$ columns) by simple parametrizations. In fact, an appropriate rescaling of $X_\CO$, $N_\H2$ and $W_\CO$ with metallicity and $U_\MW$ removes most of the trends shown in Fig.~\ref{fig:XfacGMC} and results in a simple one-to-one relationship between the rescaled variables, see Fig.~\ref{fig:XfacGMCRescale}. We approximate the relationships between the rescaled variables with a function of the form
\[
y = (1+e^{-x}/x)^{\alpha} (1 + e^{-1/x}x)^{\beta},
\]
which reduces to a power law with slope $-\alpha$ and $\beta$ in the limit $x\rightarrow{}0$ and $x\rightarrow{}\infty$, respectively.

Specifically, in case of a constant  $\CO$ line width, the scaling between $X_\CO$ and $N_\H2$ is well approximated by
\begin{align*}
y&=(Z/Z_\odot)^{0.8} U_\MW^{-0.05} X_\CO  / 6.0\times{}10^{19} \textrm{cm$^{-2}$\,K$^{-1}$\,km$^{-1}$\,s}, \\
x&=(Z/Z_\odot)^{0.83} U_\MW^{-0.06} N_\H2 /2.5\times{}10^{21} \textrm{cm$^{-2}$}, \\
\alpha&=5.5,\textrm{ and }\beta=0.91.
\end{align*}
For a virial scaling of the $\CO$ line width we find that
\begin{align*}
y&=(Z/Z_\odot)^{0.47} U_\MW^{-0.025} X_\CO  / 1.1\times{}10^{20} \textrm{cm$^{-2}$\,K$^{-1}$\,km$^{-1}$\,s}, \\
x&=(Z/Z_\odot)^{0.9} U_\MW^{-0.065} N_\H2 / 2.2\times{}10^{21} \textrm{cm$^{-2}$}, \\
\alpha&=5.5,\textrm{ and }\beta=0.445
\end{align*}
provides a good approximation.

In a similar fashion, the scaling between $X_\CO$ and $W_\CO$ can be approximated by
\begin{align*}
y&=(Z/Z_\odot)^{0.72} X_\CO  / 6.0\times{}10^{19} \textrm{cm$^{-2}$\,K$^{-1}$\,km$^{-1}$\,s}, \\
x&=(Z/Z_\odot)^{0.03} W_\CO / 75\,\textrm{K\,km\,s$^{-1}$}, \\
\alpha&=0.78,\textrm{ and }\beta=11.5, 
\end{align*}
when a constant  $\CO$ line width is assumed.
In case of a virial scaling we obtain
\begin{align*}
y&=(Z/Z_\odot)^{0.44} X_\CO  / 1.25\times{}10^{20} \textrm{cm$^{-2}$\,K$^{-1}$\,km$^{-1}$\,s}, \\
x&=(Z/Z_\odot)^{0.4} W_\CO / 30\,\textrm{K\,km\,s$^{-1}$}, \\
\alpha&=0.78,\textrm{ and }\beta=0.88.
\end{align*}

The parametrization of the relation between $X_\CO$ and $N_\H2$ makes it straightforward to include our $X_\CO$ model in numerical simulations without the full modeling described in section \ref{sect:Xmodel}. The relation between $X_\CO$ and $W_\CO$ may be used to convert observed $\CO$ intensities into $\H2$ column densities. This requires that the observations reach a spatial resolution of $\sim{}60$ pc.

These power law approximations do not provide the overall scaling of the X-factor with metallicity for an ensemble of molecular clouds with a variety of properties. For that we need to marginalize over the $\H2$ column density distribution (or the distribution of $\CO$ intensities) which itself may depend on metallicity. We will discuss this issue in more detail in the next section.
 
\subsection{Scaling with metallicity}
\label{sect:envDepXfac}

\begin{figure*}
\begin{tabular}{cc}
\includegraphics[width=80mm]{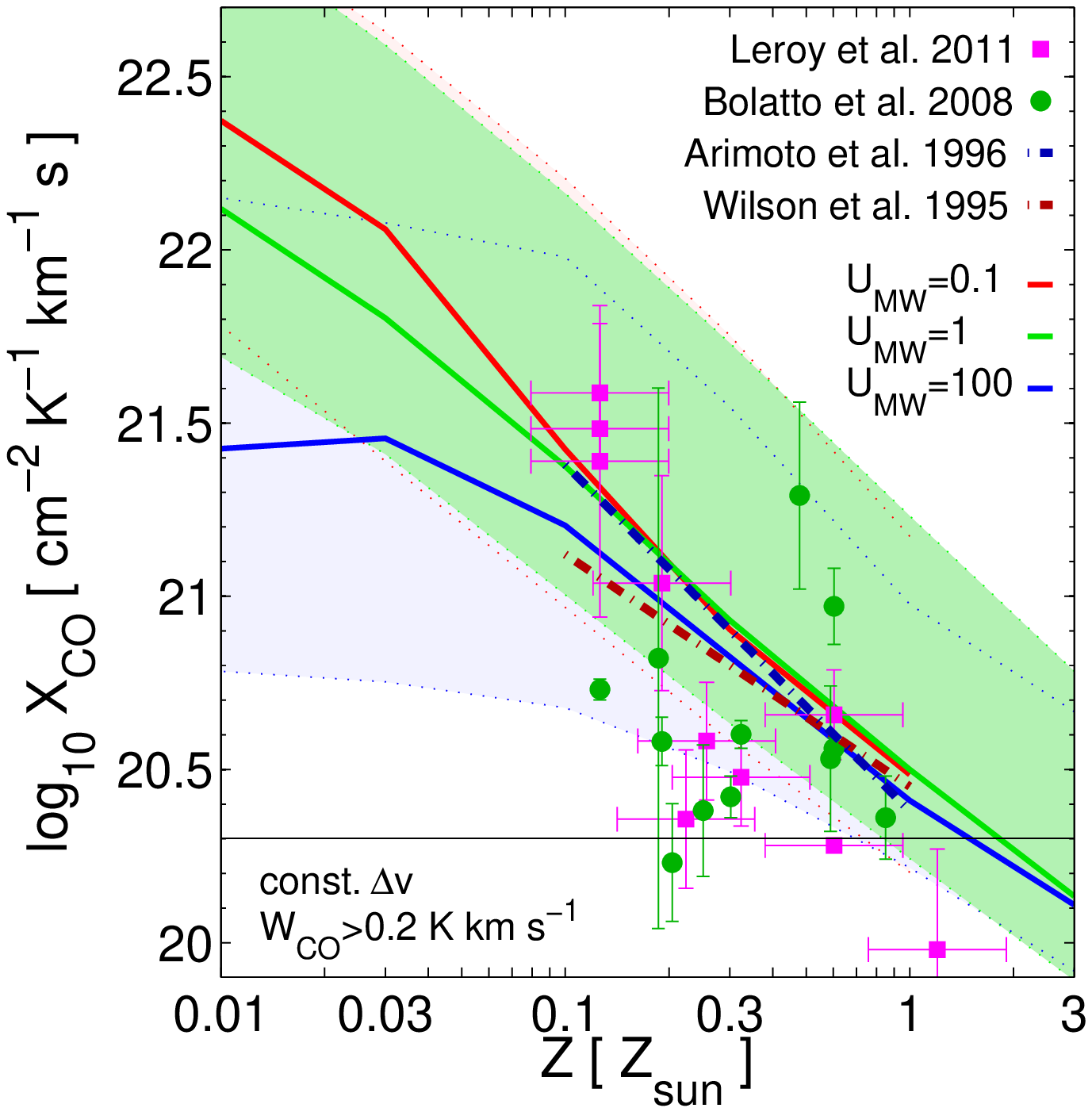} & % analysis/X_vs_Z_m0_l9_CO_new3_H2ART_LFIX20_CLUMP_Lsob_CO1_lr8_I0.2.eps
\includegraphics[width=80mm]{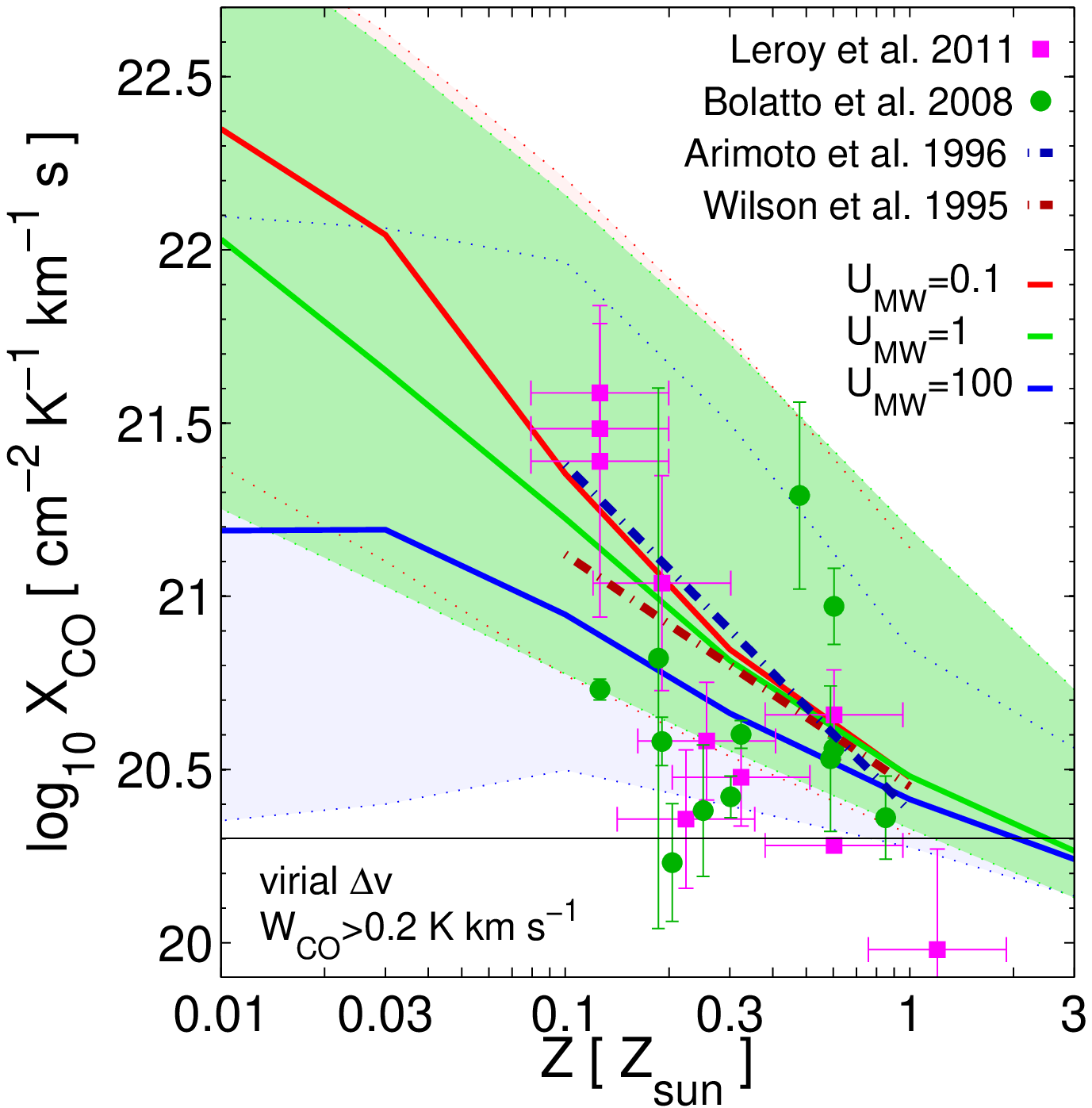}    % analysis/X_vs_Z_m2_l9_CO_new3_H2ART_LFIX20_CLUMP_Lsob_CO1_lr8_I0.2.eps 
\end{tabular}
\caption{Dependence of the X-factor on metallicity and UV radiation field on $\sim{}60$pc scales. The $X_\CO$ predictions are based on cosmological simulations with constrained ISM properties and assume a constant CO line width $\Delta{}v=3$ km s$^{-1}$ (left panel) and a virial line width scaling $\Delta{}v\propto{}\Sigma^{1/2}$ (right panel).  
The solid red, green and blue lines show the median X-factor of all cells above the $\CO$ sensitivity threshold, $W_{\CO}=0.2$ K km s$^{-1}$, for a UV radiation field of $\UMW=0.1$, $\UMW=1$, and $\UMW=100$, respectively. The light shaded areas in red, green and blue show the 16$^{\rm th}$ and 84$^{\rm th}$ percentiles of the X-factor distribution for $\UMW=0.1$, $\UMW=1$, and $\UMW=100$, respectively.
Magenta squares and green circles show X-factor measurements of individual molecular clouds by \cite{2011arXiv1102.4618L} and \cite{2008ApJ...686..948B}, respectively. 
Dot-dashed lines show observed scalings of $X_\CO$ with metallicity by \cite{1995ApJ...448L..97W} and \cite{1996PASJ...48..275A}.
%
%The dark shaded area shows the predictions of an analytical model for the $\H2$ to $\CO$ mass ratio by \cite{2011ApJ...731...25K}. 
%This model is normalized to give $X_{\CO,\MW}$ at $Z=Z_\odot$ (upper edge) or at $Z=0.7Z_\odot$ (lower edge), respectively.
%
The $\CO$ intensity threshold $W_{\CO}=0.2$ K km s$^{-1}$ roughly corresponds to the $3-\sigma{}$ intensity cut for the SMC in the sample of \cite{2011arXiv1102.4618L}. The intensity threshold is higher for other (more metal-enriched) galaxies in their sample, typically $W_{\CO}\sim{}1$ K km s$^{-1}$. Applying this higher threshold has little impact on the median X-factor for $Z\gtrsim{}0.3Z_\odot$, but it narrows the width of the X-factor distribution by $\sim{}0.2$ dex.
The horizontal line at $X_{\CO,\MW}=2\times{}10^{20}$ K$^{-1}$ cm$^{-2}$ km$^{-1}$ s corresponds to $X_{\CO,{\rm MW}}$, the canonical value of the galactic X-factor. On $\sim{}$60 pc scales there is no unique X-factor for a given $Z$ and $\UMW$, but rather a broad distribution with a median that increases with decreasing metallicity and decreasing strength of the interstellar radiation field. Given the uncertainties and the scatter in the observational data, the predictions of our X-factor model are consistent with direct measurements of $X_\CO$ in molecular clouds.}
\label{fig:XfacZ}
\end{figure*}

In Fig.~\ref{fig:XfacZ} we show how the X-factor scales with metallicity and UV radiation field on $\sim{}60$ pc scales. A clear prediction of our model is that $X_\CO$ increases with decreasing metallicity in the range $Z\sim{}0.1-3 Z_\odot$.  We compare our predictions with observations of the X-factor based on infrared (IR) dust emission by \cite{2011arXiv1102.4618L}. These observations seem to indicate a slightly steeper slope, but the deviation to our predictions is within 1-2 standard deviations of the formal fit error and hence not statistically significant. Our predictions also agree well with the slopes and normalizations found in the studies by \cite{1995ApJ...448L..97W} and \cite{1996PASJ...48..275A}.

%While at $Z\sim{}0.1Z_\odot$ the median X-factor lies somewhat below the observations, the 16$^{\rm th}$-84$^{\rm th}$ percentiles of the predicted X-factor distribution cover the \cite{2011arXiv1102.4618L} results. 

Our model predicts that the X-factor decreases with increasing UV field at sub-solar metallicity. A large UV field suppresses molecular clouds with relatively low $\H2$ column densities \citep{2011ApJ...732..115F} and, hence, clouds with large X-factors. Note this happens despite the fact that the X-factor at fixed $\H2$ column density increases with UV field (at least for $Z\gtrsim{}0.1Z_\odot$). This (moderate) UV dependence of the X-factor may contribute to the rather low $X_\CO$ values in some of the GMCs observed by \cite{2011arXiv1102.4618L}.

%We note that \cite{2011arXiv1102.4618L} employ a  $3-\sigma$ sensitivity cut in their $\CO$ maps, which corresponds to $W_\CO\geq{}1$ km s$^{-1}$ (M31, M33, LMC) and $W_\CO\geq{}0.25$ km s$^{-1}$ (SMC). 

Fig.~\ref{fig:XfacZ} also shows estimates of the X-factor based on virial masses from high resolution CO maps by \cite{2008ApJ...686..948B}. These observations do not feature strong metallicity trends, possibly due to the fact that they focus on $\CO$ bright clumps and do not account fully for $\CO$-dark molecular envelopes around those clumps. The scatter in these observations is very large. Interestingly, our X-factor model predicts a similarly large scatter. %In fact, given the large scatter in observations and theory, our X-factor model is also perfectly consistent with the observational measurements by \cite{2008ApJ...686..948B}. 

Our simulations therefore suggest that one should expect significant variations in the X-factor even at fixed metallicity and UV field. However, it is important to point out that the scatter depends on the $\CO$ sensitivity limit, with higher sensitivity (i.e., a lower limit) leading to a larger scatter. This result can be easily understood from Fig.~\ref{fig:XfacGMC} or Fig.~\ref{fig:XfacGMCRescale}. Lowering a sufficiently small sensitivity limit further will imply that more regions with lower $W_\CO$ (and hence lower $N_\H2$ and larger $X_\CO$) are included in the analysis, hence increasing the overall scatter. For instance, for $Z=Z_\odot$, $U_\MW=1$ and a $W_\CO$ threshold of $>0.2$ K km s$^{-1}$ our simulations predict a scatter (defined as half the distance between the 16\% and  84\% percentiles of $\log_{10}X_\CO$) of 0.45-0.5 dex, while it is 0.25-0.3 dex for $W_\CO>1$ K km s$^{-1}$. The scatter is not strongly dependent on the interstellar radiation field over most of the studied parameter range ($U_\MW=0.1-100$, $Z/Z_\odot=0.1-1$).

We can fit the increase of $X_\CO$ with decreasing metallicity with a pure power law, a dependence that is often assumed in the literature, 
\begin{equation}
\label{eq:fit1st}
\log_{10}X_\CO = a_1 \log_{10}(Z/Z_\odot) + a_0.
\end{equation}
We provide the fit parameters in Table \ref{tab:fitPar1}. For instance, for $U_\MW=1$, $W_\CO>0.2$ K km s$^{-1}$, and in case of a virial scaling of the $\CO$ line width the slope of the $X_\CO-Z$ relation is -0.74. What is determining this slope? Clearly, the slope of the $X_\CO-Z$ relation depends on the slope of the $X_\CO-A_V$ relation (see Fig~\ref{fig:FigXfactorModel})  and that of the $A_V-Z$ relation (see below).

\begin{table*}
\begin{center}
\caption{The dependence of $X_\CO$ on metallicity on scales of $\sim{}60$ pc}
\begin{tabular}{|r|r|r|rr|}
\tableline \tableline
$\CO$ line width & $W_\CO$ limit & $\UMW$ & $a_1$ & $a_0$  \\ \tableline
3 km s$^{-1}$ & 0.2 K km s$^{-1}$ & 1 & -0.87  & 20.49   \\
3 km s$^{-1}$ & 0.2 K km s$^{-1}$ & 100 & -0.79 & 20.41  \\
3 km s$^{-1}$ & 1 K km s$^{-1}$ & 1 & -0.81 & 20.40  \\
3 km s$^{-1}$ & 1 K km s$^{-1}$ & 100 & -0.71 & 20.39   \\
virial scaling & 0.2 K km s$^{-1}$ & 1 & -0.74 & 20.46  \\
virial scaling & 0.2 K km s$^{-1}$ & 100 & -0.53 & 20.40  \\
virial scaling & 1 K km s$^{-1}$ & 1 & -0.56  & 20.42  \\
virial scaling & 1 K km s$^{-1}$ & 100 & -0.43 & 20.41 \\
\tableline \tableline
\end{tabular}
\label{tab:fitPar1}
\tablecomments{The first three columns denote (1) the assumption about the scaling of the $\CO$ line width that enters our model, (2) the minimum $\CO$ velocity integrated intensity of a 60 pc scale resolution element in order not to be excluded from the $X_\CO$ distribution, and (3) the normalized strength of the interstellar radiation field. The parameters of equation (\ref{eq:fit1st}), i.e., first order fit parameters between $Z$ and the median of the $X_\CO$ distribution, are provided in the last two columns. The fit parameters are calculated using a least squares fit over the range $0.1Z_\odot\leq{}Z\leq{}1Z_\odot$.}
\end{center}
\end{table*} 

\begin{figure*}
\begin{center}
\begin{tabular}{c}
\includegraphics[width=160mm]{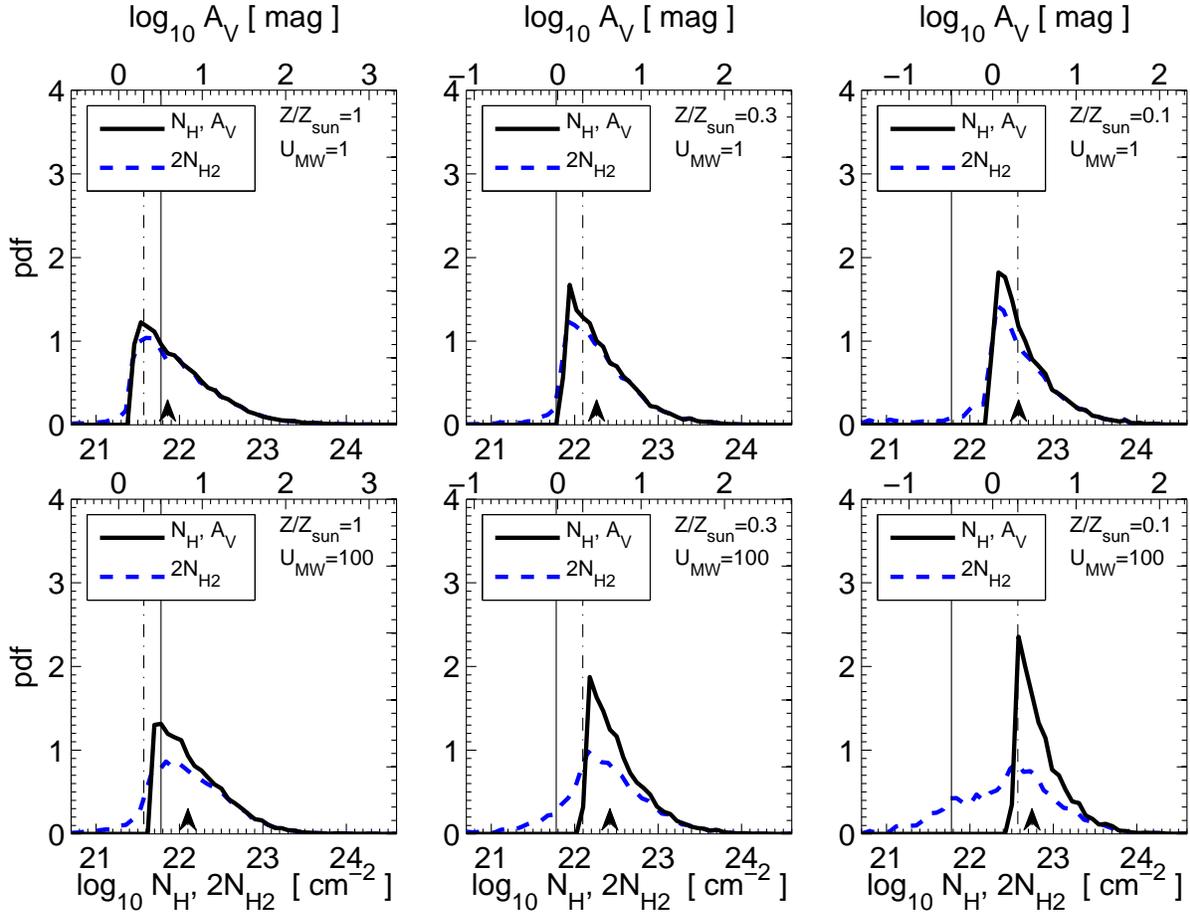}
\end{tabular}
\caption{Probability distribution functions of the mean visual extinction $A_V$, the hydrogen column density $N_{\rm H}$ and the molecular hydrogen column density $N_\H2$ on $\sim{}60$ pc scales as measured in simulations of varying metallicity and radiation field (see legend). The solid black line shows the distribution of $N_{\rm H}$ (bottom axis) and $A_V$ (top axis) of all $\sim{}60$ pc regions with a $\CO$ velocity integrated intensity $W_\CO$ greater than 0.2 K km s$^{-1}$. The sharp fall-off at low column densities is a consequence of the $W_\CO$ sensitivity threshold. The dashed blue line shows the corresponding 
distribution of $2\times{}N_\H2$ (bottom axis). All column density distributions are normalized to an integral of unity over the plotted range. 
The vertical solid line indicates $N_{\rm H}=6\times{}10^{21}$ cm$^{-2}$ and the vertical dot-dashed line corresponds to $A_V=2$. An arrow near the bottom axis shows the median of the $N_{\rm H}$ and $A_V$ probability distribution. The median $A_V$ decreases with metallicity roughly as $\propto{}Z^{0.25-0.3}$ over the considered metallicity range $Z/Z_\odot\sim{}0.1-1$. Consequently, the median column densities of molecular clouds increase with decreasing metallicity.}
\label{fig:AVNHNH2}
\end{center}
\end{figure*}

In order to study the latter we show in Fig.~\ref{fig:AVNHNH2} the (volume-weighted) probability distribution functions of the mean visual extinction $A_V$, the hydrogen column density $N_{\rm H}$, and $\H2$ column density $N_\H2$ of all $\sim{}60$ pc resolution elements above the $\CO$ sensitivity limit $0.2$ K km s$^{-1}$. This figure demonstrates that (1) the median $N_{\rm H}$ increases with decreasing metallicity, (2) the median $A_V$ decreases with decreasing metallicity, and (3) the peak in the $\H2$ surface mass distribution coincides with the peak in the hydrogen surface mass distribution. This latter point is a statement of the fact that a large fraction of the gas that is detectable in CO is hydrogen in molecular form. However, it is noteworthy that there is a significant population of ISM regions with relatively low $N_\H2/N_{\rm H}$ (and hence low $\H2$ mass fractions) that still make it above the $\CO$ sensitivity limit, especially under low $Z$ and high $U_\MW$ conditions.

A simple fit of the change of the median $A_V$ with $Z$ over the range $Z/Z_\odot=0.1-1$ gives $A_V\propto{}Z^{0.25-0.3}$. In addition, a comparison of the median $A_V$ in Fig.~\ref{fig:AVNHNH2} (e.g., $A_V\sim{}3$ for $Z=Z_\odot$, $U_\MW=1$) with the $X_\CO-A_V$ relation, Fig.~\ref{fig:FigXfactorModel}, shows that for such $A_V$ the $X_\CO-A_V$ relation has a negative slope. Therefore, when the metallicity decreases, the median $A_V$ decreases and the median X-factor increases. 
%The $X_\CO-Z$ relation that results in this way will, in general, not be a power law, but may be fitted by such over a limited range in metallicities.

A very different approach to the one presented in this paper has been pursued by
\cite{2011ApJ...731...25K}. We call this ansatz for modeling $X_\CO$, which is based on the results of PDR model calculations by \cite{2010ApJ...716.1191W}, the K/W model and discuss it in more detail in the appendix. The X-factor in the K/W model is given by
\begin{equation}
X_\CO = X_{\CO,0} e^{4\Delta{}A_V/A_V},
\label{eq:KWmodel}
\end{equation}
where $X_{\CO,0}$ is a normalization constant (we use 10$^{20}$ cm$^{-2}$ K$^{-1}$ km$^{-1}$ s), $\Delta{}A_V$ a (weak) function of metallicity of order unity, and $A_V$ the mean visual extinction of the molecular cloud. An important property of this model is that it does not contain an explicit dependence on the interstellar radiation field and, furthermore, the predicted $X_\CO-A_V$ relation is almost independent of metallicity at fixed $A_V$. Of course, in order to derive a $X_\CO-Z$ relation we need to specify $A_V(Z)$. In \cite{2011ApJ...731...25K} the basic assumption is that all molecular clouds have very similar column densities, resulting in $A_V\propto{}Z$. Under this assumption the K/W model predicts a very steep $X_\CO-Z$ relation. If, however, we post-process\footnote{We convert the density and metallicity of a simulation cell into a mean visual extinction by multiplying with the Sobolev-like length $L_{\rm sob}$, see \S\ref{sect:XfactorSims}. $\Delta{}A_V$ is a function of metallicity only and computed as described in the appendix.} our set of ART simulations with the ansatz (\ref{eq:KWmodel}), instead of using the X-factor model presented in section \ref{sect:Xmodel}, we find a  shallower dependence of the X-factor on metallicity. Again, this can be understood based on the fact that our simulations predict a weaker than linear scaling of $A_V$ with metallicity. In fact, if we use $N_{\rm H}\sim{}6\times{}10^{21}$ cm$^{-2}$ (for $Z=Z_\odot$) and the scaling $A_V\propto{}Z^{0.28}$, i.e., values based directly on what we measure in the simulations, we find that (\ref{eq:KWmodel}) results in a $X_\CO-Z$ relation similar to what we get when we post-process our ART simulation suite with the K/W model.

\begin{figure}
\begin{tabular}{c}
\includegraphics[width=80mm]{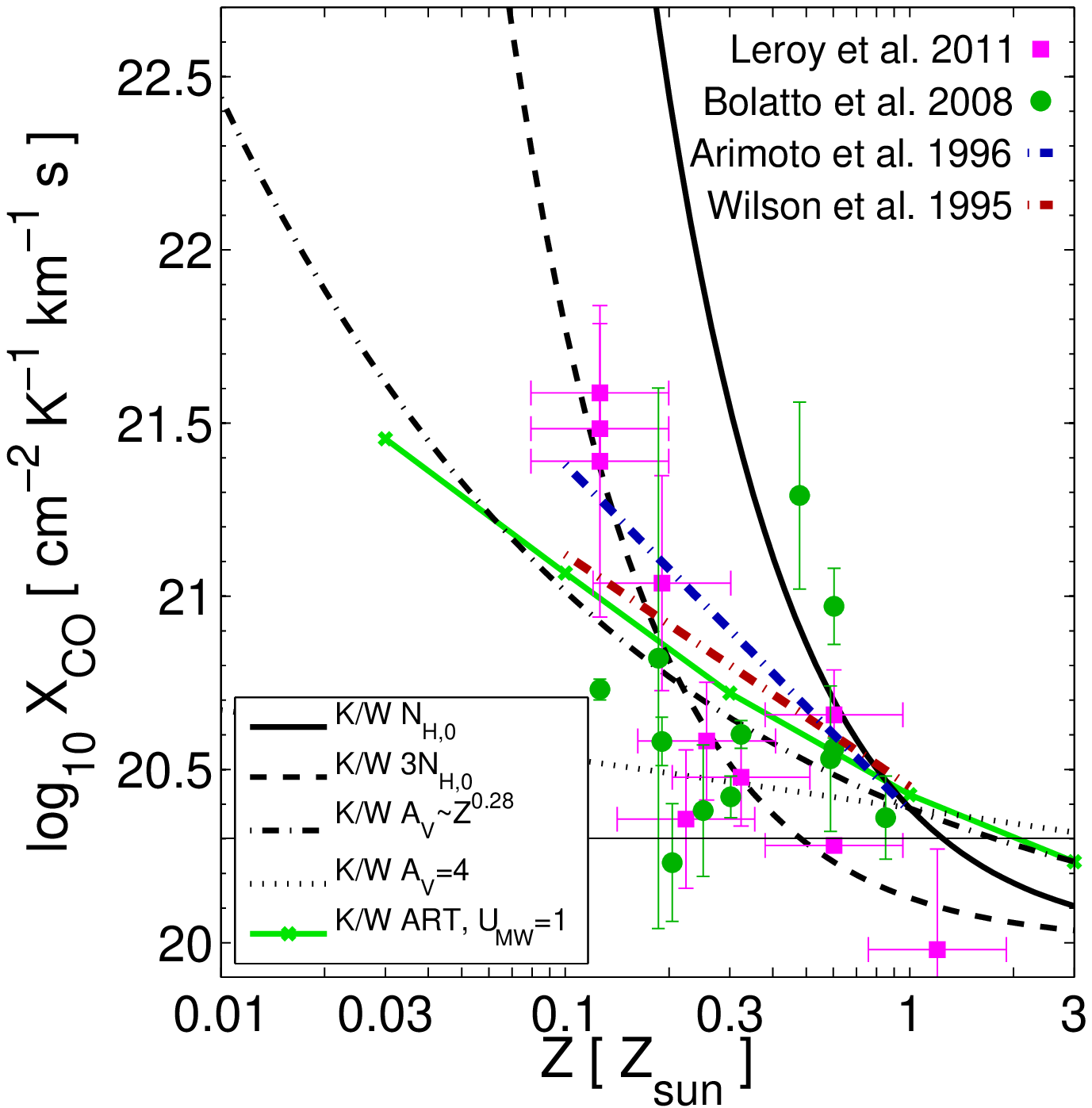}
\end{tabular}
\caption{Dependence of the X-factor on metallicity according to the K/W model (see appendix \ref{sect:KWmodel}; \citealt{2011ApJ...731...25K}, \citealt{2010ApJ...716.1191W}). Many lines and symbols are as in Fig.~\ref{fig:XfacZ}, except for the following. The thick black lines (solid, dashed, dot-dashed, and dotted) show the predictions of the K/W model for different assumptions about the scaling of a clouds mean visual extinction $A_V$ with metallicity. In particular, the black solid line assumes that hydrogen column densities of all molecular clouds (independent of metallicity) are $N_{{\rm H},0}=5.7\times{}10^{21}$ cm$^{-2}$ (corresponding to a nucleon column density $N_{{\rm H},0}/0.76=7.5\times{}10^{21}$ cm$^{-2}$). The dashed line assumes that clouds have hydrogen column densities three times larger. The dotted line corresponds to the case that all molecular clouds have the same visual extinction $A_V=4$, i.e., that $N_{\rm H}\propto{}Z^{-1}$. 
The dot-dashed lines shows $X_\CO$ based on a K/W model variant where solar metallicity molecular clouds have a column density of $N_{\rm H,0}$ and the mean visual extinction scales as $A_V\propto{}Z^{0.28}$. Finally, the solid green line with crosses shows the result of plugging the K/W X-factor model into our set of ART simulations with $U_\MW=1$ and computing the X-factor in the same way as done for Fig.~\ref{fig:XfacZ}. The outcome is a significantly shallower scaling of the X-factor with metallicity compared to models that assume a constant column density of molecular clouds. In fact, it is reasonably close to the dot-dashed line over the range $Z/Z_\odot\sim{}0.1-1$ which uses the scaling of the median $A_V$ with metallicity from the simulations, see Fig.~\ref{fig:AVNHNH2}.}
\label{fig:XfacZKrum}
\end{figure}

We note that the metallicity dependence of the X-factor has nothing to do with the fact that there are fewer carbon atoms in a gas of lower metallicity. We demonstrated in Fig~\ref{fig:xmCO} that the $\CO$ abundance is a strong function of $A_V$, but (at fixed $A_V$) not an explicit function of metallicity. The only exception occurs at very large $A_V$, but then the abundance does not matter because the line is saturated. In our model the $X_\CO-A_V$ relation does (somewhat) depend on $Z$ for large, fixed $A_V$ due to the saturation of the $\CO$ intensity which implies $X_\CO\propto{}N_\H2\propto{}A_V/Z$. However, the $X_\CO-Z$ relation does not change significantly if this $Z$ dependence is eliminated. Instead, as we have shown in this section, on GMC scales the dependence of the X-factor on metallicity is primarily a consequence of the metallicity scaling of the mean visual extinction of molecular clouds above a given $\CO$ detection limit.

\subsection{Implications for surface densities of GMCs}
\label{sect:SurfDensGMC}

Many galactic and extragalactic surveys assume a constant value of the X-factor, close to the galactic conversion factor $X_{\CO,\MW}=2\times{}10^{20}$ K$^{-1}$ cm$^{-2}$ km$^{-1}$ s, to predict $\H2$ masses or column densities from $^{12}\CO$ data. This may introduce potential biases in the inferred $\H2$ column density distributions.

\begin{figure*}
\begin{center}
\begin{tabular}{c}
\includegraphics[width=160mm]{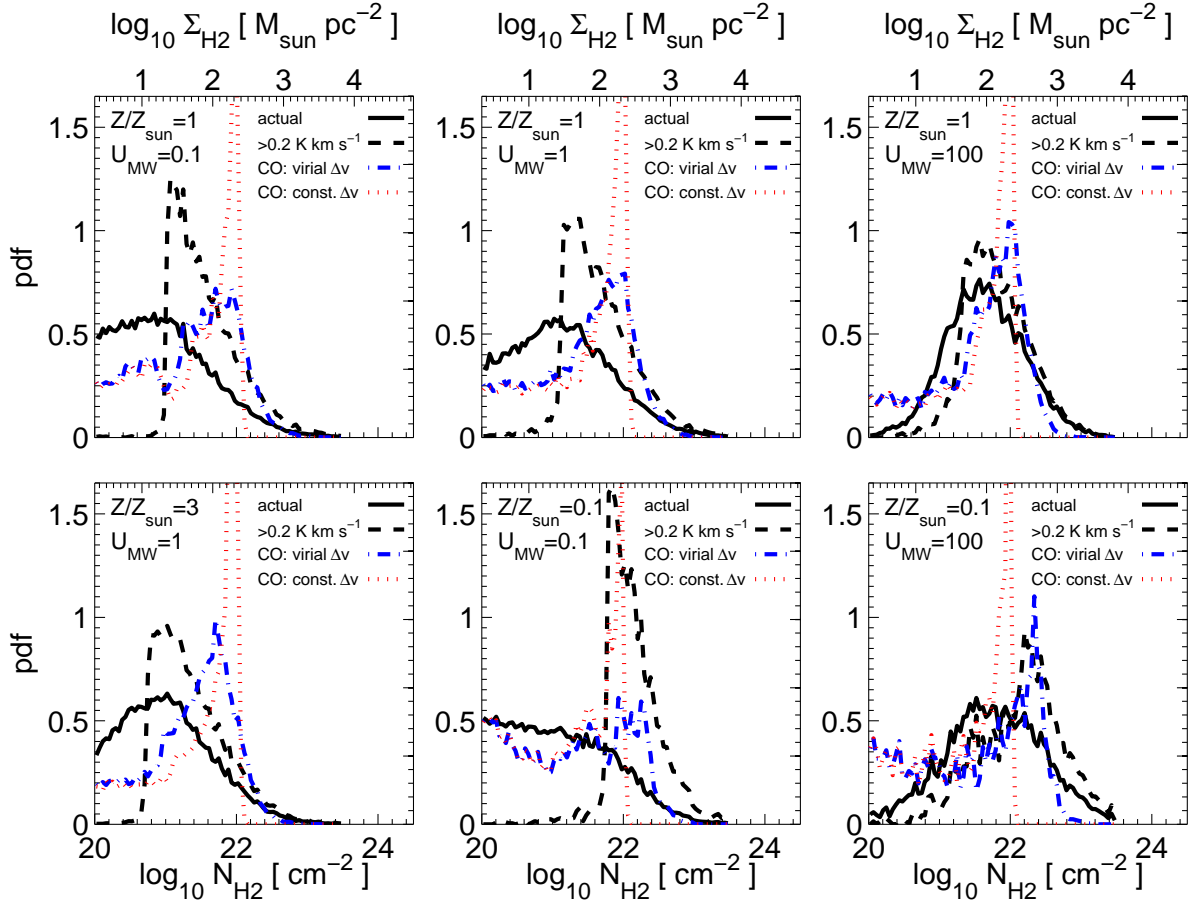} % analysis/NH2_comp_l9_CO_new3_H2ART_LFIX20_CLUMP_Lsob_CO1_lr8_M0_ICO0
\end{tabular}
\caption{Probability distribution function of the $\H2$ column density on $\sim{}60$ pc scales for simulations of varying metallicity and radiation field (see legend). The column density distributions are normalized to an integral of unity over the plotted range. 
In each panel the solid black curve shows the actual $\H2$ column density distribution of all $60$ pc regions in the simulation. The dashed black curve shows this distribution after discarding regions that have $\CO$ integrated intensities (according to our X-factor model with a virial scaling of the $\CO$ line width) below 0.2 K km s$^{-1}$. 
The dot-dashed blue (virial scaling) and dotted red (constant $\CO$ line width) curves show the $\H2$ column density distribution that is inferred from converting the $\CO$ emission as predicted by our X-factor model (see \S\ref{sect:Xmodel}) into $N_\H2$ with the help of the canonical X-factor $X_{\CO,\MW}=2\times{}10^{20}$ K$^{-1}$ cm$^{-2}$ km$^{-1}$ s. We note that these inferred $\H2$ density distributions due not change over the plotted range depending on whether resolution elements with $W_\CO<0.2$ K km s$^{-1}$ are discarded, because any such element has $N^{\rm obs}_\H2=X_{\CO,\MW}W_\CO < 4\times{}10^{19}$ cm$^{-2}$.
%
%A $\CO$ intensity threshold of $W_\CO\ge{}1$ K km s$^{-1}$ would cut the pdfs of the inferred $\H2$ column densities below $\sim{}2\times{}10^{20}$ cm$^{-2}$. 
The increase of $X_\CO$ with $N_\H2$ biases the inferred $\H2$ column density compared to the actual $\H2$ column density if a constant MW-like X-factor is assumed. This results in an apparent narrowing of the $\H2$ column density distribution and leads to a peak near $10^{22}$ cm$^{-2}$.}
\label{fig:NH2GMC}
\end{center}
\end{figure*}

To address this question we show in Fig.~\ref{fig:NH2GMC} the probability distribution function of $N_\H2$ on $\sim{}60$ pc scales for simulations with different metallicities and UV field strengths.

The solid black lines in each panel show the actual distribution of $\H2$ column densities as measured in the simulations, i.e., using the
$\H2$ density within each $\sim{}60$ pc resolution element and converting it into a column density by multiplying it with $L_{\rm sob}$, see \S\ref{sect:XfactorSims}.
The figure shows that an increase in the UV interstellar radiation field has a significant effect on the $N_\H2$ distribution. Low $\H2$ columns are suppressed and the peak of the distribution narrows and shifts toward higher $\H2$ column densities due to the environmental dependence of the HI to $\H2$ transition on $\UMW$. The hydrogen column density distribution is affected to a significantly smaller degree by a change in the radiation field. 
%Note that, at large $\UMW$, the transition between $f_\H2=0$ and $f_\H2=1$ with total gas column density is very sharp \citep{2011ApJ...728...88G} due to the role of dust in shielding the $\H2$ from the impeding UV flux. Hence, for $\UMW=100$, low $\H2$ column densities ($10^{21}$ cm$^{-2}$ or less) are only found along lines of sight with large total gas column densities.

The dashed black lines in Fig.~\ref{fig:NH2GMC} correspond to the $\H2$ distribution after discarding all $\sim{}60$ pc regions with a $\CO$ velocity integrated intensity below 0.2 K km s$^{-1}$. This $\CO$ sensitivity cut removes preferentially lines of sight with intermediate and low $\H2$ column densities, e.g., $N_\H2\lesssim{}10^{21}$ cm$^{-2}$ for $Z=1$, $U_\MW=1$, and leads to a much narrower and peakier distribution.

The blue dot-dashed and red dotted lines in Fig.~\ref{fig:NH2GMC} show the \emph{inferred} $\H2$ column densities $N^{\rm obs}_\H2=X_{\CO,\MW}W_\CO$, i.e., those derived by multiplying the $\CO$ integrated intensity with the galactic conversion factor $X_{\CO,{\rm MW}}$. Specifically, for a Milky-Way like ISM the inferred 
range\footnote{Measured as  $\mathrm{HWHM}/\sqrt{2\ln(2)}$, where HWHM is the distance from the peak to half the maximum at higher column densities.} 
of $\H2$ column densities ($\sim{}0.2$ dex for a virial scaling, $\sim{}0.05$ dex for a constant $\CO$ line width) is significantly smaller than the true width of the $\H2$ column density distribution ($\sim{}0.7$ dex). The peak position of the true $\H2$ column density distribution is at $\sim{}1.3\times{}10^{21}$ cm$^{-2}$, while the inferred distributions peak at $\sim{}1\times{}10^{22}$ cm$^{-2}$.

In contrast to the actual $\H2$ column density distribution, the inferred $\H2$ column density distributions due not change if a $\CO$ intensity limit $W_\CO>0.2$ K km s$^{-1}$ is imposed, at least for $N^{\rm obs}_\H2>4\times{}10^{19}$ cm$^{-2}$. Hence, the much narrower range of the inferred $\H2$ density distributions and the bias towards higher column densities is \emph{not} the consequence of such a limit. Instead, it arises due to the scaling of the X-factor with column density, as we now demonstrate.

The inferred $\H2$ column density $N^{\rm obs}_\H2$ is given as 
\[
N^{\rm obs}_\H2=X_{\CO,\MW}W_\CO=N_\H2 X_{\CO,\MW}/X_\CO.
\]
Figure~\ref{fig:XfacGMC} shows that in the optically thick regime the X-factor can be approximated as
\[
X_\CO = X_{\CO,\MW} \left(\frac{N_\H2}{10^{22}\textrm{cm$^{-2}$}}\right)
\]
for a constant $\CO$ line width, and as
\[
X_\CO = X_{\CO,\MW} \left(\frac{N_\H2}{10^{22}\textrm{cm$^{-2}$}}\right)^{1/2}
\]
for a virial scaling of the line width. Consequently the inferred $\H2$ column density is
\[
N^{\rm obs}_\H2 = 10^{22}\textrm{cm$^{-2}$},\textrm{and } N^{\rm obs}_\H2 = \left(10^{22}\textrm{cm$^{-2}$}N_\H2\right)^{1/2}
\]
for a constant line width and a virial scaling of the line width, respectively. 

In the optical thin regime the X-factor raises steeply (well above $X_{\CO,\MW}$) with decreasing column density and hence in general $N^{\rm obs}_\H2 < N_\H2$. This effect is particularly visible in high UV, low metallicity environments where the inferred (but not the actual!) $\H2$ column density distribution has a significant tail towards low $\H2$ column densities.

\begin{figure}
\begin{tabular}{c}
\includegraphics[width=85mm]{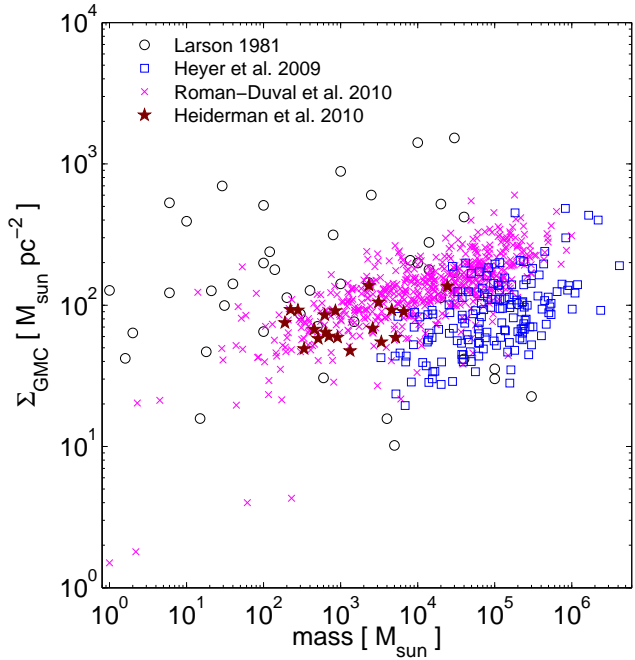} % figures/M_Sigma
\end{tabular}
\caption{Mean gas surface density of molecular clouds as a function of cloud mass from observations using non-$^{12}\CO$ tracer. \cite{1981MNRAS.194..809L} combines data from the literature, primarily $^{13}\CO$ observations, available at the time. It is thus a rather heterogeneous compilation and shows a large scatter, but no obvious trend of surface density with cloud mass. Recent $^{13}\CO$ surveys with larger statistics and better resolution \citep{2009ApJ...699.1092H, 2010ApJ...723..492R} find a trend of increasing surface density with cloud mass (or radius). These surveys also find significant variations in the cloud surface density at fixed physical scale. The LTE masses of \citet{2009ApJ...699.1092H} are multiplied by a factor two as suggested by the authors. Studies that derive gas surface density from extinction maps show reduced scatter if the surface densities are measured above a fixed extinction threshold ($A_V=2$ in \citealt{2010ApJ...723.1019H}).}
\label{fig:RSigma}
\end{figure}

The bias that we describe above adds another complication to the intense discussion of whether molecular clouds have a ``constant mean surface density'' (e.g., \citealt{1989A&A...225..517K, 1990ASSL..162..151S, 1997ApJ...474..292V, 2002ApJ...564..773E, 2002ApJ...570..734B, 2010A&A...519L...7L, 2010ApJ...723.1019H, 2010ApJ...724..687L}). Observations, that are not based on $^{12}\CO$, and hence do not suffer from the mentioned bias, 
demonstrate that it is true in a weak sense, namely as the lack of a strong correlation of the mean surface densities of clouds with their sizes or masses, see 
Fig.~\ref{fig:RSigma}. Yet, a stronger and more controversial interpretation is that all clouds have very similar mean surface densities. In fact, \cite{2010A&A...519L...7L} find, based on extinction measurements of a small sample of molecular clouds, that the mean surface densities are constant with only $\sim{}15\%$ scatter \emph{if the cloud areas are defined by a fixed extinction threshold} (but cf. \citealt{2011arXiv1107.0966G}). Using a similar approach \cite{2010ApJ...723.1019H} find that the scatter is $\sim{}30\%$ if surface densities are measured above a fixed extinction threshold of $A_V=2$.

How do these observations fit together with the result shown in the top panel of Fig.~\ref{fig:NH2GMC}, namely that of a rather broad distribution of $\H2$ column densities, or with the observations of significant variations in observed GMC surface densities shown in Fig.~\ref{fig:RSigma}?

A hint to a possible solution is that the study of \cite{2010A&A...519L...7L} finds an order of magnitude variation in the enclosed mass (and consequently surface density) if the cloud mass is measured \emph{within an aperture of fixed size} (and not within a fixed $A_V$ contour). It is thus conceivable that the ``constancy of the mean surface density'' is simply a matter of the identification method\footnote{Clouds in the MW have all rather similar metallicities and a fixed extinction threshold thus corresponds to a fixed threshold of the local surface density. Therefore, if molecular clouds had an approximately self-similar (fractal) structure \citep{1996ApJ...471..816E} or similar column density distributions (cf., \citealt{2011MNRAS.tmp.1101B}), then the mean surface density should scale with the local surface density of a given contour. This does not necessarily mean that the surface density of the entire region that is molecular is the same for each cloud.} of the molecular cloud and its characteristic properties. 

Hence, surveys that either do not properly resolve the clouds or measure masses at a fixed physical scale should find large scatter in the mean surface densities. Since we measure the surface densities on a fixed $\sim{}60$ pc scale we indeed expect to see are rather broad distribution of $\H2$ surface densities.

\section{Spatial averaging and the X-factor on galactic scales}
\label{sect:spatav}

Many extragalactic surveys use $^{12}\CO$ observations to infer spatially averaged $\H2$ column densities on $\sim{}$kpc patches of galaxies or even for galaxies as a whole (e.g., \citealt{1998ARA&A..36..189K, 2008AJ....136.2846B}). In most cases a single, constant conversion factor is assumed. In section \S\ref{sect:envDepXfac} we demonstrated and discussed the dependence of the conversion factor on metallicity and interstellar radiation field on GMC scales. We now discuss these dependences on larger averaging scales.
A related question that we want to address is by how much the X-factor can vary around its most typical value. Such variation arise from (1) the change of $X_\CO$ with $N_\H2$, and (2) the scatter of $X_\CO$ at fixed $N_\H2$ due to the degeneracy with $\H2$ fraction and hydrogen column density, see \S\ref{sect:XfacScaleGMCs}.

We compute the X-factor on large scales as the volume-weighted average\footnote{The use of a volume average instead of an area average can be justified as follows. First, cloud self-covering is presumably relatively small as the size of each resolution element ($\sim{}60$ pc) constitutes a significant fraction of the scale height of the gas disk. Second, even clouds that do spatially overlap, e.g., in lines of sight edge-on through a disk galaxy, likely have a large enough velocity difference so that their $\CO$ intensities can be added. Since the former problem only arises in (near) edge-on views the results in this section may be safely interpreted in any case as predictions for sufficiently inclined (closer to face-on) views on disk galaxies.} of the small scale $\sim{}60$ pc resolution elements,
\begin{flalign*}
\langle{}X_\CO\rangle{} &= \frac{\langle{}N_\H2\rangle{}}{\langle{}W_\CO\rangle{}}=\frac{\int{}N_\H2{}dV}{\int{}\frac{N_\H2}{X_\CO}dV},
\end{flalign*}
i.e., the spatially averaged X-factor on scale $l$ at a particular point $P$ in the simulation volume is computed as the ratio of the sum of the column densities and the $\CO$ integrated intensities of all $\sim{}60$ pc resolution elements within a box of extent $l$ centered on $P$.

\begin{figure*}
\begin{tabular}{cc}
\includegraphics[width=80mm]{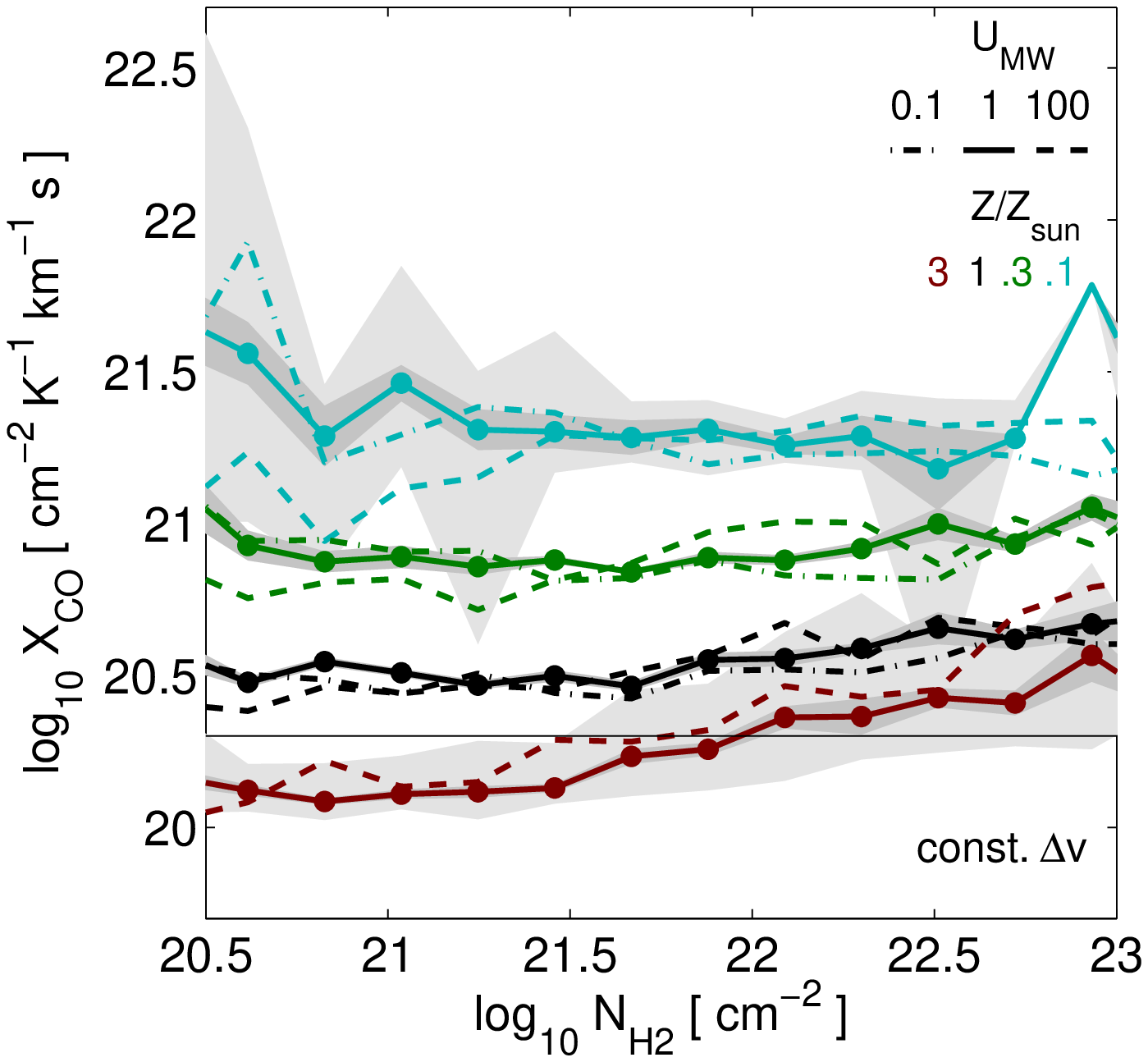} & % analysis/X_vs_NH2_m0_l5_CO_new3_H2ART_LFIX20_CLUMP_Lsob_CO1_lr8}
\includegraphics[width=80mm]{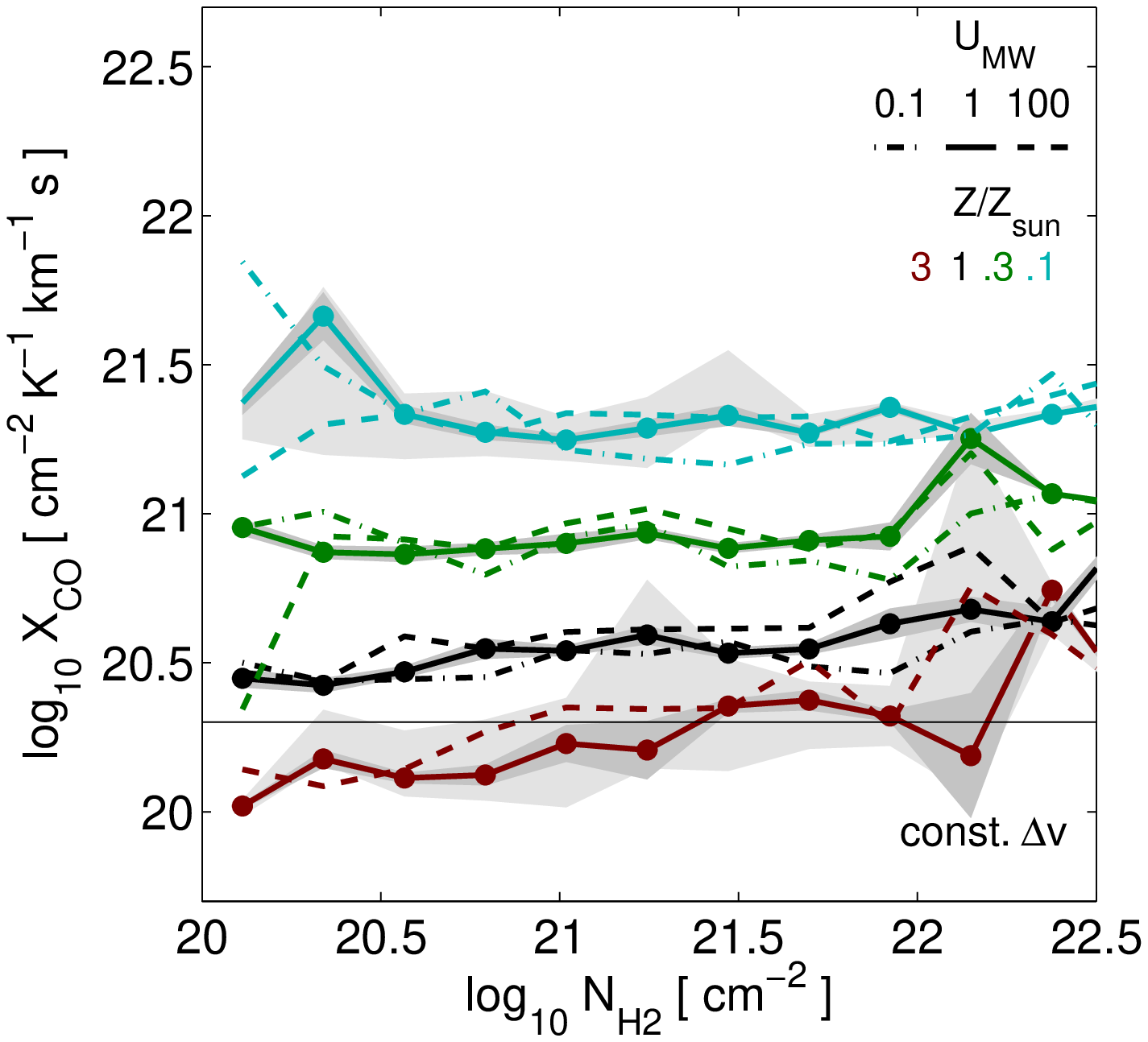} \\ % analysis/X_vs_NH2_m0_l3_CO_new3_H2ART_LFIX20_CLUMP_Lsob_CO1_lr8}
\includegraphics[width=80mm]{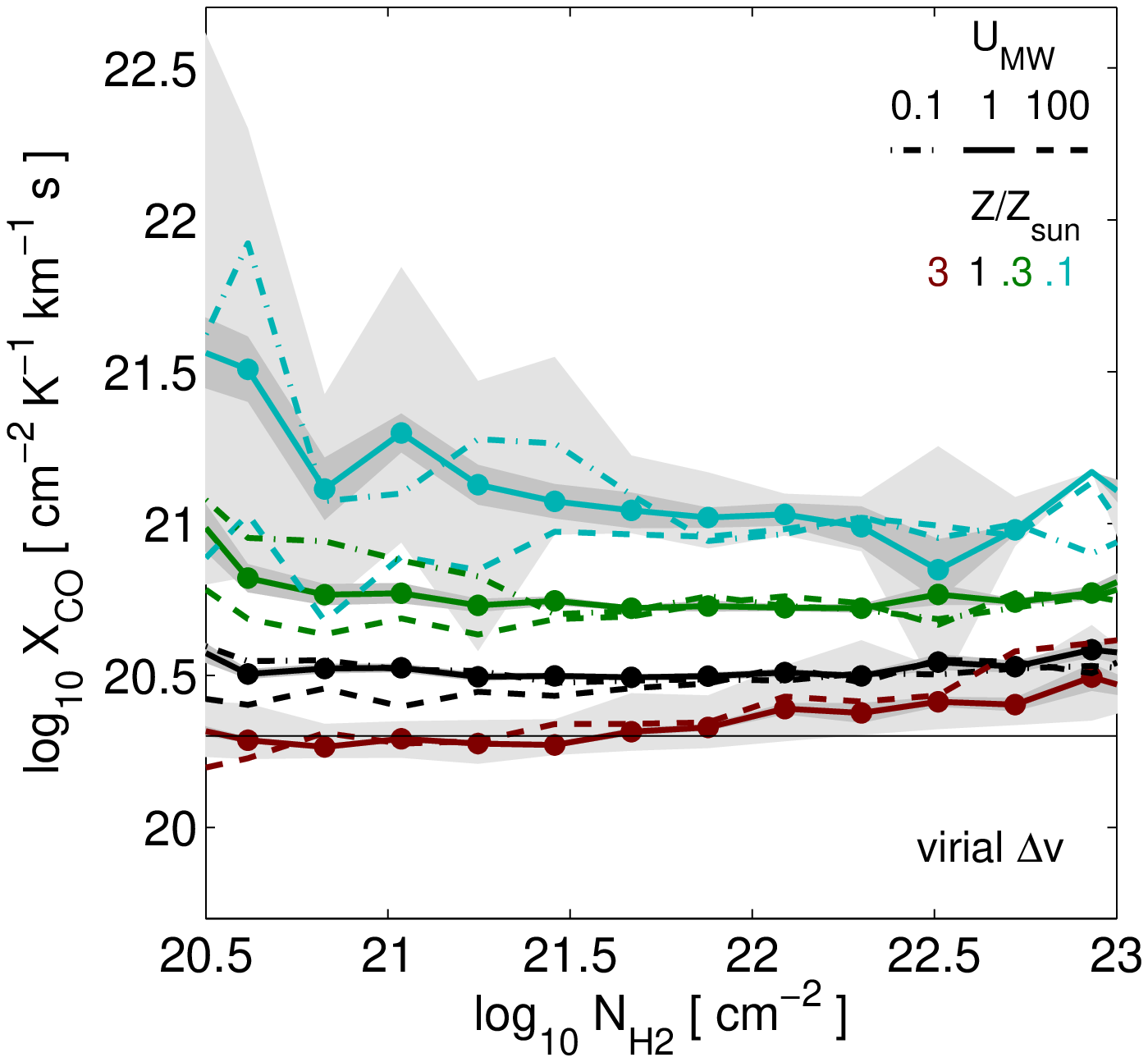} & % analysis/X_vs_NH2_m2_l5_CO_new3_H2ART_LFIX20_CLUMP_Lsob_CO1_lr8}
\includegraphics[width=80mm]{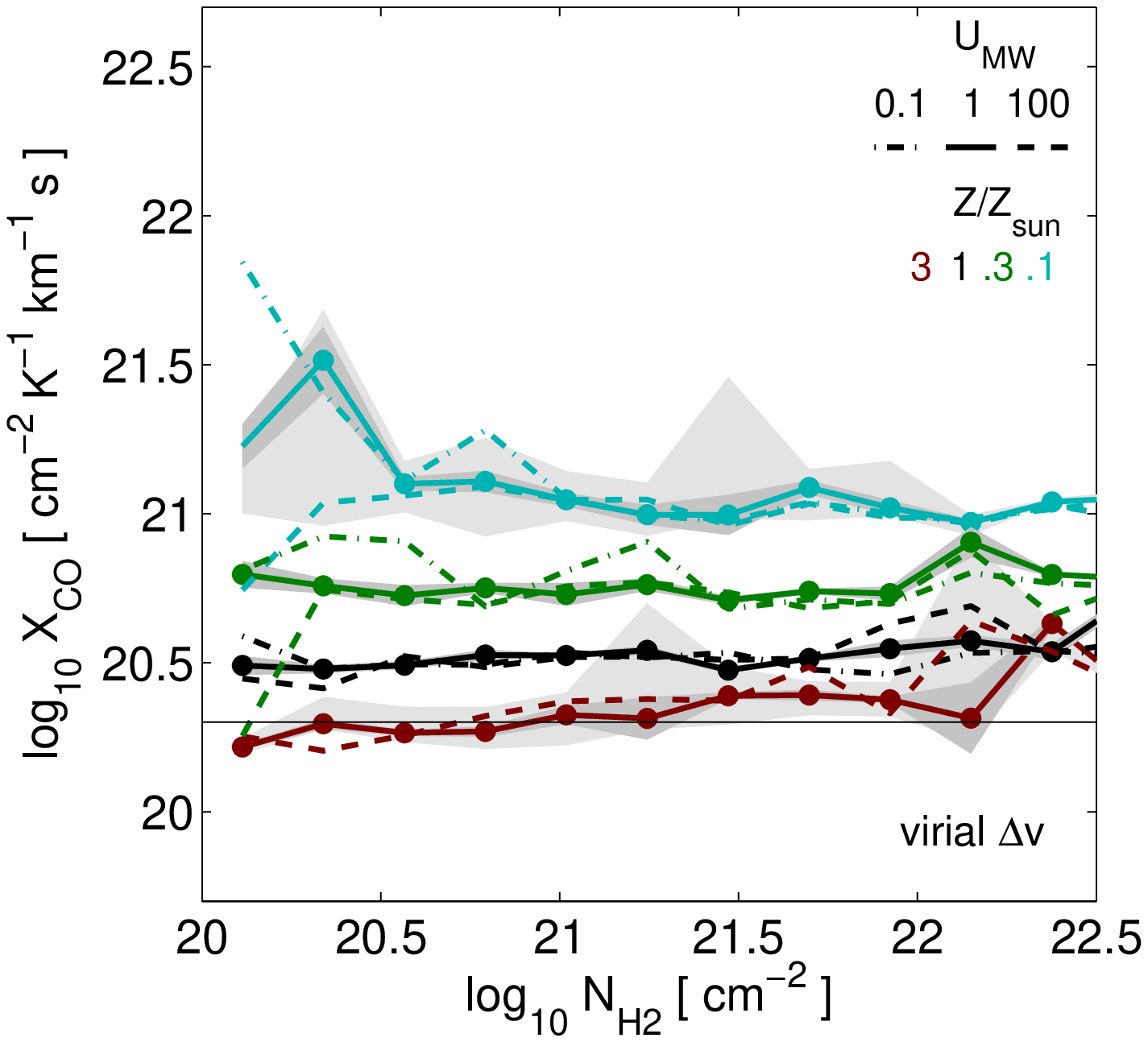}    % analysis/X_vs_NH2_m2_l3_CO_new3_H2ART_LFIX20_CLUMP_Lsob_CO1_lr8}
\end{tabular}
\caption{\small X-factor vs $\H2$ column density on galactic scales. 
(Top row) model predictions of the X-factor assuming a constant line width ($\Delta{}v=3$ km s$^{-1}$). (Bottom row) analogous predictions assuming a virial line width scaling ($\Delta{}v\propto{}\Sigma^{1/2}$), see \S\ref{sect:Xmodel}.
(Left column) spatial averaging on scales of 1 kpc. (Right column) spatial averaging on scales of 4 kpc. 
The curves correspond to cosmological simulations with constrained ISM properties. The UV radiation field varies between $\UMW=0.1$ (dot-dashed lines), $\UMW=1$ (solid lines), and $\UMW=100$ (dashed lines). The metallicity varies (from bottom to top; see legend) between $Z=3Z_\odot$, $Z=Z_\odot$, $Z=0.3Z_\odot$, and $Z=0.1Z_\odot$. Each curve connects the median $X_\CO$ for a given $\H2$ column density. The 16$^{\rm th}$ and 84${\rm th}$ percentiles of the $X_\CO$ distribution for $\UMW=1$ are shown by light gray shaded areas (for $Z=3Z_\odot$ and $Z=0.1Z_\odot$). The standard deviation of the average $X$-factor is shown by the dark shaded regions. Both, percentiles and standard deviation, are only trustworthy if the given $\H2$ column density bin contains 5 or more data points (indicated by filled circles). The galactic X-factor $X_{\CO,\MW}=2\times{}10^{20}$ K$^{-1}$ cm$^{-2}$ km$^{-1}$ is shown by a horizontal solid line. The X-factor on $\gtrsim{}$kpc scales depends primarily on metallicity and only weakly on the $\H2$ column density or the strength of the UV interstellar radiation field.}
\label{fig:XfacLS}
\end{figure*}

Figure~\ref{fig:XfacLS} shows the resulting $X_\CO$ as a function of $\H2$ column density on 1 kpc and 4 kpc scales. Compared with Fig.~\ref{fig:XfacGMC} the spatial averaging on kpc scales and above reduces the variation of $X_\CO$ with $N_\H2$, especially for low $N_\H2$. For instance, at solar metallicity and on kpc scales, the X-factor changes by less than a factor of two when $N_\H2$ changes by 2 orders of magnitude between $10^{21}$ cm$^{-2}$ and $10^{23}$ cm$^{-2}$. In contrast, on $\sim{}60$ pc scales the corresponding change is a factor $\sim{}100$.

\begin{figure*}
\begin{center}
\begin{tabular}{c}
\includegraphics[width=165mm]{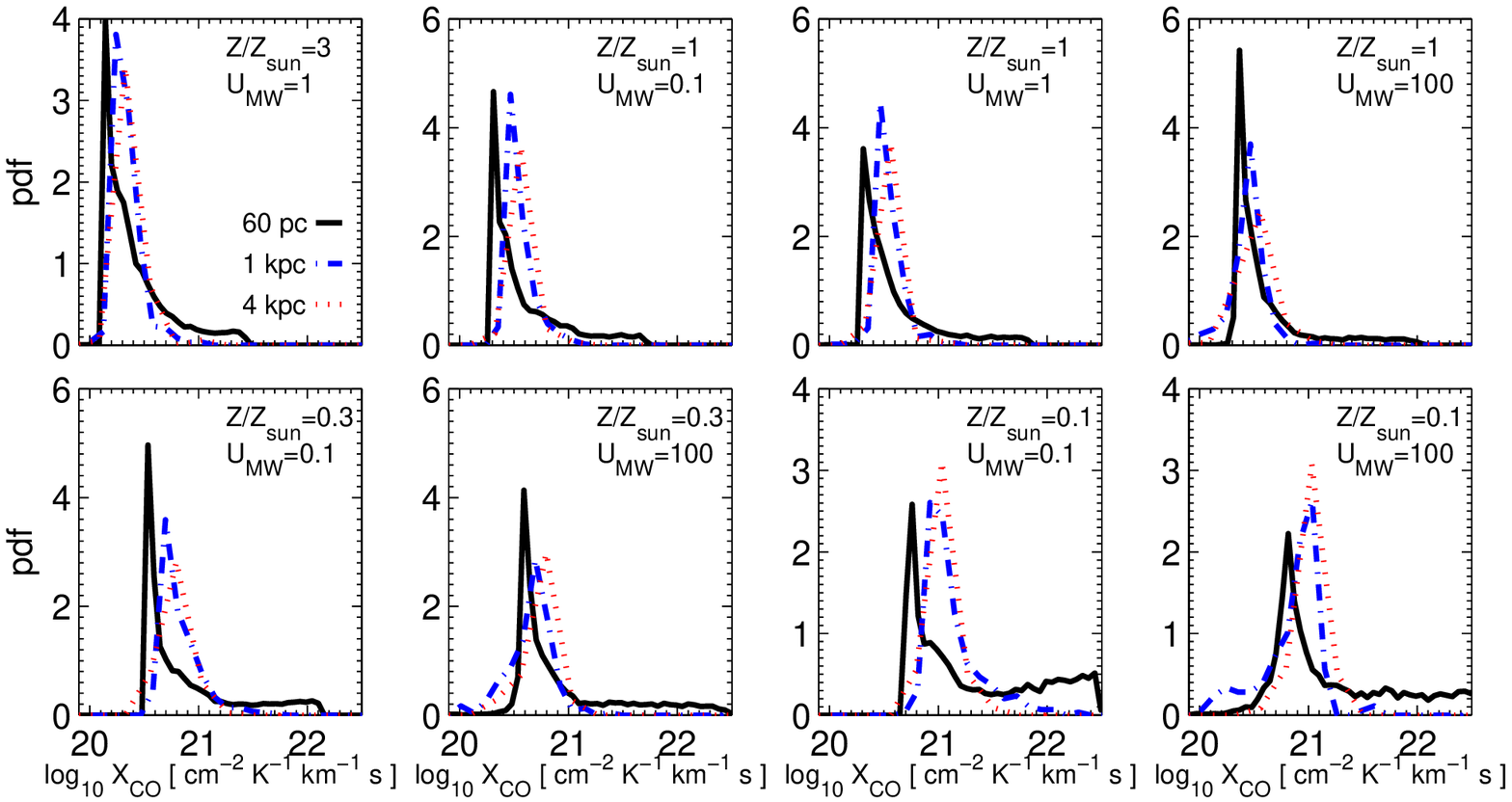} % analysis/Xhist2_m2_l3_CO_new3_H2ART_LFIX20_CLUMP_Lsob_CO1_lr8_M0_ICO0.2.eps
\end{tabular}
\caption{Distribution of $X_\CO$ as function of scale and ISM properties. Each panel corresponds to a simulation with constrained ISM properties (see legend). Different lines correspond to different spatial averaging scales: 60 pc (solid black line), 1 kpc (dot-dashed blue line), 4 kpc (dotted red line). Only patches of the ISM with  $\CO$ intensities above 0.2 K km s$^{-1}$ are included in the plotted $X_\CO$ distribution. The underlying X-factor model assumes a virial scaling of the $\CO$ line width. The $X_\CO$ distribution becomes narrower and more symmetric when measured on $\gtrsim{}$kpc averaging scales compared with GMC scales. Also the peak of the distribution shifts toward larger values of $X_\CO$.}
\label{fig:Xhist}
\end{center}
\end{figure*}

This demonstrates that the median (or the mean) of the $X_\CO$ distribution becomes less dependent on $N_\H2$ as one goes to larger scales, but it does not tell us much about the scatter of $X_\CO$ on such scales. We therefore show in Fig.~\ref{fig:Xhist} the $X_\CO$ distribution on both $\gtrsim{}$kpc and on $\sim{}60$ pc scales. We only include those 60 pc, 1 kpc, or 4 kpc ISM patches that have a $\CO$ velocity integrated intensity of at least 0.2 K km s$^{-1}$.

The figure highlights two important effects of spatial averaging. First, the $X_\CO$ distribution on $\gtrsim{}$kpc scales is considerably narrower than the one measured on scales of GMCs. Also the peak shifts toward higher $X_\CO$ values. Secondly, the distribution of $\log_{10}X_\CO$ becomes more symmetric and bears a closer resemblance to a normal distribution. Hence, even if $X_\CO$ is not constant on small scales and can, in fact, vary over an order of magnitude or more, the spatial averaging ensures that variations of the X-factor on $\gtrsim{}$ kpc scales are much smaller. For instance, the X-factor varies by typically less than a factor $\sim{}2$ around its peak value on $\gtrsim{}$kpc scales.

\begin{figure*}
\begin{tabular}{ccc}
\includegraphics[width=80mm]{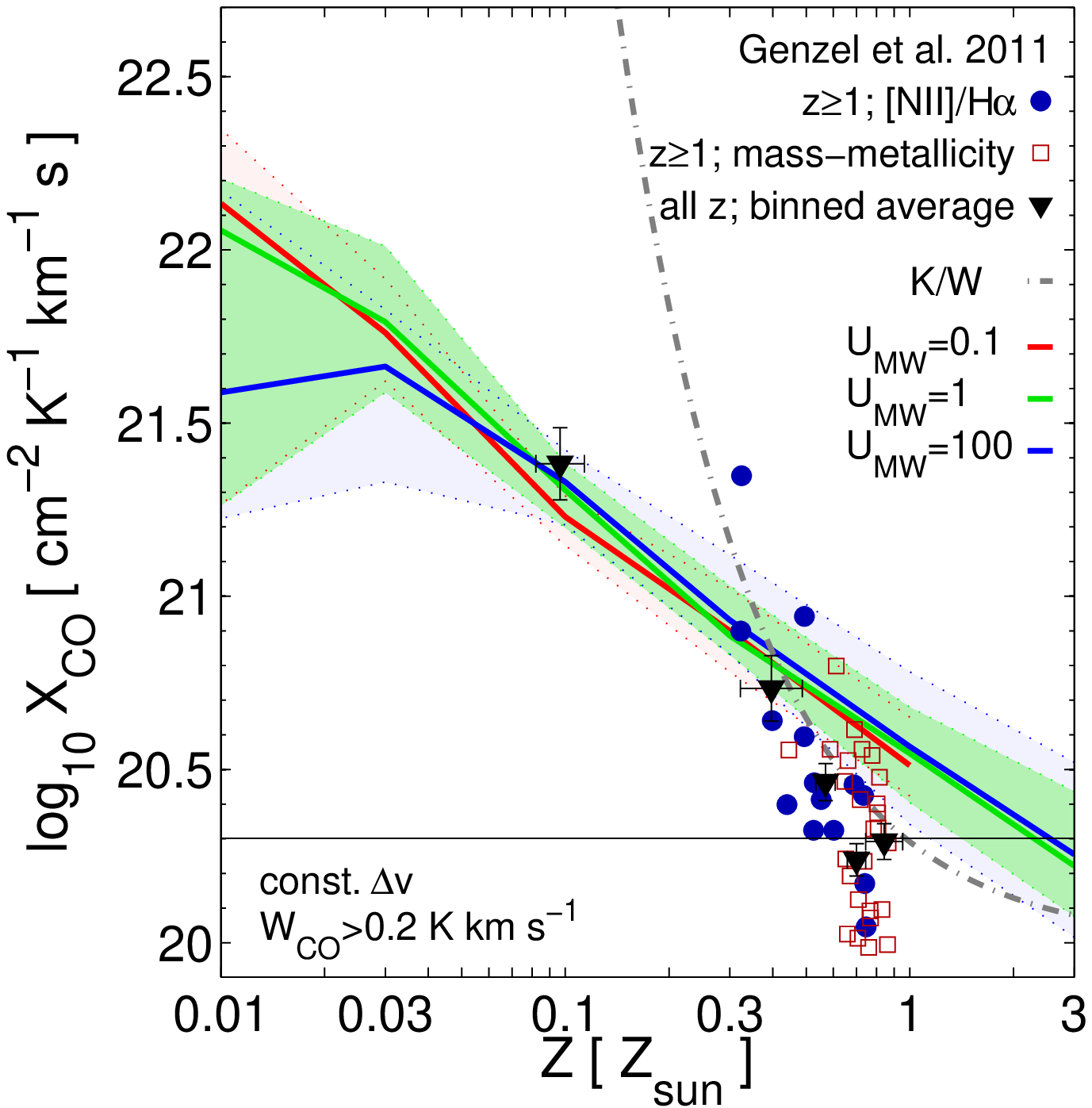} & % analysis/X_vs_Z_m0_l3_CO_new3_H2ART_LFIX20_CLUMP_Lsob_CO1_lr8_I0.2.eps}
\includegraphics[width=80mm]{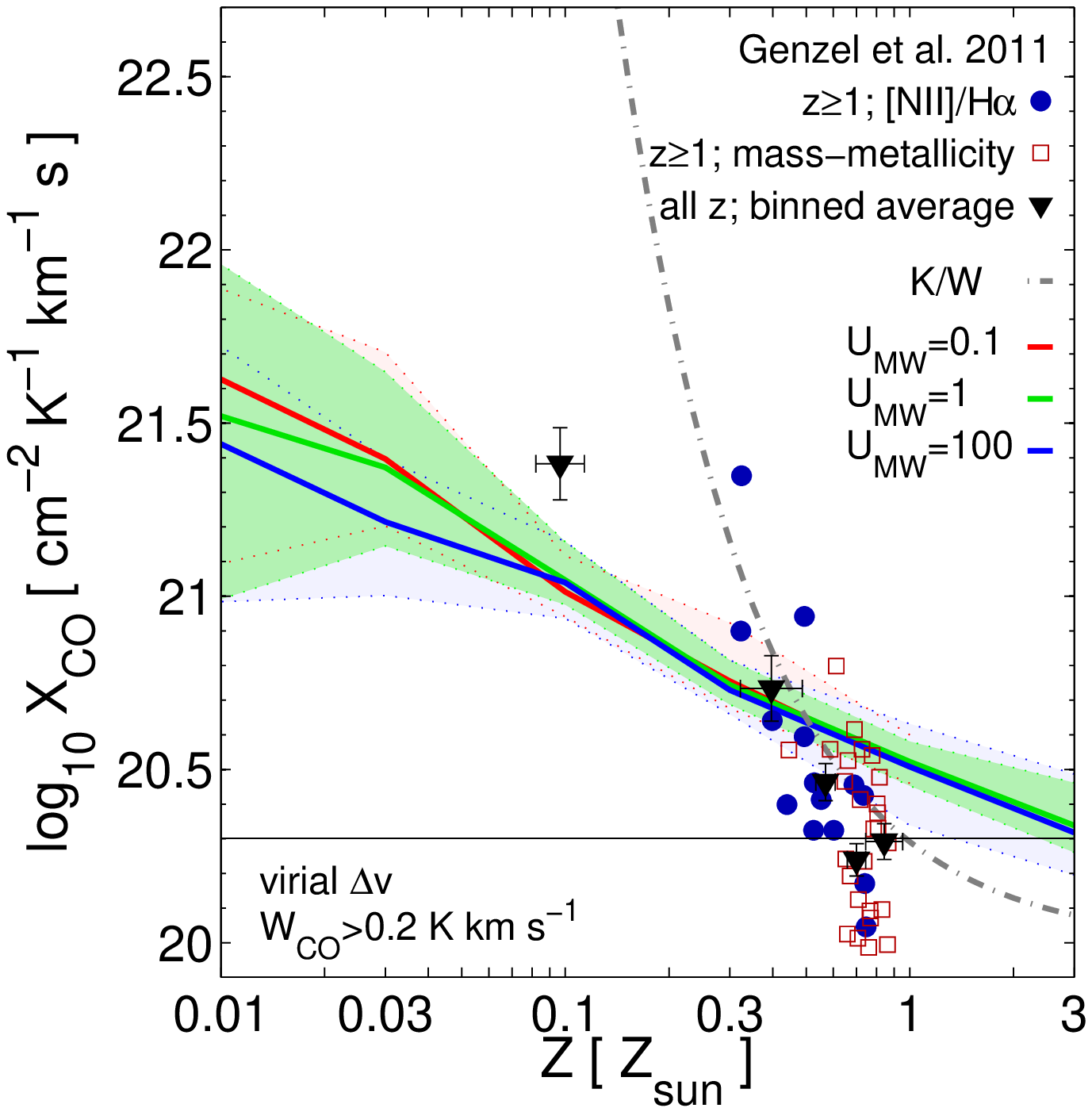} % analysis/X_vs_Z_m2_l3_CO_new3_H2ART_LFIX20_CLUMP_Lsob_CO1_lr8_I0.2.eps}
\end{tabular}
\caption{Dependence of the X-factor on metallicity and UV radiation field on 4 kpc scales.  The $X_\CO$ predictions are based on cosmological simulations with constrained ISM properties and assume a constant CO line width $\Delta{}v=3$ km s$^{-1}$ (left panel), and a virial line width scaling $\Delta{}v\propto{}\Sigma^{1/2}$ (right panel). Blue circles, red squares, and black triangles indicate $X_\CO$ estimates that have been derived assuming a non-evolving, almost linear $\Sigma_{\H2}-\Sigma_{\rm SFR}$ relation out to $z\sim{}3$ \citep{2011arXiv1106.2098G}. All other symbols and labels are as in Fig.~\ref{fig:XfacZ}. Our X-factor model predicts a strong dependence of $X_\CO$ on metallicity, although less steep compared with the predictions of \cite{2011arXiv1106.2098G} or the K/W model (see appendix). The scaling is close to a power law with exponent -0.76 (constant $\CO$ line width), and exponent -0.52 (virial line width scaling), respectively. Fit parameters are provided in Table \ref{tab:fitPar2}.}
\label{fig:XfacZLS}
\end{figure*}

In Fig.~\ref{fig:XfacZLS} we quantify the dependence of the X-factor on metallicity and UV radiation field on 4 kpc scales. Similar to the results on GMC scales, see section \ref{sect:envDepXfac}, we find that $X_\CO$ is strongly metallicity dependent. However, the UV field (in the range $\UMW=0.1-100$) plays now no important role. This can be understood from the fact that a change in the $\H2$ column density distribution hardly affects the X-factor, see Fig.~\ref{fig:XfacLS}. The UV field becomes relevant only at at $Z<0.1Z_\odot$, due to the decrease of the X-factor with increasing UV field at fixed $\H2$ column density for a low metallicity ISM as discussed previously.

The scaling of $X_\CO$ with metallicity is close to a power law. The parameters of a first order fit (in $\log$-$\log$ space) can be found in Table \ref{tab:fitPar2}. Quantitatively, the power law exponent does not seem to be a strong function of spatial scale (compare with Fig.~\ref{fig:XfacZLS} with Fig.~\ref{fig:XfacGMC} and Table \ref{tab:fitPar1} with Table \ref{tab:fitPar2}). The value of the power law exponent depends on assumptions about the $\CO$ line width and the detection threshold in $\CO$ intensity. We typically find power law indices $a_1$ for the median $X_\CO-Z$ relation in the range $\sim{}[-0.5,-0.8]$. 

The power law slope can be understood from the results presented in \S \ref{sect:ParamModel}. Since our galaxy models have a homogeneous metallicity distribution, the metallicity rescaling of $X_\CO$, $N_\H2$ and $W_\CO$ that removes the most of the metallicity dependence on small scales also works on galactic scales. In fact, Fig. \ref{fig:XfacLS} shows that $X_\CO$ is almost independent of $N_\H2$ at fixed metallicity. Hence, $y\sim{}Z^{\gamma}X_\CO$ (see \S\ref{sect:ParamModel}) is approximately independent of both $Z$ and $N_\H2$ and, therefore, $X_\CO\propto{} Z^{-\gamma}$. This scaling argument predicts $a_1=-\gamma\sim{}-0.8$ in case of a constant $\CO$ line width and $a_1\sim{}-0.5$ for a virial scaling. These expectations agree well with the ones obtained from a direct fit (Table  \ref{tab:fitPar2}).

The metallicity dependence that we find is shallower than the dependence expected from attributing offsets of $z>1$ star forming galaxies in the $\Sigma_{\H2} -\Sigma_{SFR}$ relation merely to the metallicity dependence of $X_\CO$ \citep{2011arXiv1106.2098G}, which requires a power law index of $\sim{}[-1.3, -1.9]$. This could imply that some part of the observed offsets is not due to $X_\CO$ variations, but, possibly, due to an actual deviation from the $\Sigma_{\H2} -\Sigma_{SFR}$ relation, as is observed more dramatically in mergers \citep{2010MNRAS.407.2091G} or in local galaxies with high specific star formation rates \citep{2011MNRAS.415...61S}. A dependence of the $^{12}\CO _{2\rightarrow{}1}$ (and/or $^{12}\CO_{3\rightarrow{}2}$) to $^{12}\CO_{1\rightarrow{}0}$ conversion factor on metallicity may be an alternative possibility. 
%
%A potential caveat that one has to keep in mind is that the $z\gtrsim{}1$ sample of \cite{2011arXiv1106.2098G} only spans a relatively narrow range in metallicities.
%Finally, our model also predicts a much milder increase of the X-factor with decreasing metallicity compared with the model of \cite{2011ApJ...731...25K}.

\begin{table*}
\begin{center}
\caption{The dependence of $X_\CO$ on metallicity on scales of 4 kpc}
\begin{tabular}{|r|r|r|rr|rrr|}
\tableline \tableline
$\CO$ line width & $W_\CO$ limit & $\UMW$ & $a_1$ & $a_0$ \\ \tableline
3 km s$^{-1}$ & 0 K km s$^{-1}$ & 1 & -0.82 & 20.51 \\
3 km s$^{-1}$ & 0 K km s$^{-1}$ & 100  & -0.83 & 20.55  \\
3 km s$^{-1}$ & 0.2 K km s$^{-1}$ & 1 & -0.76 & 20.53   \\
3 km s$^{-1}$ & 0.2 K km s$^{-1}$ & 100 & -0.76 & 20.56  \\
3 km s$^{-1}$ & 1 K km s$^{-1}$ & 1 & -0.75 & 20.54  \\
3 km s$^{-1}$ & 1 K km s$^{-1}$ & 100 & -0.71 & 20.60 \\
virial scaling & 0 K km s$^{-1}$ & 1 & -0.66 & 20.49 \\
virial scaling & 0 K km s$^{-1}$ & 100 & -0.65 & 20.47  \\
virial scaling & 0.2 K km s$^{-1}$ & 1 & -0.52 & 20.51  \\
virial scaling & 0.2 K km s$^{-1}$ & 100 & -0.53 & 20.49  \\
virial scaling & 1 K km s$^{-1}$ & 1 & -0.47 & 20.52  \\
virial scaling & 1 K km s$^{-1}$ & 100 & -0.50 & 20.51 \\
\tableline \tableline
\end{tabular}
\label{tab:fitPar2}
\tablecomments{
As Table \ref{tab:fitPar2}, but for spatial scales of 4 kpc.
}
\end{center}
\end{table*}

\section{Summary and Conclusions} 

A proper physical interpretation of observational data based on $\CO$ emission requires a thorough understanding and modeling of the conversion factor, the X-factor, between molecular hydrogen and the $J=1\rightarrow{}0$ emission line of carbon monoxide. We presented a novel approach to study the properties of the X-factor based on a combination of high resolution ($\sim{}0.1$ pc) MHD simulations of the ISM \citep{2011MNRAS.412..337G} as a ``subgrid'' model and gas distribution of $\sim 60$~pc scales derived from self-consistent cosmological hydrodynamical simulations. This approach is feasible, because the spatial resolution of our cosmological simulations ($\sim{}60$ pc) is reasonably well matched to the box sizes of the ISM simulations ($\sim{}20$ pc). In fact, this scale, which roughly corresponds to the scale of GMCs, is a natural scale to separate physical processes occurring on galactic and cosmological scales from those relevant for star formation. 

The main advantage of our approach is that it takes into account the complicated interplay of chemical evolution, heating, and cooling in a turbulent ISM on small scales and is thus a step forward compared to, e.g., one-dimensional steady state PDR models or to ad hoc models of small scale physics. 
A caveat is, of course, that it also suffers from any inaccuracies or missing physics in the underlying small scale simulations, but we are hopeful that these problems can be alleviated in the future. 
Our approach extents previous work that study the X-factor on ISM scales (e.g., \citealt{2011MNRAS.412.1686S, 2011MNRAS.tmp..943S}) to galactic scales and can be easily used in cosmological simulations, where all cell properties (densities, metallicities, radiation fields) are known, and thus provide self-consistent boundary conditions for the sub-grid modeling. 

The main predictions of our model and the main results of this paper are as follows.
\begin{enumerate}
\item The X-factor on GMC scales depends sensitively on metallicity, dust extinction, $\H2$ column density, while two orders of magnitude variations of the UV radiation field lead only to moderate changes of $X_\CO$ at a fixed hydrogen column. The changes in $X_\CO$ are even smaller at a fixed $\H2$ column density, because a higher UV field tends to reduce the $\H2$ column density at a given hydrogen column, which offsets to some extent the increase of $X_\CO$ at fixed hydrogen column.

\item In galaxies with solar metallicity the X-factor on GMC scales is predicted to be $\sim{}2-4\times{}10^{20}$ K$^{-1}$ cm$^{-2}$ km$^{-1}$ s (variations due to dependence of $X_\CO$ on $N_\H2$), in agreement with the canonically assumed galactic conversion factor, for a wide range of gas column densities ($\sim{}50-500$ $M_\odot$) and with relatively small scatter at fixed $N_\H2$ ($<0.3$ dex).  Our model also predicts an increase of the X-factor with decreasing metallicity that is in quantitative agreement with direct measurements of $X_\CO$ from IR dust emission.

\item The simulations predict an $\H2$ column density distributions with a peak around $\sim{}30-100$ $M_\odot$ pc$^{-2}$ for an ISM with solar metallicity, in agreement with GMCs observations in the Milky Way.
 
\item We show that GMC column densities inferred from $^{12}\CO$ ($J=1\rightarrow{}0$) observations via a constant galactic conversion factor are biased toward higher column densities and show a strong peak in their distribution, see Fig.~\ref{fig:NH2GMC}. In addition, such an approach makes the spurious prediction of many lines of sight with low $\H2$ columns in ISM regions that are exposed to a large UV field. These observational biases are a consequence of relying on a constant galactic conversion factor that ignores the scaling of $X_\CO$ with $\H2$ column density.

\item Spatial averaging, from GMC to $\sim{}$ kpc scales (and beyond), decreases the dependence of $X_\CO$ on the $\H2$ (or total gas) column and the scatter in the X-factor. The column density dependence only remains at very low or large $\H2$ columns, see. Fig~\ref{fig:XfacLS}. The UV radiation field plays only a small role on $\sim{}$kpc scales, and hence, to a good approximation, the metallicity is the primary driver of the X-factor on $\gtrsim{}$kpc scales.

\item On kpc scales and above the X-factor scales with metallicity approximately as a power law with an exponent in the range $[-0.5,-0.8]$, depending on assumptions about the $\CO$ line width and the applied $\CO$ intensity detection threshold. The power law exponent on GMC scales is similar $[-0.5,-0.9]$.

\end{enumerate}

Recent $\CO$ emission studies \citep{2010ApJ...714L.118D, 2010ApJ...713..686D, 2010MNRAS.407.2091G, 2010Natur.463..781T, 2011ApJ...734L..25E} have improved our knowledge of the relation between star formation and its fuel, molecular gas, out to redshift 3. 
With instruments of the next generation, e.g., the square kilometer array, it will further be possible to detect $\CO$ bright gas at even higher redshifts. Going back in time these galaxies presumably differ strongly in their properties from their present-day (or even their $z\sim{}1-3$) counterparts. Since all these studies observe $\CO$ we need to understand whether this gives unbiased estimates of the $\H2$ density in the environments probed by the observations. Hence, a clear understanding and precise modeling of the systematic trends of the X-factor with metallicity and $\H2$ column density (among other factors) will remain a crucial challenge in order to properly interpret these future observations.\\

\acknowledgements 
The authors thank M. Krumholz for insightful and critical comments on an early draft of this work and for providing the IDL routine of the \cite{2011ApJ...731...25K} model. They thank M. Krumholz and D. Narayanan for stimulating discussions during the ``Star Formation in Galaxies: From Recipes to Real Physic'' workshop at the Aspen Center for Physics, the ``2011 Aspen Meeting on Galaxy Evolution'', and the ``2011 Santa Cruz Galaxies Workshop''. RF thanks S. Glover, A. Saintonge, and R. Shetty for discussions about the X-factor during the Ringberg meeting ``Star Formation in Galaxies: The Herschel Era''. The authors wish to express their gratitude towards the organizers of these workshops. This work was supported in part by the DOE at Fermilab, by the NSF grants AST-0507596 and AST-0708154, and by the Kavli Institute for Cosmological Physics at the University of Chicago through the NSF grant PHY-0551142 and an endowment from the Kavli Foundation. The simulations used in this work have been performed on the Joint Fermilab - KICP Supercomputing Cluster, supported by grants from Fermilab, Kavli Institute for Cosmological Physics, and the University of Chicago. This work made extensive use of the NASA Astrophysics Data System and {\tt arXiv.org} preprint server.

\appendix

\section{The Krumholz / Wolfire model}
\label{sect:KWmodel}

\begin{figure*}
\begin{center}
\begin{tabular}{c}
\includegraphics[width=80mm]{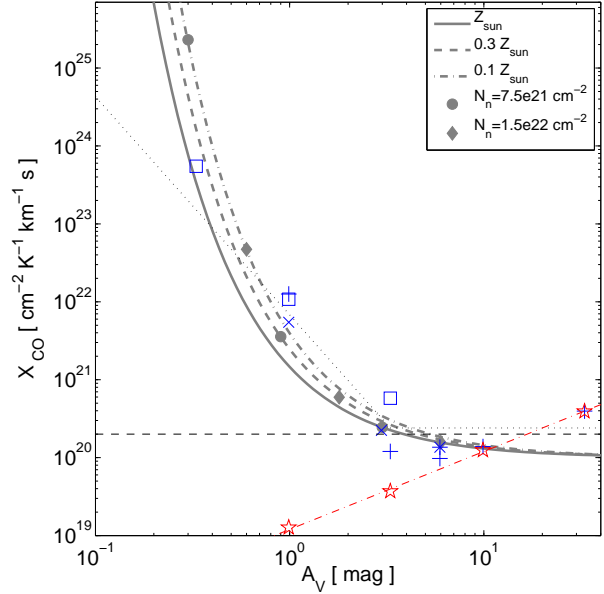}
\end{tabular}
\caption{X-factor of the $J=1\rightarrow{}0$ $^{12}\CO$ transition as function of mean extinction in the K/W model. The solid, dashed, and dot-dashed gray lines are the predictions of the K/W model for solar, 0.3 solar, and 0.1 solar metallicity, respectively. Filled circles and diamonds show the predicted X-factor for a nucleon column density $N_{\rm n}=N_{\rm H}/0.76=7.5\times{}10^{21}$ cm$^{-2}$ ($\sim{}60$ $M_\odot$ pc$^{-2}$) and $1.5\times{}10^{22}$ cm$^{-2}$ ($\sim{}120$ $M_\odot$ pc$^{-2}$), respectively. The other symbols show the results of driven turbulence, magneto-hydrodynamics simulations presented in \cite{2011MNRAS.412..337G}. The star symbols correspond to a simulation without UV radiation, the other symbols are for simulations with $\UMW=1$, see Fig.~\ref{fig:FigXfactorModel}. The horizontal, dashed line indicates the galactic X-factor $X_{\CO,{\rm MW}}=2\times{}10^{20}$ cm$^{-2}$ K$^{-1}$ km$^{-1}$ s. The dotted line indicates the trend of X-factor with $A_V$ found in \cite{2011MNRAS.412..337G} for $\UMW=1$. The thin dot-dashed line in the bottom right shows the expected scaling $X_\CO\propto{}A_V$ when the $\CO$ line is optically thick and the line width fixed.
For a Milky-Way like ISM ($U_\MW=1$, $Z=Z_\odot$) the predictions of the K/W model agree reasonably well with the results of the ISM simulations by \cite{2011MNRAS.412..337G} and, hence, with the predictions of our X-factor model presented in section \ref{sect:Xmodel}, except possibly at $A_V<0.4$ and $A_V>10$. However, the K/W model, which is not explicitly dependent on the strength of the UV field, differs noticeably from the predictions of our X-factor model for $\UMW\neq{}1$. Also, it differs from the results of the no UV field run by \cite{2011MNRAS.412..337G}.}
\label{fig:XKW}
\end{center}
\end{figure*}

In this section we give a short outline of the Krumholz/Wolfire (K/W) X-factor model and compare its predictions with those presented in section \ref{sect:Xmodel}. The K/W model was introduced by \cite{2011ApJ...731...25K} and is largely based on the results of PDR simulations by \cite{2010ApJ...716.1191W}.
 
The latter authors studied the case of a spherical cloud with an overall 1/r density profile embedded in an isotropic radiation field. The cloud consists of cold  ($T<300$ K) gas clumps of density $n_c$ that contain most of the cloud mass and a warm ($T\sim{}8000$ K) interclump medium with large volume filling factor. For a given mean visual extinction $A_V$,  metallicity $Z$ and impinging radiation field $U_\MW$ Wolfire et al. compute (1) the radius $R_\CO$ at which the optical depth of the $J=1\rightarrow{}0$ transition to the cloud surface is unity, and (2) the radius $R_\H2$ at which the $\H2$ mass fraction is 50\%. Hence, they arrive at a prediction for the ratio between the cloud mass within $R_\H2$ and the cloud mass within $R_\CO$
\begin{equation}
\label{eq:massRatio}
\frac{M(<R_\H2)}{M(<R_\CO)}=\exp(4\,[0.53 - 0.045\ln(U_\MW / n_c [\rm{cm^{-3}}]) - 0.097 \ln(Z/Z_\odot)]/A_V).
\end{equation}
The dependence of the mass ratio on $U_\MW / n_c [\rm{cm^{-3}}]$ can be eliminated with some additional modeling (\citealt{2008ApJ...689..865K,2009ApJ...693..216K})
\begin{equation}
\label{eq:UMWovern}
U_\MW / n_c [\rm{cm^{-3}}] = 0.044(1+3.1 [Z/Z_\odot]^{0.365})/4.1.
\end{equation}
Based on the expression for the mass ratio it is now possible to construct a simple estimate for the X-factor 
\[
X_\CO=\frac{N_\H2}{W_\CO}=\frac{M_\H2}{L_\CO}=0.76\frac{M(<R_\H2)}{M(<R_\CO)}\frac{M(<R_\CO)}{L_\CO},
\]
where $L_\CO$ is the $\CO$ luminosity and the factor 0.76 accounts for the presence of Helium in the cloud. Observations of $\CO$ bright regions indicate that the last factor is roughly independent of the cloud environment, i.e. metallicity and radiation field \citep{2008ApJ...686..948B}, but one should keep in mind that the observed scatter is large.
Hence, for clouds that contain $\CO$ optically thick sub-regions (for optically thin clouds the last factor is undefined and the method breaks) the X-factor should scale with $M(<R_\H2)/M(<R_\CO)$. We fix the normalization by enforcing that $X_\CO\sim{}X_{\CO,\MW}$ for a nucleon column density $N_{\rm n}=N_{\rm H}/0.76=10^{22}$ cm$^{-2}$ ($\sim{}80$ $M_\odot$ pc$^{-2}$) and $Z=Z_\odot$:
\begin{equation}
\label{eq:KW}
X_\CO=10^{20} \frac{M(<R_\H2)}{M(<R_\CO)} \,[\textrm{cm$^{-2}$ K$^{-1}$ km$^{-1}$ s}].
\end{equation}
The K/W model, defined by equations (\ref{eq:massRatio}), (\ref{eq:UMWovern}), and (\ref{eq:KW}), provides an estimate of $X_\CO$ as function
of $A_V$ and metallicity.

Fig. \ref{fig:XKW} shows the predictions of the K/W model. The X-factor depends primarily on visual extinction, i.e., at a fixed visual extinction the dependence on metallicity is relatively weak. Quantitatively, at $U_\MW=1$, it agrees nicely with the results from the ISM simulations by \cite{2011MNRAS.412..337G} over the visual extinction range $0.4\lesssim{}A_V\lesssim{}2$ and thus with the results of our model presented in Fig.~\ref{fig:FigXfactorModel} over such a range. A noticeably difference between the K/W model and our X-factor model appears at high visual extinction. Here, the K/W model predicts roughly constant X-factor at $A_V>3$, asymptotically approaching $\sim{}10^{20}$ cm$^{-2}$ K$^{-1}$ km$^{-1}$ s, while our model predicts that the X-factor should depend on metallicity (the precise scaling depends on assumptions about the $\CO$ line width). We note that the simulations by \cite{2011MNRAS.412..337G}  are also indicative of such a metal-dependent scaling (see also \citealt{2011MNRAS.tmp..943S}). The other main difference between the K/W model and our model is that the X-factor in the K/W model is not explicitly dependent on the strength of the interstellar radiation field, while in our model the X-factor varies significantly with $U_\MW$ at a fixed visual extinction, especially at $A_V\lesssim{}3$.

In order to provide $X_\CO$ as a function of metallicity alone a further, and crucial, assumption about the visual extinction of molecular clouds has to be made. \cite{2011ApJ...731...25K} assume (private communication) that the nucleon column density of all molecular clouds is $\sim{}7.5\times{}10^{21}$ cm$^{-2}$ and, therefore, that the visual extinction of molecular clouds scales with metallicity $A_V\propto{}Z$. Fig. \ref{fig:XKW} explains why this assumption leads to a very steep increase of the X-factor with decreasing metallicity. If the column densities of molecular clouds were, e.g., two times larger ($\sim{}1.5\times{}10^{22}$ cm$^{-2}$) a two times lower metallicity would be required to reach the same X-factor, resulting in a less steep $X_\CO-Z$ relation over the observed metallicity range $Z\sim{}0.1-1 Z_\odot$. Hence, the slope\footnote{Strictly speaking, the $X_\CO-Z$ dependence in the K/W model does not obey a power law, but it can be fit with such a relation over a limited range in metallicities.} of the $X_\CO-Z$ relation as predicted by the K/W model depends on assumptions about the average column densities of molecular clouds. 

Also, it is not entirely obvious why typical column densities of molecular clouds should not increase with decreasing metallicities in order to compensate to some extent for the loss in shielding. Observations of constant mass surface densities in $\CO$ bright clumps, such as found by \cite{2008ApJ...686..948B}, do not rule out this possibility since at low metallicity most of the column density may stem from CO dark molecular gas or even from atomic hydrogen surrounding those clumps. Consequently, if the typical visual extinction of molecular clouds detectable in CO does not scale linearly with metallicity, the X-factor will increase more gradually with decreasing metallicity. For instance, in the extreme case that all CO detectable clouds have similar visual extinction, the K/W model would predict that $X_\CO$ varies only very weakly with metallicity.

\section{Shielding functions}
\label{sect:ShieldingFunctions}

\begin{figure*}
\begin{center}
\begin{tabular}{cc}
\includegraphics[width=80mm]{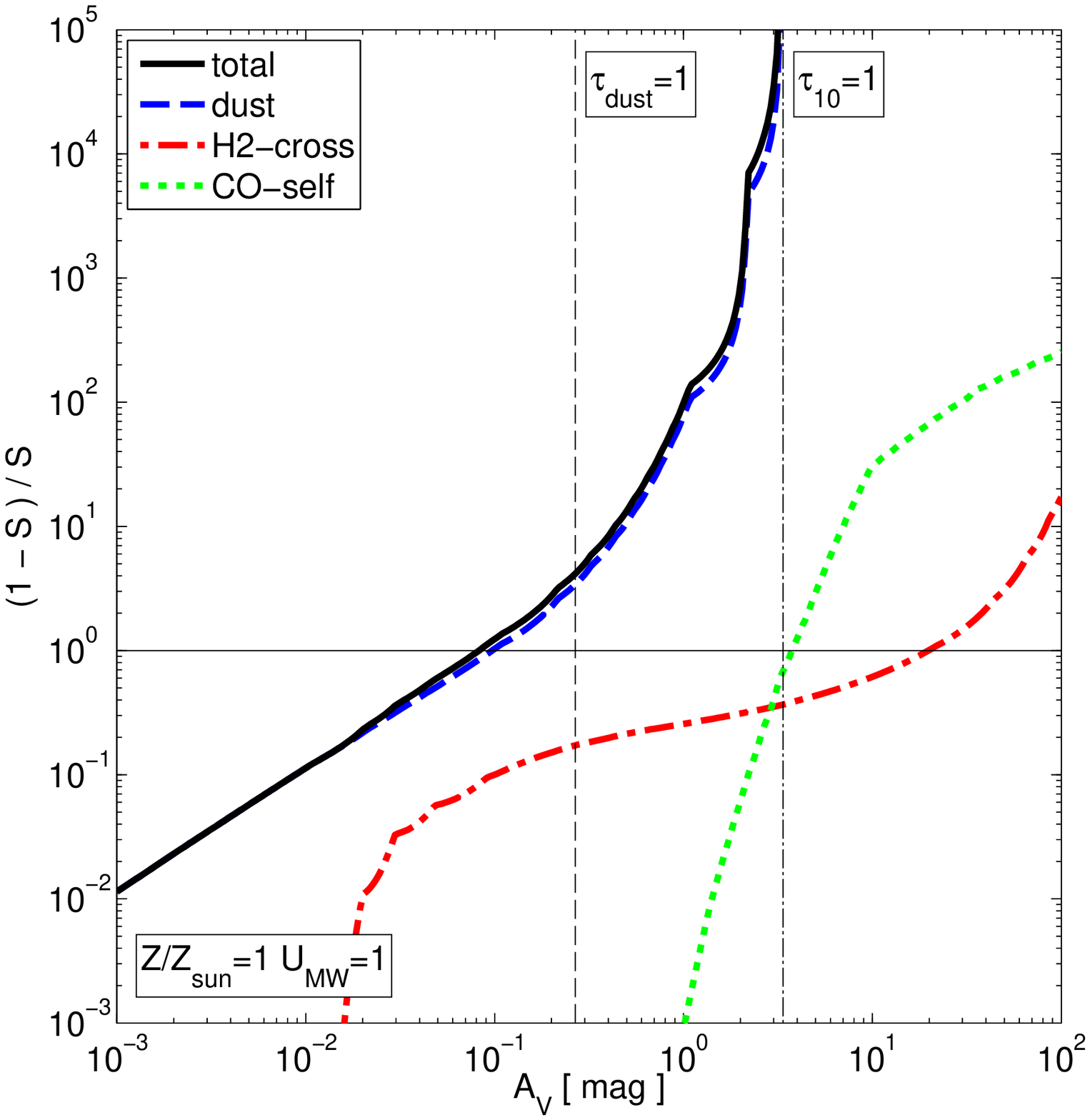} &
\includegraphics[width=80mm]{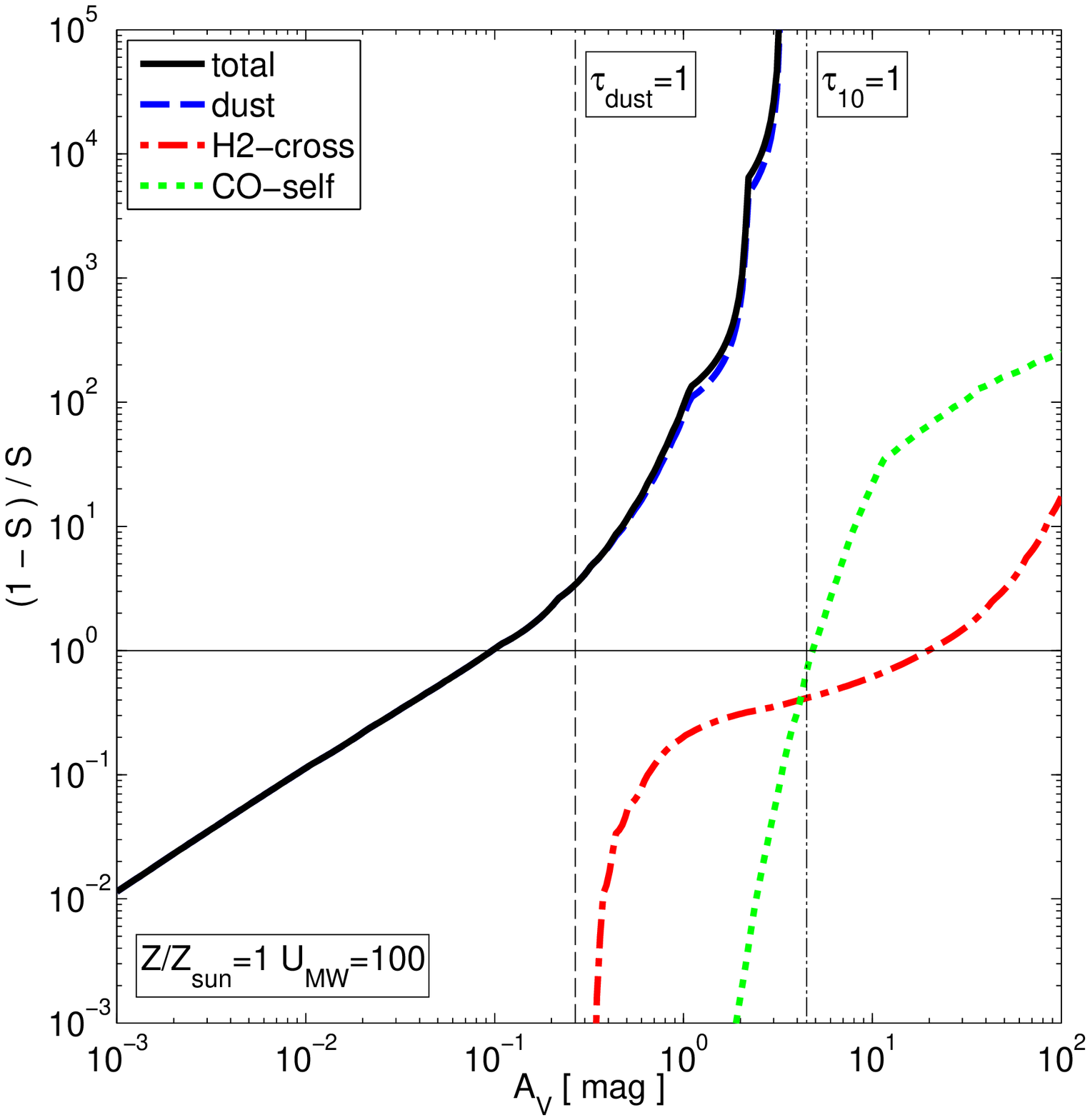} \\
\includegraphics[width=80mm]{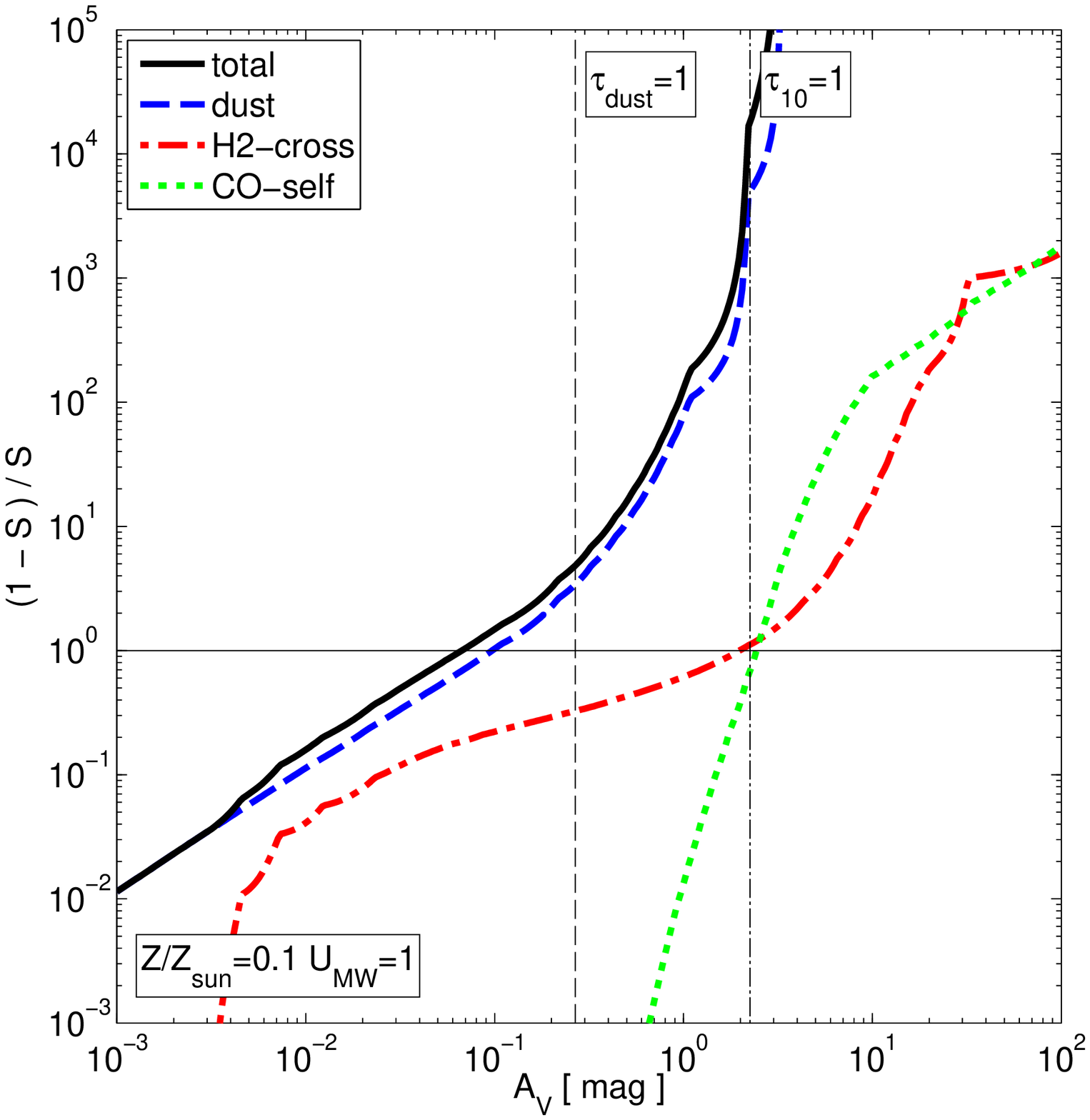} &
\includegraphics[width=80mm]{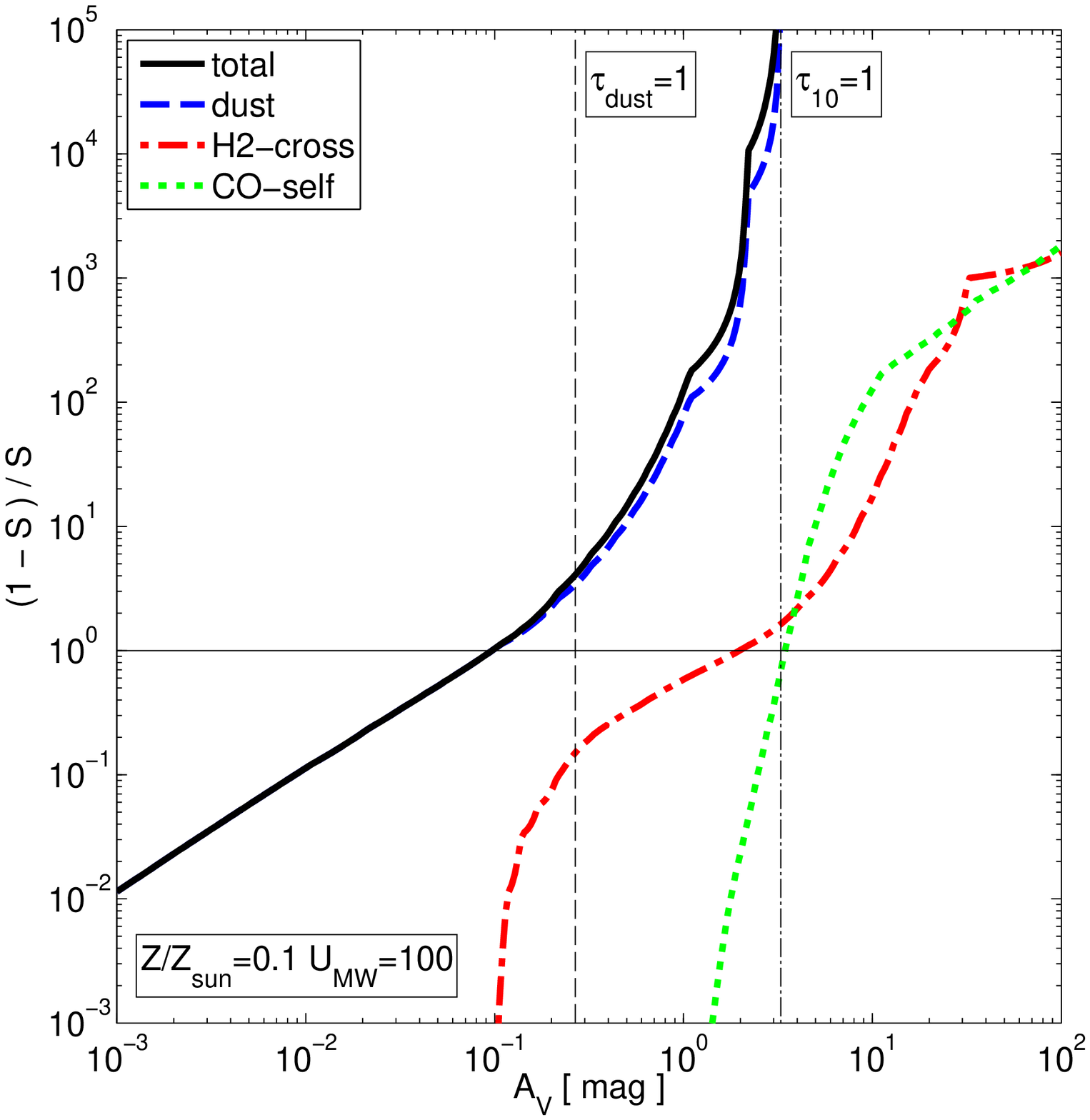} \\
\end{tabular}
\caption{$\CO$ shielding factors that enter the X-factor model, \S\ref{sect:Xmodel}, via equation (\ref{eq:AVUMW2}) for four different ISM conditions: (from top left to bottom right) $Z=Z_\odot$ \& $U_\MW=1$, $Z=Z_\odot$ \& $U_\MW=100$, $Z=0.1 Z_\odot$ \& $U_\MW=1$, and $Z=0.1 Z_\odot$ \& $U_\MW=100$. Shown are the shielding factors $S$ for dust shielding (dashed blue line), $\H2$ cross shielding (dot-dashed red line), $\CO$ self-shielding (dotted green lines) and the total shielding (product of the former three). $S=0$ (or $(1-S)/S=\infty$) corresponds to complete shielding from the UV interstellar radiation field, $S=1$ (or $(1-S)/S=0$) is the no shielding case. The horizontal solid black line indicates $S=0.5$. The vertical dashed (dot-dashed) line indicates the visual extinction that corresponds to a dust optical depth (optical depth in the CO J=1-0 line) of unity. The overall shielding of $\CO$ is dominated by dust shielding.}
\label{fig:ShieldFunc}
\end{center}
\end{figure*}

The X-factor has been measured by \cite{2011MNRAS.412..337G} for $U_\MW=1$, but not for any other non-vanishing radiation field. The model presented in \S\ref{sect:Xmodel} arrives at an estimate of the $\CO$ abundance for $U_\MW\neq{}1$ based on the assumption of photodissociation equilibrium of $\CO$. This  approach, equation (\ref{eq:AVUMW2}), takes various mechanisms into account that shield $\CO$ from the interstellar radiation field. Here we show these shielding factors explicitly for different ISM conditions and as a function of the mean visual extinction $A_V$. 

We compute for each $A_V$, $Z$, and $U_\MW$ the $\CO$ abundances using the model in \S\ref{sect:Xmodel}, and the $\H2$ abundances according to the fitting formula in \cite{2011ApJ...728...88G}. Gas densities are obtained from $A_V$ and $Z$ assuming a line-of-sight length of 20 pc. We then convert $\CO$ and $\H2$ abundances into shielding column densities by multiplying them with the coherence length $L_c=1$ pc. Finally, we use the tabulated shielding functions of \cite{1996A&A...311..690L} to calculate the corresponding shielding factor as function of $A_V$. Note that the $\CO$ self-shielding factor is only known a posteriori, i.e., after equation (\ref{eq:AVUMW2}) is solved for the $\CO$ abundance.

Fig.~\ref{fig:ShieldFunc} shows the different shielding contributions. For a sufficiently small coherence length ($L_c\lesssim{}1$ pc), the shielding is dominated by dust shielding.

\bibliographystyle{apj}

%\bibliography{paper1}

\end{document}